\documentclass[times]{MyArticle}

\usepackage[hyphens]{url} 
\usepackage[colorlinks,bookmarksopen,bookmarksnumbered,citecolor=blue,urlcolor=blue]{hyperref}

\usepackage[dvipsnames,svgnames,table]{xcolor}
\usepackage{tabularx}
\usepackage{algorithm}
\usepackage{algpseudocode}
\usepackage{amsthm}

\usepackage{graphicx}
\usepackage{comment}
\usepackage{tikz}
\usetikzlibrary{shapes.geometric, arrows, positioning, intersections, arrows.meta}
\usepackage{ circuitikz }
\usepackage{pgfplots}
\usepackage{amsmath}
\usepackage{float}
\usepackage{siunitx}
\usepackage{subcaption}
\makeatletter
\def\url#1{}
\def\biburl#1{}
\makeatother

\usepackage{lineno}
\pgfplotsset{compat=1.18}
\begin{document}

 
\title{Boundary treatment algorithms for meshfree RANS turbulence modeling}
 
\author{Mohan Padmanabha\affil{1,2}\corrauth, 
J\"org Kuhnert\affil{1}, 
Nicolas R. Gauger\affil{2}, 
Pratik Suchde\affil{1,3} } 

\address{\affilnum{1} Fraunhofer ITWM, Department "Grid-Free Methods", Division "Processes and Materials", Fraunhofer-Platz 1, Kaiserslautern, 67663, Germany \break %
\affilnum{2}Chair for Scientific Computing, RPTU University Kaiserslautern-Landau, Paul-Ehrlich-Strasse 34, Kaiserslautern, 67663, Germany \break %
\affilnum{3} Faculty of Science, Technology and Medicine, University of Luxembourg, Luxembourg}

\corraddr{E-mail: mohan.padmanabha@itwm.fraunhofer.de, pratik.suchde@gmail.com}

\begin{abstract}
In this paper, we propose improved wall-treatment strategies for meshfree methods applied to turbulent flows. The goal is to enhance wall-function handling in simulations of high-Reynolds-number turbulent flows and to better understand the performance of first-order turbulence models within these frameworks. While wall-function techniques are well established for mesh-based methods, their implementation in meshfree methods faces unique challenges. The main difficulties arise from scattered point distributions and dynamic point movement in Lagrangian frameworks. To address these issues, we evaluate a baseline closest-neighbor approach alongside two novel techniques: the nearest-band neighbor (NBN) method and the shifted boundary (SB) method. The NBN method enforces wall functions on a band of interior points, helping to maintain uniform point selection. On the other hand, the SB method virtually moves boundary points to a fixed wall-normal distance, eliminating the spatial noise associated with point movement. We evaluate these methods using first-order turbulence closures: Spalart--Allmaras, $k-\varepsilon$, and $k-\omega$ turbulence models. These methods are numerically validated on 1D Couette flow, a turbulent flat plate, and a 3D NACA 0012 airfoil at high Reynolds numbers. Results demonstrate that both novel methods significantly outperform the standard closest-neighbor approach on canonical flat geometries. For flat plates, the SB method provides superior stability and perfectly smooth $y^+$ distributions. However, when applied to the complex, curved NACA 0012 profile, the NBN method proves to be significantly more robust and geometrically flexible, accurately predicting skin friction and pressure coefficients. In contrast, the SB method exhibits setbacks in numerical diffusion and premature flow separation on curved geometries. This is due to uncorrected normal-vector shifting and adverse pressure gradients. This work establishes the NBN method as a reliable, robust foundation for simulating turbulent flows over practical aerodynamic bodies using meshfree methods, while extending the highly stable SB framework to curved geometries in future research and showcasing a foundation for simulating wall-bounded turbulent flows at high Reynolds numbers using meshfree collocation methods. 
\end{abstract}

 
\maketitle
 
\vspace{-6pt}
 

\section{Introduction}\label{sec1}
When simulating turbulent flows close to walls, there is always a trade-off between accuracy and computational cost. Detailed approaches like Direct Numerical Simulation (DNS) or Large Eddy Simulation (LES) can capture all or most turbulence scales, but they are far too computationally demanding for most engineering applications. A more practical alternative is the Reynolds-Averaged Navier--Stokes (RANS) method, which, when combined with suitable turbulence models, offers a good balance between cost and accuracy. To handle the complexity near walls and avoid resolving the boundary layer directly, traditional wall functions are often used in mesh-based methods \cite{marusic_wall-bounded_2010}. These functions help bridge the gap between the unresolved boundary layer and the rest of the flow by relating the wall shear stress to the velocity at the first computational cell or point away from the wall~\cite{kalitzin_near-wall_2005}.

Conventional CFD methods rely on structured or unstructured meshes to discretize the computational domain. However, generating high-quality meshes for complex geometries can be time-consuming and prone to errors. Managing moving boundaries, especially in Lagrangian simulations, is particularly challenging as it often requires frequent remeshing, which can affect both the accuracy and efficiency of the simulation~\cite{liu_introduction_2010}.

Meshfree methods have emerged as an attractive alternative to conventional mesh-based approaches. Instead of fixed grids, these approaches use scattered points with no topological connection to discretize the computational domain \cite{belytschko_meshless_1996,nguyen_meshless_2008,chen_meshfree_2017}. The point cloud-based approach enables automatic generation and adaptation of the domain discretization \cite{suchde_point_2023}. The Lagrangian point cloud approach allows dynamic changes in the domain to be captured naturally \cite{suchde_fully_2019}. 
Meshfree methods provide distinct advantages over conventional mesh-based approaches, particularly when dealing with free surface flows and multiphase phenomena. Meshfree methods can handle large deformations and complex interfaces dynamically without the requirement of mesh regeneration \cite{patel_meshless_2020,monaghan_smoothed_2012,nguyen_meshless_2008}. Ongoing research continues to enhance meshfree formulations, including improved strategies for velocity-pressure coupling~\cite{matsuda_least-squares_2025}. Recent advancements in meshfree methodologies have allowed diverse multi-physics applications, including free surface interfacial flows \cite{gotoh_state---art_2018}, multiphase flows \cite{hu_multi-phase_2006}, cavitation prediction \cite{lyu_towards_2023}, and fluid-structure interaction \cite{zhang_efficient_2022}. 

Significant progress has been made in meshfree methods to incorporate complex viscous and turbulent flow models. For instance, vortex particle methods have been applied to investigate flow bifurcations and wake dynamics past stalled airfoils \cite{rossi_multiple_2018}. Comparative studies have shown the capabilities of different particle models, such as Smoothed Particle Hydrodynamics (SPH) and Diffused Vortex Hydrodynamics, for viscous flow simulations \cite{colagrossi_particle_2016}. Despite these advancements, most work remains restricted to low-Reynolds-number regimes \cite{violeau_numerical_2007}. Recent developments have demonstrated LES implementations using meshfree methods with appropriate filtering techniques \cite{mayrhofer_dns_2015,arai_large_2013,meringolo_large_2023,antuono_smoothed_2021}. However, the high computational costs of these approaches often limit their practical feasibility for high-Reynolds-number flows.

A critical research gap exists in simulating high-Reynolds-number turbulent flows with meshfree methods, with only a limited number of studies attempting to address this challenge \cite{naghian_numerical_2017,nakayama_wall-layer_2022,bao_smoothed_2023}. Within the SPH community, recent efforts have attempted to bridge this gap by coupling the $k-\varepsilon$ turbulence model with wall functions \cite{bao_smoothed_2023}. Additionally, other studies have proposed near-wall compensation schemes to improve boundary consistency in RANS models \cite{wang_weakly_2025}. However, these developments remain predominantly limited to the integral-based SPH framework, leaving the challenge largely unaddressed for meshfree collocation approaches. In particular, the main limitation lies in boundary layer treatment and wall function implementation. Similar challenges in boundary treatment have been observed in other non-conforming methods, where improper handling can lead to excessive diffusion near walls \cite{stoter_nitsches_2021}. This challenge arises from the fundamental differences between meshfree and mesh-based discretization approaches. While traditional mesh-based methods provide a well-defined wall-adjacent cell or element, meshfree methods require new strategies for identifying wall-adjacent points on which boundary conditions and wall functions can be enforced. In this paper, we address this critical gap in wall boundary treatment for meshfree simulations of turbulent flows, particularly at high Reynolds numbers. Our study also examines the influence of different turbulence models when combined with various wall-treatment methods. 

Our main goal is to design, implement, and validate novel wall-treatment methodologies specifically tailored for meshfree collocation approaches within a Lagrangian URANS framework. This study examines three distinct methods for wall function implementation: (i) The first is a closest neighbor approach, which serves as the baseline method for selecting wall-adjacent points. This provides the closest analogy to how wall functions are handled in mesh-based methods. (ii) The second is a novel nearest band neighbor method, designed to select a band of interior neighbors based on their distance from the wall. (iii) The third is a novel shifted boundary method, which virtually shifts the boundary points to a specified distance from the wall. These three boundary treatments are evaluated and compared using the Spalart--Allmaras, $k-\varepsilon$, and $k-\omega$ turbulence models. Validation is performed using established benchmark cases, including flow over a flat plate and flow around a wing with a NACA 0012 airfoil.

All the methods presented in this paper are implemented within the in-house MESHFREE commercial code \cite{MESHFREE}, which is based on the Generalized Finite Difference Method (GFDM). The meshless GFDM approach considered here uses scattered, non-uniformly distributed point clouds without any background mesh structure. All simulations, except for the 1D Couette flow case, are performed in 3D, making the methods suitable for a wide range of industrial applications.

The remainder of this paper is organized as follows: Section 2 presents the preliminary theoretical background on the Generalized Finite Difference Method (GFDM), flow governing equations, turbulence models, and traditional wall functions. Section 3 explains the implementation of the three wall-treatment methods, including the novel nearest band neighbor method and the shifted boundary method. Section 4 describes the validation test cases and presents results, comparing the performance of the different wall-treatment methods across various turbulence models. Finally, Section 5 presents the conclusions drawn from this research.

\section{Preliminaries}\label{sec2}
In computational fluid dynamics, the momentum equation is a fundamental principle for describing fluid motion. This paper focuses on the Reynolds-Averaged Navier--Stokes (RANS) equations, which utilize turbulence models based on the eddy viscosity concept to represent turbulent flow behavior. This section discusses two modeling techniques. Low Reynolds number models effectively resolve the dynamics within the viscous sub-layer of the boundary layer, but are computationally demanding. Conversely, high Reynolds number models use wall functions to link the near-wall region with the outer flow, enabling coarser discretization. This approach reduces computational costs while maintaining accuracy in predictions of turbulent boundary layers, which is the main focus of this paper. Additionally, the Generalized Finite Difference Method (GFDM) offers a flexible framework for discretizing the governing equations, including the momentum equation, without the constraints of structured mesh generation. This flexibility facilitates the effective analysis of complex geometries and flow conditions. These aspects will be explored below. 

\subsection{GFDM discretization}
\label{sec:GFDM}
The simulations and analyses this work are carried out using the Lagrangian mesh-free method based on the Generalized Finite Difference Method (GFDM) \cite{liszka_finite_1980}. The simulation domain is decomposed into $N$ numerical points (also referred to as particles), which include points in the interior and on the boundary of the domain. Point placement is based on the advancing front technique \cite{lohner_advancing_1998, suchde2023point}. Point density (or refinement) depends on a smoothing function called the resolution $h$. Points are scattered throughout the domain, and the density of these points is determined by three parameters: $r_{\mathrm{min}}$, $h$, and $r_{\mathrm{max}}$. To avoid point clustering, no two points are allowed within a distance of $r_{\mathrm{min}}h$ from each other. Furthermore, it is ensured that there is at least one point within a sphere of radius $r_{\mathrm{max}}h$. We refer to \cite{suchde_fully_2019, suchde2023point} for details on how these inter-point distances are enforced.

GFDM is a generalized finite difference approach, where the derivative of a function $f$ at a point $i$ is defined as a linear combination of function values at neighboring points \cite{suchde_meshfree_2018, halada2025overview} :
\begin{equation*}
    \partial^* f (\vec{x_i}) \approx \tilde{\partial}^*_i f = \sum_{j \in S_i} c^*_{ij} f_j \,.  
\end{equation*}
Here $i$ is the index of the point where the derivative is being computed, and $j$ is the index of neighboring points of point $i$. Here, '$*$' represents the differential operator being applied, such as $x, y, \Delta$. The symbol $\partial^*$ denotes the continuous derivative and $\tilde{\partial}^*$ represents the discrete derivative at $i$. The stencil coefficients $c^*_{ij}$ are computed using the weighted least squares approximation \cite{halada2025overview}.

\subsection{Navier--Stokes equation}
Consider a fluid flow governed by the conservation equations of mass and momentum:
\begin{align}
\frac{\partial \rho \vec{u}^*}{\partial t} + \nabla \cdot (\rho \vec{u}^*) &= 0 \,,
\label{eq:mass_conservation} \\
\frac{\partial \rho \vec{u}^*}{\partial t} + \nabla \cdot (\rho \vec{u}^* \otimes \vec{u}^{*}) &= -\nabla p + \nabla \cdot \boldsymbol{\sigma} + \vec{g} \,. 
\label{eq:momentum_conservation}
\end{align}
Here, $\vec{u}^*$ is the instantaneous velocity, $p$ is the pressure, $\vec{g}$ represents external forces, and $\boldsymbol{\sigma}$ is the stress tensor, defined by: 
\begin{equation}
    \boldsymbol{\sigma} = \lambda (\nabla \cdot \vec{u}^*) \delta_{ij} + 2 \nu \rho \mathbf{S}^*\,,
    \quad\text{where}\quad
    \mathbf{S}^* = \frac{1}{2} \left( \nabla \vec{u}^* + \nabla \vec{u}^{*T} \right)\,. 
\end{equation}
Here, $\lambda$ is the volume viscosity, and $\nu$ is the kinematic viscosity. The strain tensor $\mathbf{S}^*$ is computed using the velocity $\vec{u}^*$.

The instantaneous velocity $\vec{u}^*$ can be decomposed into mean velocity $\vec{U}$  and fluctuating velocity $\vec{u}$ as $\vec{u}^* = \vec{U} + \vec{u}$. By applying Reynolds decomposition and averaging the quantities, we derive the Reynolds-Averaged Navier--Stokes (RANS) equation:
\begin{equation}
    \frac{D \vec{U}}{D t} = -\frac{1}{\rho} \nabla p + \nu \nabla^2 \vec{U} - \nabla \cdot \overline{\vec{u} \vec{u}^{T}} + \vec{g}\,.
\end{equation}
Here, $\frac{D}{D t}$ is the total derivative, given by $\frac{D}{D t} = \frac{\partial}{\partial t} + \vec{U} \cdot \nabla$.
The averaged components are denoted by $\overline{\cdot}$, when performed over the decomposed Navier--Stokes equations, gives rise to the Reynolds stress tensor $\overline{\vec{u} \vec{u}^{T}}$, which introduces the closure problem. To address this closure problem, Boussinesq's eddy viscosity assumption \cite{boussinesq_essai_nodate} is used, where momentum transfer from the turbulent eddies is modeled using eddy viscosity:
\begin{equation}
    \overline{\vec{u} \vec{u}^{T}} = 2 \nu_{\mathrm{t}} \mathbf{S} + \frac{2}{3} k \delta_{ij}\,.
    \label{eq:EddyViscosityModel}
\end{equation}
Here, $\nu_{\mathrm{t}}$ is the turbulent viscosity, which can be obtained using different methods such as algebraic model \cite{prandtl_7_1925} or first-order turbulence models \cite{wilcox_formulation_2008, aupoix_extensions_2003, chien_predictions_1982}. $\mathbf{S}$ is the strain tensor computed using the mean velocity $\vec{U}$, and $k$ is the kinetic energy. 

\subsection{Turbulence models}
\label{sec:TurbulenceModels}
Using the eddy viscosity model, Eq.\, \eqref{eq:EddyViscosityModel}, the RANS equation can be closed. However, the turbulent viscosity $\nu_{\mathrm{t}}$ must be computed. In this paper, we consider three first-order eddy viscosity models: the one-equation Spalart--Allmaras (SA) turbulence model and the two-equation $k-\varepsilon$ and $k-\omega$ turbulence models, which will be used to compute turbulent viscosity.

\subsubsection{Turbulence model Spalart--Allmaras}\hfill

The Spalart--Allmaras model (SA) uses one transport equation, for $\tilde{\nu}$, which is then used to compute the turbulent viscosity $\nu_{\mathrm{t}}$. Many variations of the original SA model \cite{spalart_one-equation_1992} have been proposed. Here, we consider the model without the transition term \cite{aupoix_extensions_2003}, given by:
\begin{equation}
    \frac{D \tilde{\nu}}{D t} = C_{\mathrm{b}1} \tilde{S} \tilde{\nu} - \left(C_{\mathrm{w}1} f_{\mathrm{w}}\right)\left(\frac{\tilde{\nu}}{d}\right)^2 + \frac{1}{\sigma} \nabla \cdot \left((\nu + \tilde{\nu}) \nabla \tilde{\nu}\right) + \frac{C_{\mathrm{b}2}}{\sigma} (\nabla \tilde{\nu})^2\,.
    \label{eq:spalart_allmaras}
\end{equation}
Here, $\tilde{S}$ is computed using the rotation and strain tensors, $C_{\mathrm{b}1}$, $C_{\mathrm{w}1}$, $\sigma$, and $C_{\mathrm{b}2}$ are the model constants, and $f_{\mathrm{w}}$ is a coefficient that depends on $\tilde{S}$ and the distance from the nearest boundary $d$. The turbulent viscosity $\nu_{\mathrm{t}}$ required for the eddy viscosity model is obtained from:
\begin{equation}
    \nu_{\mathrm{t}} = \tilde{\nu} f_{\mathrm{v1}}.
\end{equation}
Here, $f_{\mathrm{v1}}$ is a scalar factor used to scale the values of $\tilde{\nu}$, and the value of $\tilde{\nu}$ approaches $0$ at the wall.

\subsubsection{Turbulence model \texorpdfstring{$k-\varepsilon$}{k-epsilon}}\hfill

Originally proposed by Chien \cite{chien_predictions_1982}, the $k-\varepsilon$ turbulence model is similar to the SA model, but with two transport equations: one each for the kinetic energy $k$ and dissipation rate $\varepsilon$:
\begin{align}
\frac{D k}{D t} &= P - \varepsilon + \nabla \cdot \left( \rho \left( \nu + \frac{\nu_{\mathrm{t}}}{\sigma_k} \right) \nabla k \right) \,, \\
\frac{D \varepsilon}{D t} &= C_{\varepsilon1} f_1 \frac{\varepsilon}{k} P - C_{\varepsilon2} f_2 \frac{\varepsilon^2}{k} + \nabla \cdot \left( \rho \left( \nu + \frac{\nu_{\mathrm{t}}}{\sigma_{\varepsilon}} \right) \nabla \varepsilon \right) \,.
\end{align}
Here, $P$ is the production term computed using $P = \rho \nu_{\mathrm{t}} S^2$. The model-based coefficients are $\sigma_k$, $C_{\varepsilon1}$, $C_{\varepsilon2}$,  $f_2$, and $\sigma_{\varepsilon}$. The turbulent viscosity $\nu_{\mathrm{t}}$ is obtained from the kinetic energy $k$ and dissipation rate $\varepsilon$:
\begin{equation}
    \nu_{\mathrm{t}} = C_{\mu} \frac{k^2}{\varepsilon}\,.
\end{equation}
The values of $k$ and $\varepsilon$ are scaled using the coefficient $C_{\mu} = 0.09$ \cite{chien_predictions_1982}. The kinetic energy $k$ and the dissipation rate $\varepsilon$ approach $0$ at the wall.

\subsubsection{Turbulence model \texorpdfstring{$k-\omega$}{k-Omega}}\hfill

The second two-equation eddy viscosity model considered here is $k-\omega$. Several variations of the original model have been proposed. The model that we have considered here is based on \cite{wilcox_formulation_2008}. The transport equations for the kinetic energy $k$ and specific dissipation rate $\omega$ are given by:
\begin{align}
\frac{D k}{D t} &= P - \beta^* \omega k + \nabla \cdot \left( \rho \left( \nu + \sigma_k \frac{k}{\omega} \right) \nabla k \right)\,, \\
\frac{D \omega}{D t} &= \frac{\gamma \omega}{k} P - \beta \omega^2 + \nabla \cdot \left( \rho \left( \nu + \sigma_{\omega} \frac{k}{\omega} \right) \nabla \omega \right) + \frac{\sigma_{\mathrm{d}}}{\omega} \nabla k \cdot \nabla \omega\,.
\end{align}
Here, similar to the $k-\varepsilon$ model, $P$ is the production term; and $\beta^*$, $\beta$, $\sigma_k$, $\sigma_{\omega}$, and $\sigma_{\mathrm{d}}$ are the model coefficients. The turbulent viscosity is obtained from $k$ and $\omega$:
\begin{equation}
    \nu_{\mathrm{t}} = \frac{k}{\omega} \,.
    \label{eq:turbulent_viscosity}
\end{equation}
At the wall, the kinetic energy is $k_{\mathrm{w}} = 0$ as in the $k-\varepsilon$ model. The specific dissipation rate $\omega$ tends to infinity as it approaches the wall. This boundary condition cannot be specified for numerical implementation; hence, the proposed boundary condition \cite{wilcox_formulation_2008} is:
\begin{equation}
    \omega_{\mathrm{w}} = \frac{6 \nu_{\mathrm{w}}}{\beta_0 d^2} \,.
\end{equation}
Here, $\beta_0 = 0.0708$; $\nu_{\mathrm{w}}$ is the viscosity at the wall, and $d$ is the distance between the wall and the closest interior point.

\subsection{Wall-bounded turbulent flow and wall functions}
\label{sec:WallBoundedTurbulentFlow}
Many industrial applications involve wall-bounded flow, where turbulence generated by the wall has a significant influence on the flow profile. The velocity variations from the no-slip boundary to the free-stream flow are large, resulting in high velocity gradients, especially close to the wall. To resolve this flow numerically, a finer discretization is required close to the wall. In RANS simulations, resolving the viscous sublayer directly using fine discretization corresponds to low-Reynolds-number formulations. While the boundary layer can be captured accurately with this approach, the finer discretization is computationally expensive. To avoid high computation costs and to use coarser discretization close to the boundary, wall corrections are performed to determine approximate values using the laws of the wall. This approach of using wall functions is associated with high-Reynolds-number RANS models.

\subsubsection{The laws of the wall}\label{sec:LawsofWall}\hfill

The flow close to the no-slip wall in the boundary layer can be categorized into three different layers based on the behavior of the flow \cite{prandtl_7_1925, tennekes_first_1972}, known as the laws of the wall. The type of layer is determined based on the dimensionless wall-normal coordinate $y^+$ and the dimensionless wall velocity $U^{+}$. These quantities, $y^{+}$ and $U^{+}$ are computed using the friction velocity $U_{\tau}$, which is in turn computed using the wall shear stress $\tau_{\mathrm{w}}$.
\begin{equation}
y^{+}=\frac{y_{\mathrm{p}}U_{\tau}}{\nu}  
   \quad\text{and}\quad 
U^{+}= \frac{U_{\mathrm{p}}}{U_{\tau}} 
\quad\text{where}\quad
U_{\tau} = \sqrt{ \frac{\tau_{\mathrm{w}}}{\rho_{\mathrm{w}}} }\,. 
\end{equation}
Here, $\nu$ and $\rho_{\mathrm{w}}$ are the kinematic viscosity and the density of the fluid respectively, and $U_{\mathrm{p}}$ is the tangential velocity computed using $U_{\mathrm{p}} = \|U_{\mathrm{rel}} - (U_{\mathrm{rel}} \cdot \mathbf{n})\,\mathbf{n}\|$ by subtracting the wall-normal projection and and $U_{\mathrm{rel}}$ is the relative velocity obtained by substracting the wall velocity $U_{\mathrm{w}}$ and $y_{\mathrm{p}}$ represent the velocity and the distance from the wall at the location of interest in the fluid domain.

The three layers comprising the wall region are the viscous sub-layer, buffer layer, and inertial sub-layer (or log layer). The viscous sub-layer ranges from $0 < y^{+} < 5$, where the turbulent effects are negligible compared to the viscous effects. The viscous sub-layer follows the relation $U^{+}=y^{+}$. The buffer layer ranges from $5 < y^{+} < 30$, where turbulent and viscous effects both vary significantly, making it difficult to derive a generalized model for turbulence. The inertial sub-layer, or log layer, ranges from $30 < y^{+}$ to the end of the boundary layer. In the log layer, the turbulent effects are dominant, and the viscous effects are negligible. The asymptotic relation of $y^{+}$ and $U^{+}$ exhibits logarithmic behavior:    
\begin{equation}
    U^{+}= \frac{1}{\kappa} \ln(E y^{+}) \,. 
    \label{eq:LogBehavior}
\end{equation}
where $\kappa=0.41$ is the von Kármán constant, and $E=9.8$.  Due to the logarithmic relation, additional modifications (wall functions) are required to capture the appropriate behavior of the flow near the wall.

\section{Methodology}
\label{sec:Methodology}
Turbulence behavior in 3D simulations using meshfree collocation methods has received limited attention. The aim of this paper is to examine the behavior of first-order turbulence models, Spalart--Allmaras, $k-\varepsilon$, and $k-\omega$ in the mesh-free collocation architecture, especially with the aspect of high Reynolds number models, with the implementation of wall functions.

Wall functions and their implementation in mesh-based methods are well established, where the modification to the equation is carried out for the first internal points/cells/elements attached to the boundary. Selecting these points is straightforward, and wall function application has been studied extensively  \cite{launder_numerical_1974, WallFunction_craft_2002,Bredberg_Wall_2000}. In meshfree methods,  the location where the wall function should be applied is not obvious. In this work, we propose and analyze three approaches to applying the wall function. 

In each approach, for each boundary point, either a single interior point or a set of interior points is selected for wall function application. Wall function modifications are based on the distance from the first interior point to the wall. The computed corrections are then incorporated into the momentum and turbulence equations. 

To determine the velocity correction, the respective $y^{+}$ and $U^{+}$ must be computed at the first interior point's distance from the wall. To compute $y^{+}$ at that location, $U_{\tau}$ is required, which is also unknown. To obtain $y^{+}$,  we iterate the following equations until convergence is achieved. 
\begin{equation}
    U_{\tau}= \frac{U_{\mathrm{p}}\kappa}{\ln(Ey_{\mathrm{p}}^{+})}  \,
   \quad\text{where $y_{\mathrm{p}}^{+}$ is computed using:}\quad 
    y_{\mathrm{p}}^{+} = \frac{y_{\mathrm{p}}U_{\tau}}{\nu} \,.
    \label{eq:UtauandYplusCompute}
\end{equation}
Here, $y_{\mathrm{p}}$ and $U_{\mathrm{p}}$ represent the distance and the velocity magnitude at the first interior point $p$ from the wall. The constants are $E=9.8$ and $\kappa = 0.41$, and $\nu$ is the kinematic viscosity of the fluid. $y_{\mathrm{p}}^{+}$ is the $y^{+}$ at the location $\mathrm{p}$. If the turbulent kinetic energy $k$ is available, we initialize $U_{\tau}$ using $U_{\tau} = C_{\mu}^{1/4} \sqrt{k_{\mathrm{p}}}$ and iterate until convergence is achieved. Based on the obtained $y_{\mathrm{p}}^{+}$ value, the effective wall shear stress $\tau_{\mathrm{w}}$ is computed using:
\begin{equation}
    \tau_{\mathrm{w}} = \frac{\rho U_{\tau} U_{\mathrm{p}} \kappa}{\ln(Ey^{+})} \,. 
    \label{eq:WallShearStress}
\end{equation}
Here, the shear stress $\tau_{\mathrm{w}}$ is added as a source term to the momentum equation. The correction can also be made by computing the effective wall viscosity $\nu_{\mathrm{w}}$ using:
\begin{equation}
    \nu_{\mathrm{w}} = \nu \left( \frac{y^{+}\kappa}{\ln(Ey^{+})} -1 \right) \,.
    \label{eq:WallViscosityCompute}
\end{equation}
Wall functions must be adopted based on where the first interior point next to the boundary in the turbulent boundary layer is located. If the point lies in the viscous layer, no wall function is used; if the point lies in the log layer, the wall function defined above is used. If the point lies in the buffer layer, there is no single solution available; hence, different strategies can be applied \cite{kader_temperature_1981} \cite{launder_numerical_1974}. The standard method \cite{launder_numerical_1974} is to use a switching strategy where points up to the midpoint of the buffer region are treated with the viscous law, requiring no wall function. If the point $p$ lies above the midpoint of the buffer layer, then the log layer function is used. Another strategy is the blending function \cite{kader_temperature_1981}, which provides a smooth transition from the viscous to the log region. In this paper, we use Reichardt’s wall function \cite{reichardt_vollstandige_1951}:
\begin{equation}
			U^+ = \frac{1}{\kappa}\ln(1+\kappa y^+) + 7.8\left(1 - \exp\left(-\frac{y^+}{11}\right) - \frac{y^+}{11}\exp\left(-\frac{y^+}{3}\right)\right) .
            \label{eq:Reichardts_law}
\end{equation}
Reichardt's wall function smoothly blends all three regions of the wall law into a single continuous expression, eliminating the need to switch between different equations.

\subsubsection*{Turbulence models wall treatment }~\\
\label{sec:TurbulenceWallFunction}
Along with the momentum equation, the turbulence models must also be modified. Wall functions have been studied extensively, especially in mesh-based methods \cite{tennekes_first_1972} \cite{kalitzin_near-wall_2005}.

The wall function for Spalart--Allmaras is straightforward, as the modifications apply only to the momentum equation and not to the Spalart--Allmaras transport equation. The modification to the momentum equation follows the approach shown in the section \ref{sec:LawsofWall} based on the location of the first interior point $p$.

With two-equation turbulence models $k-\omega$ and $k-\varepsilon$, modifications are applied to $k_{\mathrm{p}}$ and $\omega_{\mathrm{p}}$ or $\varepsilon_{\mathrm{p}}$ at interior points $\mathrm{p}$ close to the boundary selected for applying wall function. 

\subsubsection*{The standard wall function}~\\
For the standard wall function \cite{launder_numerical_1974}, the modification is carried out by setting the value of the turbulent kinetic energy $k_{\mathrm{p}}$ directly on the first interior point $\mathrm{p}$ based on the computed shear velocity $U_{\tau}$ and the coefficient $C_{\mu}$. The value of $U_{\tau}$ is computed based on the location of the first interior point from the boundary $\mathrm{p}$ as:
\begin{equation}
k_{\mathrm{p}} = \frac{U_{\tau}^{2}}{\sqrt{ C_{\mu} }}. 
\end{equation}
The values of $\omega_{\mathrm{p}}$ and $\varepsilon_{\mathrm{p}}$ at the first interior point are modified based on the value of $k_{\mathrm{p}}$. The modifications performed on the equations of $\omega_{\mathrm{p}}$ and $\varepsilon_{\mathrm{p}}$ are: 
\begin{equation}
\omega_{\mathrm{p}} = \frac{\sqrt{ k_{\mathrm{p}} }}{C_{\mu}^{1/4}\kappa y_{\mathrm{p}}} \quad \text{and}\quad  \varepsilon_{\mathrm{p}} = \frac{C_{\mu}^{3/4} k_{\mathrm{p}}^{3/2}}{\kappa y_{\mathrm{p}}}. 
\end{equation}
Where $C_{\mu} = 0.09$, $\kappa = 0.41$, and the values of $k_{\mathrm{p}}$ and $y_{\mathrm{p}}$ at the first interior point $p$ are used. 

\subsubsection*{The Launder--Spalding wall function}~\\
The other wall function methodology considered is the Launder--Spalding wall function \cite{launder_numerical_1974}. 
Here, the value of $k_{\mathrm{p}}$ is not directly changed at the first interior point, as in the standard wall-function method. Instead, the production term in the transport equation for $k_{\mathrm{p}}$ is modified to obtain the production-dissipation relation as:
\begin{equation} 
P_{k_\mathrm{p}} = U_{\tau}^{2} \frac{U_{\tau}}{\kappa y_{\mathrm{p}}}. 
\label{eq:ProductionForKequation}
\end{equation}
Where $P_{k_\mathrm{p}}$ is the production term in the transport equation for turbulent kinetic energy $k_{\mathrm{p}}$. For $\omega_{\mathrm{p}}$ or $\varepsilon_{\mathrm{p}}$, the values are directly computed based on the value of $k_{\mathrm{p}}$ at the first interior point, similar to the standard wall function. 

\subsubsection*{Implementing wall functions in meshfree methods}~\\
In this paper, we propose three methods to incorporate wall functions and boundary treatment for meshfree collocation methods. The three methods differ in how the first interior points from the boundary are selected to apply the wall function.

Without sufficient resolution to resolve the boundary layer, wall functions must be applied for the velocity correction at the first interior point(s) adjacent to the wall. To enforce wall function corrections, we must select the first interior point(s). Due to the presence of scattered points in meshfree collocation methods, the choice of the first interior point is ambiguous, as there is no topology or any direct connection to the boundary. A second challenge is that the points move with the flow (in a Lagrangian sense); hence, point selection must be performed at each time step. We consider the closest neighbor method and two novel methods proposed in this paper to address the problems in meshfree methods. The first method adapts the classical mesh-based approach to the mesh-free situation by choosing the closest neighbor to the boundary point. The two novel methods discussed are the nearest-band neighbor method and the shifted-boundary method. Each method is evaluated with meshfree collocation using the turbulence models discussed earlier.

\textbf{Remark: } In several figures below, the distance between boundary points and the closest interior points is exaggerated for ease of visualization.

\subsection{Closest neighbor (CN) method}
\label{sec:ClosestNeighbor}

The first method is based on the approach used in mesh-based methods \cite{tennekes_first_1972}. Here, for each boundary point, we apply the wall function to its closest interior point, see Figure~\ref{fig:CN_illustration}. First, we flag the boundaries where the wall functions are required. For each boundary point on the flagged boundaries, we select the closest interior point for wall function application. We repeat this process at each time step. Since the points move with the flow in a Lagrangian sense, the closest neighbor points must be recomputed at every time step. For the boundary points used to select interior points for the wall function, we change their boundary condition from a no-slip condition to a Neumann (zero-gradient) condition.

\begin{figure}[!htb]
    \centering
    \begin{minipage}[b]{0.45\textwidth}
        \centering
            \begin{tikzpicture}
                \pgfmathsetseed{1213} 
                
                \draw[thick, name path=curve] (0,0) to[out=30,in=150] (4,0);
                
                \foreach \x in {0,1,...,9} {
                    \pgfmathsetmacro{\t}{\x/9} 
                    \path[name path=vert] ({\t*4}, -1) -- ({\t*4}, 1); 
                    \draw [name intersections={of=curve and vert, by=p}, red, fill=red!50] (p) circle (2pt);
            
                    \pgfmathsetmacro{\offset}{rand*0.2 + 0.8} 
                    \path[blue, fill=blue!50] ($(p) + (rand*0.05, \offset)$)  coordinate (bp\x) circle (2pt);
            
                    \pgfmathsetmacro{\offset}{rand*0.22 + 1.3} 
                    \path[blue, fill=blue!50] ($(p) + (rand*0.05, \offset)$) circle (2pt);
            
                    \pgfmathsetmacro{\offset}{rand*0.27 + 1.8} 
                    \path[blue, fill=blue!50] ($(p) + (rand*0.05, \offset)$) circle (2pt);
            
                    \pgfmathsetmacro{\offset}{rand*0.21 + 2.3} 
                    \path[blue, fill=blue!50] ($(p) + (rand*0.05, \offset)$) circle (2pt);
                }
            \end{tikzpicture}
    \end{minipage}
    \hspace{0.05\textwidth} 
    \begin{minipage}[b]{0.45\textwidth}
        \centering
            \begin{tikzpicture}
                \pgfmathsetseed{1213} 
                \draw[thick, name path=curve] (0,0) to[out=30,in=150] (4,0);
                
                \foreach \x in {0,1,...,9} {
                    \pgfmathsetmacro{\t}{\x/9} 
                    \path[name path=vert] ({\t*4}, -1) -- ({\t*4}, 1); 
                    \draw [name intersections={of=curve and vert, by=p}, red, fill=red!50] (p) circle (2pt);
            
                    \pgfmathsetmacro{\offset}{rand*0.2 + 0.8} 
                    \path[blue, fill=blue!50] ($(p) + (rand*0.05, \offset)$)  coordinate (bp\x) circle (2pt);
            
                    \pgfmathsetmacro{\offset}{rand*0.22 + 1.3} 
                    \path[blue, fill=blue!50] ($(p) + (rand*0.05, \offset)$) circle (2pt);
            
                    \pgfmathsetmacro{\offset}{rand*0.27 + 1.8} 
                    \path[blue, fill=blue!50] ($(p) + (rand*0.05, \offset)$) circle (2pt);
            
                    \pgfmathsetmacro{\offset}{rand*0.21 + 2.3} 
                    \path[blue, fill=blue!50] ($(p) + (rand*0.05, \offset)$) circle (2pt);
            
                    \ifnum\x=2
                        \path (p) coordinate (targetRed);
                    \fi
                    \ifnum\x=2
                        \path (bp\x) coordinate (targetBlue);
                    \fi
            
                    \ifnum\x=1
                        \path (p) coordinate (targetRed1);
                    \fi
                    \ifnum\x=1
                        \path (bp\x) coordinate (targetBlue1);
                    \fi
                    \draw[-latex, thick] (p) -- (bp\x);
                }

            \end{tikzpicture}
    \end{minipage}
    \caption{Illustration of the closest neighbor selection method in 2D, Red points are boundary points,  blue points are interior points, and the solid line is the boundary wall. The arrow represents the identification of the closest interior neighbor point for each boundary point.}
    \label{fig:CN_illustration}
\end{figure}

Since the wall function is applied to a single closest neighbor for each boundary point, we refer to this method as the closest neighbor (CN) method. The implementation of the CN method within a Lagrangian meshfree solver is shown in Algorithm \ref{alg:ClosestNeighbor}.  

\begin{algorithm}
\caption{Algorithm for Closest neighbor(CN) method}\label{alg:ClosestNeighbor}
\begin{algorithmic}[1]
\While{time step loop}
    \State Lagrangian point cloud movement \Comment{see \cite{suchde_point_2018}}
    \State Update neighbor tree
    \State Add/delete points to prevent holes/clustering \Comment{section \ref{sec:GFDM}}
    \State \textit{For each wall point, flag the closest interior neighbor point} \Comment{section \ref{sec:ClosestNeighbor}}
    \ForAll{$i \in$ Point cloud}
        \State Setup of momentum equation for velocity
        \If{$i$ is Flagged CN} 
            \State \textit{Impose the computed wall stress as source term to momentum equation} \Comment{equation \eqref{eq:WallShearStress}}
        \ElsIf{$i$ is wall points}
            \State \textit{Set appropriate boundary condition}
        \EndIf
    \EndFor
    \State Solve mass and momentum conservation equations
    \ForAll{$i \in$ Point cloud}
        \State Setup of turbulence equation $\{\mathrm{SA},k-\varepsilon \text{ or } k-\omega\}$ \Comment{section \ref{sec:TurbulenceModels}}
        \If{$i$ is Flagged CN} 
            \State \textit{Update turbulent quantities based on wall function} \Comment{section \ref{sec:TurbulenceWallFunction}}
        \ElsIf{$i$ is wall points}
            \State \textit{Set appropriate boundary condition}
        \EndIf
    \EndFor
    \State Solve turbulence equations and update turbulent viscosity $\nu_{t}$ \Comment{section \ref{sec:TurbulenceModels}}
    \State Post-processing calculations
\EndWhile
\end{algorithmic}
\end{algorithm}

This method has several drawbacks. In certain flow scenarios, such as downdrafts or shear, some interior points may move closer to the boundary points than other neighboring interior points. As illustrated in Figure~\ref{fig:GapInClosestNeighbor}, a single interior point might be the closest interior neighbor of multiple boundary points. At the same time, other interior points near the boundary might not be the closest interior point for any boundary point. Thus, the wall function is applied only to the single closest point, ignoring the other neighboring points. This leads to non-uniform application of the wall function near the boundary. The lack of uniform wall function coverage leads to incorrect solutions during gradient or Laplace operations. Because points move in a Lagrangian sense, such gaps might propagate or appear even when there was no gap in the previous time step. This situation is amplified in 3D simulations and for curved boundaries. This drawback necessitates the development of alternative approaches to apply the wall functions, which motivates the next two approaches proposed.

\begin{figure}[!htb]
    \centering
    \begin{tikzpicture}
    \pgfmathsetseed{1213} 
    
    \draw[thick, name path=curve] (0,0) to[out=30,in=150] (4,0);
    
    \foreach \x in {0,1,...,9} {
        \pgfmathsetmacro{\t}{\x/9} 
        \path[name path=vert] ({\t*4}, -1) -- ({\t*4}, 1); 
        \draw [name intersections={of=curve and vert, by=p}, red, fill=red!50] (p) circle (2pt);
        \coordinate (R\x) at (p); 
        \ifnum\x=1
            \pgfmathsetmacro{\offset}{rand*0.2 + 0.8}
            \path[blue, fill=blue!50] (0.65, 1.35) circle  (2pt); 
            \coordinate (B\x) at (0.65, 1.35);
        \else\ifnum\x=2
            \pgfmathsetmacro{\offset}{rand*0.2 + 0.8}
            \path[blue, fill=blue!50] (1.05, 1.3) circle  (2pt); 
            \coordinate (B\x) at (1.05, 1.3);

        \else\ifnum\x=5
            \pgfmathsetmacro{\offset}{rand*0.2 + 0.8}
            \path[blue, fill=blue!50] (2.25, 1.2) circle (2pt); 
            \coordinate (B\x) at (2.25, 1.2);

        \else\ifnum\x=6
            \pgfmathsetmacro{\offset}{rand*0.2 + 0.8}
            \path[blue, fill=blue!50] (2.7, 1.4) circle (2pt); 
            \coordinate (B\x) at (2.7, 1.4);

        \else\ifnum\x=7
            \pgfmathsetmacro{\offset}{rand*0.2 + 0.8}
            \path[blue, fill=blue!50] (3.1, 1.4) circle (2pt); 
            \coordinate (B\x) at (3.1, 1.4);

        \else
            \pgfmathsetmacro{\offset}{rand*0.2 + 0.8} 
            \pgfmathsetmacro{\myrand}{rand*0.05} 
            \path[blue, fill=blue!50] ($(p) + (\myrand, \offset)$)  circle (2pt);
            \coordinate (B\x) at ($(p) + (\myrand, \offset)$);

        \fi\fi\fi\fi\fi

        \pgfmathsetmacro{\offset}{rand*0.22 + 1.3} 
        \path[blue, fill=blue!50] ($(p) + (rand*0.05, \offset)$) circle (2pt);

        \pgfmathsetmacro{\offset}{rand*0.27 + 1.8} 
        \path[blue, fill=blue!50] ($(p) + (rand*0.05, \offset)$) circle (2pt);

        \pgfmathsetmacro{\offset}{rand*0.21 + 2.3} 
        \path[blue, fill=blue!50] ($(p) + (rand*0.05, \offset)$) circle (2pt);

        } 
        
    \draw[-latex, thick] (R0) -- (B0);
    \draw[-latex, thick] (R1) -- (B0);
    \draw[-latex, thick] (R2) -- (B2);
    \draw[-latex, thick] (R3) -- (B2);
    \draw[-latex, thick] (R4) -- (B5);
    \draw[-latex, thick] (R5) -- (B5);
    \draw[-latex, thick] (R6) -- (B5);
    \draw[-latex, thick] (R7) -- (B8);
    \draw[-latex, thick] (R8) -- (B8);
    \draw[-latex, thick] (R9) -- (B9);
\end{tikzpicture}
\caption{Closest neighbor method. Illustration of non-uniform coverage of the selection of the closest neighbor points in the interior points. Each boundary point (red dot) looks for the closest neighbor interior point (blue dot), leading to regions where there are no interior points selected, causing numerical instabilities.}
\label{fig:GapInClosestNeighbor}
\end{figure}

\subsection{Nearest-band neighbor (NBN) method}
\label{sec:NearestBandNeighbor}

To overcome the non-uniform wall function application, we propose the application of the wall function to not just the closest interior neighbor of a boundary point but also to all interior neighbors within a specified distance $\delta h$, as shown in Figure \ref{fig:NeighborSelectionByHeight}. Here, $h$ is the resolution, and $r_{\mathrm{min}}h < \delta h < h$, where $r_{\mathrm{min}}$ is the minimum separation factor (as a fraction of 
$h$) where the points are deleted for being too close to the boundary (see Section~\ref{sec:GFDM}). We refer to this approach as the nearest-band neighbor method (NBN), since interior points are selected within a band close to the boundary. The appropriate distance $\delta h$ by which the points need to be selected will be studied in Section~\ref{sec:Results} for uniform point selection and numerical accuracy. The boundary points used to select interior points for the wall function are given a Neumann (zero-gradient) boundary condition similar to that of the closest neighbor method.

\begin{figure}[!htb]
    \centering
    \begin{tikzpicture}
    \pgfmathsetseed{1219} 
    \draw[thick, name path=curve] (0,0) to[out=30,in=150] (4,0);
    \draw[dotted, thick, name path=shifted_curve] (-0.2,0.95) to[out=30,in=150] (4.2,0.95);
    \foreach \x in {0,1,...,9} {
        \pgfmathsetmacro{\t}{\x/9} 
        \path[name path=vert] ({\t*4}, -1) -- ({\t*4}, 1); 
        \draw [name intersections={of=curve and vert, by=p}, red, fill=red!50] (p) circle (2pt);
        \coordinate (R\x) at (p); 
        \ifnum\x=2
            \pgfmathsetmacro{\offset}{rand*0.2 + 0.8}
            \path[blue, fill=gray] (0.95, 1.1) circle  (2pt); 
            \coordinate (B\x) at (0.95, 1.1);

        \else\ifnum\x=5
            \pgfmathsetmacro{\offset}{rand*0.2 + 0.8}
            \path[blue, fill=gray] (2.1, 1.2) circle (2pt); 
            \coordinate (B\x) at (2.1, 1.2);

        \else
            \pgfmathsetmacro{\offset}{rand*0.2 + 0.8} 
            \pgfmathsetmacro{\myrand}{rand*0.05} 
            \path[blue, fill=gray] ($(p) + (\myrand, \offset)$)  circle (2pt);
            \coordinate (B\x) at ($(p) + (\myrand, \offset)$);

        \fi\fi

        \pgfmathsetmacro{\offset}{rand*0.22 + 1.3} 
        \path[blue, fill=blue!50] ($(p) + (rand*0.05, \offset)$) circle (2pt);

        \pgfmathsetmacro{\offset}{rand*0.27 + 1.8} 
        \path[blue, fill=blue!50] ($(p) + (rand*0.05, \offset)$) circle (2pt);

        \pgfmathsetmacro{\offset}{rand*0.21 + 2.3} 
        \path[blue, fill=blue!50] ($(p) + (rand*0.05, \offset)$) circle (2pt);

        }
        \draw [|<->|, >=latex,dotted] (4.3,0.9) -- (4.05,-0.05);
    
        \node [font=\small, rotate around={-30:(0,0)}] at (4.5,0.15) {$\delta h$};    
\end{tikzpicture}
\caption{Nearest-band neighbor method. Selection of neighbor interior points based on specified distance $\delta h$. Due to non-uniformity of the point distribution, the selection of interior points (gray dot) for application of wall functions is made based on the distance $\delta h$ (shown by dotted line) to the boundary, allowing for uniform selection of points. Other interior points are marked with blue dots.}
\label{fig:NeighborSelectionByHeight}
\end{figure}

As with the CN method described above, in the NBN method, we must select of nearby points at each time step due to the Lagrangian movement of points. The implementation of the NBN method within a Lagrangian meshfree solver is shown in Algorithm \ref{alg:NearestBandNeighbor}.

\begin{algorithm}
\caption{Algorithm for Nearest-Band Neighbor(NBN) method}\label{alg:NearestBandNeighbor}
\begin{algorithmic}[1]
\While{time step loop}
    \State Lagrangian point cloud movement \Comment{see \cite{suchde_point_2018}}
    \State Update neighbor tree
    \State Add/delete points to prevent holes/clustering \Comment{section \ref{sec:GFDM}}
    \State \textit{Flag all interior points $ < \delta h$ from wall points} \Comment{section \ref{sec:NearestBandNeighbor}}
    \ForAll{$i \in$ Point cloud}
        \State Setup of momentum equation for velocity
        \If{$i$ is Flagged NBN} 
            \State \textit{Impose the computed wall stress as source term to momentum equation} \Comment{equation \eqref{eq:WallShearStress}}
        \ElsIf{$i$ is wall points}
            \State \textit{Set appropriate boundary condition}
        \EndIf
    \EndFor
    \State Solve mass and momentum conservation equations
    \ForAll{$i \in$ Point cloud}
        \State Setup of turbulence equation $\{\mathrm{SA},k-\varepsilon \text{ or } k-\omega\}$ \Comment{section \ref{sec:TurbulenceModels}}
        \If{$i$ is Flagged NBN} 
            \State \textit{Update turbulent quantities based on wall function} \Comment{section \ref{sec:TurbulenceWallFunction}}
        \ElsIf{$i$ is wall points}
            \State \textit{Set appropriate boundary condition}
        \EndIf
    \EndFor
    \State Solve turbulence equations and update turbulent viscosity $\nu_{t}$ \Comment{section \ref{sec:TurbulenceModels}}
    \State Post-processing calculations
\EndWhile
\end{algorithmic}
\end{algorithm}

\subsection{Shifted-boundary (SB) method}
\label{sec:ShiftedBoundary}

In both methods above, for each boundary point, there is an explicit selection process to choose the interior points with reference to the boundary point, where the wall function will be applied. We now propose another novel approach for applying the wall function that avoids this selection process for the interior points altogether. In this method, we virtually shift the boundary points by a distance $\alpha h$ in the normal direction towards the interior. These shifted points are treated as interior points, and wall functions are applied to them. Since these boundary points are treated as virtual shifted points while the physical points remain on the boundary, no additional selection process for interior points is required. This process of virtually shifting the boundary points in the normal direction towards the interior of the domain is illustrated in the Figure~\ref{fig:ShiftedBoundaryMethod}.  We refer to this approach as the shifted-boundary (SB) approach. 

Since the virtual shifted points are treated as interior points, they move with the flow when the velocity is non-zero. The points act as Lagrangian points and move with the flow. To maintain the shifted distance at $\alpha h$ for the virtually shifted points, we treat these points in a semi-Lagrangian manner, allowing movement only in the tangential direction while fixing them in the normal direction. These restrictions require special treatment so the virtual points behave as Lagrangian points in the tangential direction and Eulerian in the normal direction to the boundary plane. This leads to the modification of the gradient and Laplace approximations in the normal direction for the momentum and turbulence equations. We compute this using the classical finite difference approach based on the shift distance. Computing the diffusion operator with the classical finite difference method requires additional information on the interior side of the virtual shifted point. To provide this information, we introduce an additional virtual layer of thickness $\beta h$  solely for approximating the values needed for the diffusion operator. The approximation of the tangential component requires no modification, since the points are Lagrangian in the tangential direction; the gradient and Laplace approximations are computed using the GFDM method. The differential operator modification is only carried out for the boundary points, which are virtually shifted. Other points in the domain require no modification.

\begin{figure}[!htb]
\centering
\begin{tikzpicture}
    \pgfmathsetseed{1213} 
    
    \draw[thick, dotted, name path=curve] (0,0) to[out=30,in=150] (4,0);
    
    \draw[thick, name path=curve2] (0.0,-0.3) to[out=30,in=150] (4,-0.3);
    
    \draw[gray, very thick, dotted, name path=curve4] (-0.25,0.15) to[out=30,in=150] (4.25,0.15);
    
    \foreach \x in {0,1,...,9} {
        \pgfmathsetmacro{\t}{\x/9} 
        \path[name path=vert] ({\t*4}, -1) -- ({\t*4}, 1); 
        \draw [name intersections={of=curve and vert, by=p}, orange, fill=orange!50] (p) circle (2pt);
    
        \pgfmathsetmacro{\offset}{rand*0.2 + 0.5} 
        \path[blue, fill=blue!50] ($(p) + (rand*0.05, \offset)$) circle (2pt);
    
        \pgfmathsetmacro{\offset}{rand*0.22 + 1.0} 
        \path[blue, fill=blue!50] ($(p) + (rand*0.05, \offset)$) circle (2pt);
    
        \pgfmathsetmacro{\offset}{rand*0.27 + 1.5} 
        \path[blue, fill=blue!50] ($(p) + (rand*0.05, \offset)$) circle (2pt);
    
        \pgfmathsetmacro{\offset}{rand*0.21 + 2.0} 
        \path[blue, fill=blue!50] ($(p) + (rand*0.05, \offset)$) circle (2pt);
    
        \pgfmathsetmacro{\n}{9} 
        \pgfmathsetmacro{\shift}{0.15 - (0.3 * \x / \n)} 
        \pgfmathsetmacro{\t}{\x / \n} 
        \path[name path=vert2] ({\t * 4 + \shift}, -1) -- ({\t * 4 + \shift}, 1); 
        \draw [name intersections={of=curve2 and vert2, by=p2}, red, fill=red!50] (p2) circle (2pt);
    }
        \draw [>|-|<, >=latex,dotted ] (-0.3,0.02) -- (0.0,-0.48);
        \draw [|<->|, >=latex,dotted] (4.4,0.09) -- (4.06,-0.35);
    
        \node [font=\small, rotate around={30:(0,0)}] at (-0.38,-0.37) {$\alpha h$};
        \node [font=\small, rotate around={-30:(0,0)}] at (4.5,-0.3) {$\beta h$};
    \end{tikzpicture}
    \caption{Illustration of the shifted-boundary method in 2D with additional shifted points. The boundary points (red dots) are virtually shifted (orange dots) to a distance of $\alpha h$ and are treated as interior points but fixed at a height of $\alpha h$ (dotted line) in the normal direction during the simulation. The momentum thickness approximation is carried out at the height of $\beta h$ (gray dotted line).}
    \label{fig:ShiftedBoundaryMethod}
\end{figure}

While all earlier figures illustrate the 2D scenario for ease of visualization, the formulation below, and the numerical results in Section \ref{sec:Results}, consider the general 3D scenario. For the virtually shifted points, while solving the flow and turbulence equations, the gradient approximation is split into normal and tangential components:
\begin{equation}
    \nabla f = (\nabla f \cdot \vec{n}) \vec{n} + \nabla_T f \,,
\end{equation}
where $f$ is a variable for which the gradient needs to be computed, $\vec{n}$ is the unit normal vector. The normal component of the gradient is $(\nabla f \cdot \vec{n}) \vec{n}$. The tangential component of the gradient, denoted by $\nabla_T f$, is computed using GFDM.  For GFDM computation of the tangential differential operator, we restrict neighborhood selection to the boundary points and calculate the gradient in the tangential plane \cite{suchde2019meshfree}.

For the computation of the gradient in the normal direction, the standard finite difference scheme is used:
\begin{equation}
    (\nabla f \cdot \vec{n}) \approx \frac{f_{\text{shifted}} - f_{\text{boundary}}}{\alpha h} \,,
\end{equation}
where $f_{\text{shifted}}$ is the value at the shifted point, $f_{\text{boundary}}$ is the value at the boundary point, and $\alpha h$ is the shift distance. Thus, the gradient computation at the shifted points is given by:  
\begin{equation}
  \nabla f = \nabla_T f +\left(\frac{f_{\text{shifted}}-f_{\text{boundary}}}{\alpha h}\cdot \vec{n}\right) \vec{n}.
  \label{eq:GradSplitOfTangentandNormalDirection}
\end{equation}
We compute the Laplace operator similarly by splitting it into the normal and tangential components $\Delta f = \Delta_{n} f + \Delta_T f$. Where $\Delta_{n} f $ is the normal component and $\Delta_T f$ is the tangential component. 
The computation of the tangential part is carried out using GFDM by considering only boundary points, similar to the gradient operator. In the normal direction, the computation is performed using a simple finite difference scheme as: 
\begin{equation}
  \Delta_{n} f = \left(\frac{f_{\text{momentum}}-f_{\text{shifted}}}{\beta h - \alpha h} - \frac{f_{\text{shifted}}-f_{\text{boundary}}}{\alpha h}\right) \frac{2}{\beta h}.
  \label{eq:LaplaceSplitOfTangentandNormalDirection}
\end{equation}
Here $\Delta_n f$ represents the Laplacian in the normal direction, $\alpha h$ is the shift distance, and $f_{\text{momentum}}$ is the function value approximated at the height of $\beta h$ from the boundary location. The approximation of $f_{\text{momentum}}$ is carried out using only the interior points, excluding the virtual shifted points and boundary points. The impact of the parameters $\alpha$ and $\beta$ on numerical accuracy will be presented in Section~\ref{sec:Results}. The implementation of the SB method within a Lagrangian meshfree solver is shown in Algorithm \ref{alg:ShiftedBoundary}. 

\begin{algorithm}
\caption{Algorithm for Shifted Boundary(SB) method}\label{alg:ShiftedBoundary}
\begin{algorithmic}[1]
\While{time step loop}
    \State Lagrangian point cloud movement \Comment{see \cite{suchde_point_2018}}
    \State Update neighbor tree
    \State Add/delete points to prevent holes/clustering \Comment{section \ref{sec:GFDM}}
    \ForAll{$i \in$ Point cloud}
        \State Setup of momentum equation for velocity
        \If{$i$ is wall points}
            \State \textit{Virtual shift points $\alpha h$ in the normal direction} 
            \State \textit{Compute the differential operator in normal and tangential direction for velocity } \Comment{equations \eqref{eq:GradSplitOfTangentandNormalDirection} \eqref{eq:LaplaceSplitOfTangentandNormalDirection}}
            \State \textit{substitute the newly computed differential operators in momentum equation}
        \EndIf
    \EndFor
    \State Solve mass and momentum conservation equations 
    \ForAll{$i \in$ Point cloud}
        \State Setup of turbulence equation $\{\mathrm{SA},k\text{-}\varepsilon \text{ or } k\text{-}\omega\}$ \Comment{section \ref{sec:TurbulenceModels}}
        \If{$i$ is wall points}
            \State \textit{Compute the differential operator in normal and tangential direction for turbulent quantities} \Comment{equations \eqref{eq:GradSplitOfTangentandNormalDirection} \eqref{eq:LaplaceSplitOfTangentandNormalDirection}}
            \State \textit{Set up the turbulent equations according to the wall functions} \Comment{section \ref{sec:TurbulenceWallFunction}}
        \EndIf
    \EndFor
    \State Solve turbulence equations and update turbulent viscosity $\nu_{t}$ \Comment{section \ref{sec:TurbulenceModels}}
    \State Post-processing calculations
\EndWhile
\end{algorithmic}
\end{algorithm}

\section{Results}\label{sec:Results}

We now investigate the performance of the three proposed boundary treatment methods: the closest neighbor method, the nearest-band neighbor method, and the shifted-boundary method. The boundary treatment methods are evaluated on first-order turbulence models, namely the Spalart--Allmaras (SA), $k-\varepsilon$, and $k-\omega$ models with wall functions. Common validation test cases are chosen such that the results obtained can be readily validated against available data. Initially, we focus on a one-dimensional (1D) scenario involving Couette flow, which provides a straightforward implementation and testing environment. Following this, we examine more complex simulations, including flow over a flat plate and flow around a NACA 0012 airfoil, both analyzed in three dimensions to encompass different potential flow scenarios. All the turbulence models and boundary treatment methods introduced above were implemented in the in-house commercial software MESHFREE \cite{MESHFREE}. The time integration in MESHFREE is performed using the segregated, predictor-corrector approach ~\cite{Brown2001, Chorin1968}. Initially, the integration starts with the second-order Lagrangian movement of the points~\cite{Suchde2018_PCM}. The velocity predictions are computed using the momentum equations via a first-order implicit scheme. Followed by the projection step from the pressure Poisson equation to ensure mass conservation and to obtain the corrected velocity, followed by a pressure update \cite{suchde_meshfree_2018}. The turbulent equations are computed with an implicit time integration scheme. Further details regarding the time integration scheme can be referenced in \cite{Drumm2008, Jefferies2015, Kuhnert2014, Michel2021, suchde_meshfree_2018}. 

\subsection{1D simulation case of Couette flow}

The 1D case of Couette flow is selected as the first test case to simplify the governing equations of both the turbulence models and the Navier--Stokes equations by eliminating the convection term and the velocity-pressure coupling. For the computation of differential operators, a classical finite-difference scheme is used. We consider a long channel with a height $H$ (see Figure~\ref{fig:1DCouetteFlow}). The bottom plate of the channel remains fixed, while the top plate moves at a constant velocity $U_{\mathrm{m}}$. Because the motion of the upper plate drives the flow, the velocity variation occurs solely in the normal direction. Since velocity varies only in one direction, a 1D simulation along the height $H$ of the channel is sufficient. We can neglect the convection term. Furthermore, a zero-pressure gradient case is considered. 

\begin{figure}[!htb]
\centering
\resizebox{0.4\textwidth}{!}{%
\begin{circuitikz}
\tikzstyle{every node}=[font=\small]

    \draw[thick] (0,0) -- (10,0.0);
    \foreach \x in {0,0.5,...,10}
    {
        \draw (\x,0) -- (\x+0.5,-0.5);
    }
    \draw[line width=2pt] (0,4) rectangle (10,4.2);

    \draw [<->, >=Stealth,line width=0.75mm] (6.5,0.0) -- (6.5,4.0);
    \node [font=\Large, rotate around={0:(0,0)}] at (6.8,2) {H};
    
    \draw [->, >=Stealth,line width=1.5mm, color = gray] (3.5,4.6) -- (6.5,4.6);
    \node [font=\Large, rotate around={0:(0,0)}] at (4.7,5.0) {$U_{\text{m}}$};
    \node [font=\Large, rotate around={0:(0,0)}] at (4.7,-1.0) {$U_{\text{w}}$};
    
    \draw [->, >=Stealth,line width=1.5mm, color = gray] (0.5,2.2) -- (3.5,2.2);
    \node [font=\Large, rotate around={0:(0,0)}] at (1.5,2.8) {Fluid Flow};
    
    \draw [->, >=Stealth,line width=0.75mm] (-1.0,-1.0) -- (-1.0,0.0);
    \node [font=\Large, rotate around={0:(0,0)}] at (-0.7,0.0) {Y};
    
    \draw [->, >=Stealth,line width=0.75mm] (-1.0,-1.0) -- (0.0,-1.0);
    \node [font=\Large, rotate around={0:(0,0)}] at (0.3,-1.0) {X};
    
\end{circuitikz}
}%
\caption{1D Couette Flow schematic illustrating the configuration with a fixed bottom plate with velocity $U_{\mathrm{w}} = 0 \, \frac{\text{m}}{\text{s}}$ and a moving top plate. The flow is driven by the top plate $U_{\mathrm{m}}$, including the height $H$ and the moving wall velocity $U_{\mathrm{m}} = 12.8 \, \frac{\text{m}}{\text{s}}$. The 1D simulation is carried out along the channel height $H$. } 
\label{fig:1DCouetteFlow}
\end{figure}

The simulation parameters are chosen to replicate those from the work of Telbany et al. \cite{telbany_velocity_1980}, where channel height $H = 0.066 \, \text{m}$ and the top moving wall velocity $U_{\mathrm{m}} = 12.8 \, \frac{\text{m}}{\text{s}}$. The fluid properties considered are those of air, with a density of $1 \, \text{kg/m}^3$ and viscosity of $1.388 \times 10^{-5} \, \text{m}^2/\text{s}$, assuming incompressibility. Three turbulence models, SA, $k-\varepsilon$, and $k-\omega$, are implemented, with both the closest neighbor method (In 1D, the closest neighbor method and the nearest-band neighbor method are identical) and shifted-boundary methods. For the 1D discretization, points are distributed equidistantly with a discretization size $dx$ along the height $H$.

The results obtained from the closest neighbor method and shifted boundary method are validated against experimental data from \cite{telbany_velocity_1980}. The velocity profiles across different turbulence models are compared to experimental results. The 1D domain is discretized into N equidistant grid points with grid size $dx$. 

The analysis begins with the closest boundary method, which is straightforward to execute. As described in the section \ref{sec:Methodology}, the first interior node is selected for applying the wall function. The number of discretization points is fixed to $N = 40$ so that the first interior point lies on the log layer. Both the standard and Launder--Spalding wall function methods \cite{launder_numerical_1974,tennekes_first_1972} are employed for the $k-\varepsilon$ and $k-\omega$ models, and the standard wall function is used for the Spalart--Allmaras model. The simulation is carried out until the residues such as the velocity ($U$) and turbulent viscosity ($\nu_t$) are lower than $10^{-8}$. The initial conditions are fixed based on the moving wall's velocity $U_{\mathrm{m}}$, with boundary conditions specified as mentioned above. 

\begin{figure}[!htb]
    \centering
    \includegraphics[width=0.6\textwidth]{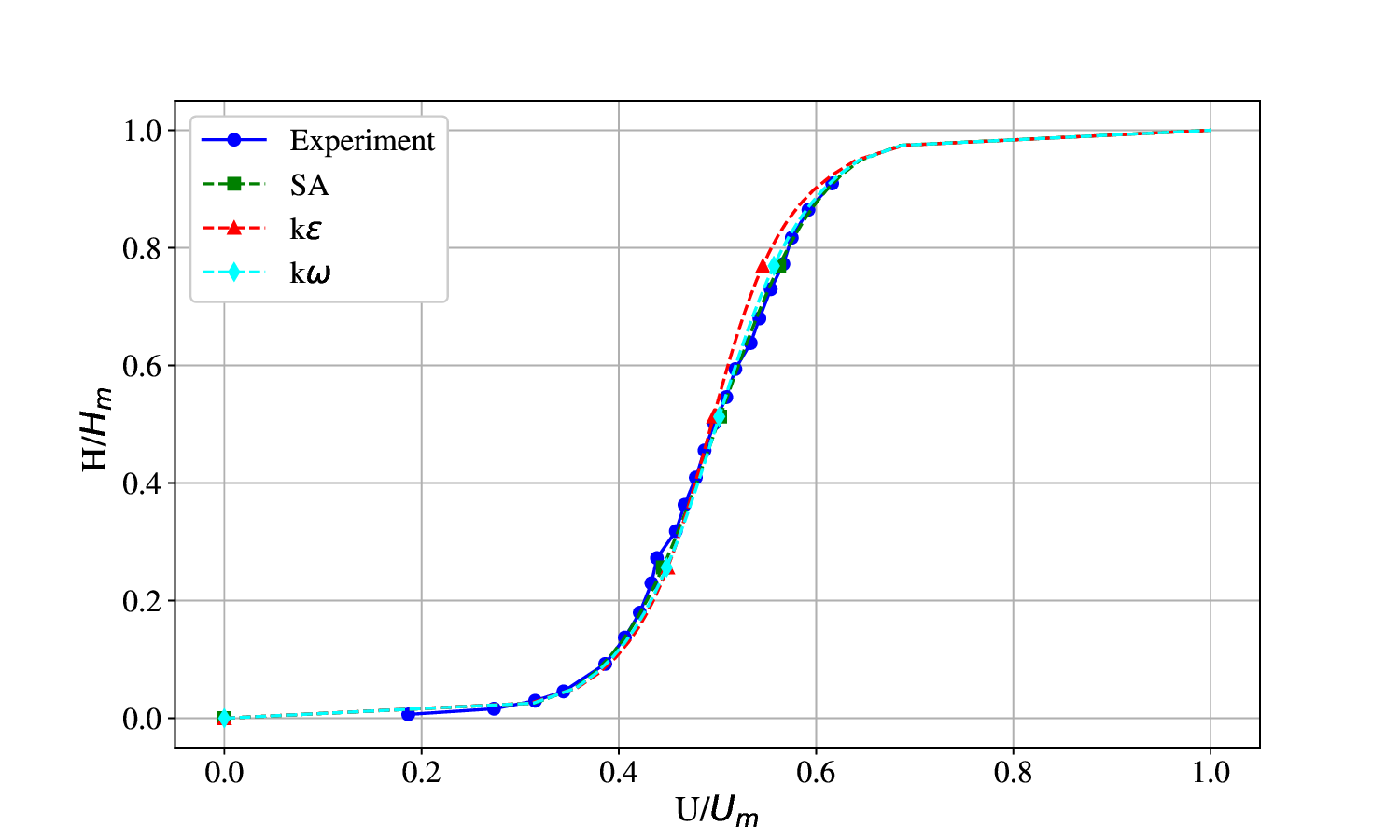}
    \caption{1D Couette Flow results comparison using the closest neighbor method against experimental results. Velocity profiles show the performance of $k-\omega$ (cyan line), $k-\varepsilon$ (red line), and Spalart--Allmaras (green line) turbulence models against the experimental results (blue line) from \cite{telbany_velocity_1980}, particularly near the wall, showcasing the accuracy of the closest neighbor approach.}
    \label{fig:1DNearest}
\end{figure}

When comparing the performance of different turbulence models against experimental results, we observe that all the turbulence models follow the velocity profile of the experimental results. This is illustrated in Figure~\ref{fig:1DNearest}. A slight variation can be observed with the turbulence model, especially with the $k-\varepsilon$ model, towards the moving wall, demonstrating that the $k-\omega$ model is slightly more accurate near the wall compared to the $k-\varepsilon$  model \cite{menter_two-equation_1994}.

The simulation setup for the shifted-boundary method is similar to that of the closest neighbor method, with the distinction that the boundary points are shifted from the wall by a specified distance. Here, the shifted distance $\alpha h$ is set to $0.5 dx$, while the momentum boundary thickness $\beta h$ is $1.0 dx$. Various values of $\alpha h$ and $\beta h$ are tested to evaluate their effects on the simulation results. The number of points for the SB method is reduced to $N=20$ to make sure that the shifted point lies in the log layer. 

\begin{figure}[!htb]
    \centering
    \includegraphics[width=0.6\textwidth]{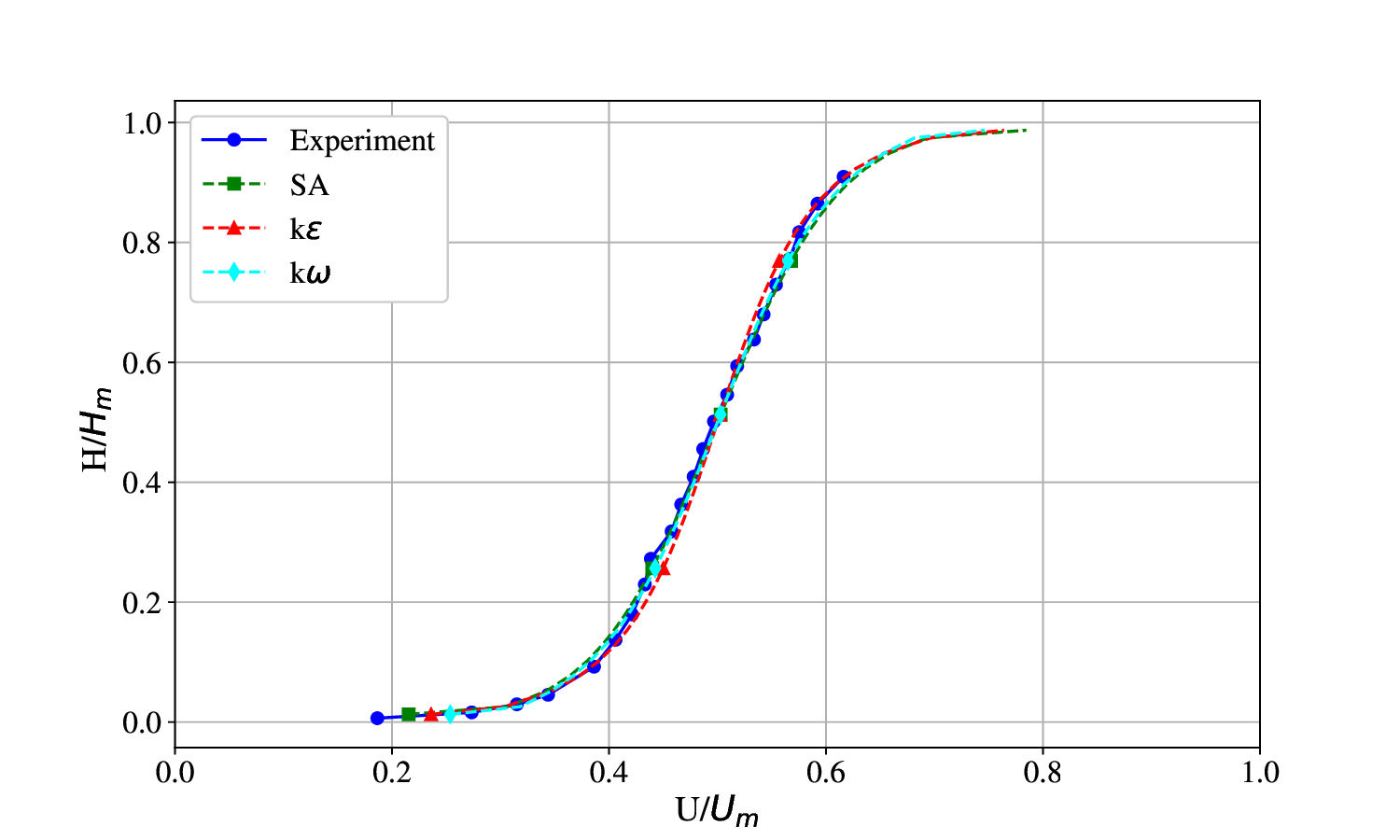}
    \caption{1D Couette Flow results comparison with the shifted-boundary method. Velocity profiles from the simulation are compared with experimental data, demonstrating the effectiveness of the shifted-boundary treatment.}
    \label{fig:1DShifted}
\end{figure}

The comparison of the velocity profiles of different turbulence models can be seen in Figure~\ref{fig:1DShifted}. The results obtained from the shifted boundary method produce similar results to those of the closest boundary method. We observe that the velocity profile of $k-\varepsilon$ with the SB method matches well with the experimental results, similar to that of the CN method. The minor oscillations observed in the CN method are not observed here. Boundary points are not visible in Figure~\ref{fig:1DShifted}, because we have virtually shifted the boundary points to treat them as interior points at a distance $0.5 dx$ from the wall.     

Both the shifted-boundary and closest neighbor methods yield good agreement with the experimental results. The number of points in the SB method is reduced from $N=40$ to $N=20$ to maintain approximately the same $y^+$ value, but the velocity profiles still match well with experimental results. We also see that the $k-\varepsilon$ turbulence model with the SB method produces less oscillation compared to that of the CN method. As the boundary treatment methods produced promising results with 1D simulation, the next step is to test the method in a 3D environment to assess the behavior of the boundary treatment methods.

\subsection{Flow over a flat plate}

The next simulation case involves turbulent flow over a flat plate, a widely studied test case in fluid dynamics. This simulation is configured in either 2D or 3D, utilizing symmetry boundary conditions along the sidewalls. The focus of the present work is to adapt the boundary treatment and turbulence model implementation for industrial applications, which predominantly require 3D simulations. Thus, we only consider a 3D geometry here, which can help identify potential challenges or limitations in actual applications. The simulation setup consists of a flat plate of length $2 \, \text{m}$ placed in a rectangular channel with a height of $1 \, \text{m}$ (see Figure~\ref{fig:FlatPlateayout}). 

\begin{figure}[!htb]
\centering
\resizebox{0.6\textwidth}{!}{%
\begin{circuitikz}
\tikzstyle{every node}=[font=\small]
    \draw[line width=2pt] (0,0) rectangle (23.3,10);

    \draw[gray, line width=2pt] (5.0,3.0) rectangle (28.3,13);
    \draw[gray, line width=2pt] (0,0) -- (5,3);
    \draw[gray, line width=2pt] (23.3,10) -- (28.3,13);
    \draw[gray, line width=2pt] (0,10) -- (5,13);
    \draw[gray, line width=2pt] (23.3,0) -- (28.3,3);
    \draw[gray, line width=2pt] (3.3,0) -- (8.3,3);

    \draw [<->, >=Stealth,line width=0.75mm] (0,9.5) -- (23.3,9.5);
    \draw [|<->|, >=Stealth,line width=0.75mm] (-0.5,0.0) -- (-0.5,10);
    \draw [<->, >=Stealth,line width=0.75mm] (1,10) -- (6,13);
    \draw [|<->|, >=Stealth,line width=0.75mm] (3.3,-0.5) -- (23.3,-0.5);

    \node [font=\huge, rotate around={0:(0,0)}] at (13.3,-1) {2 m};
    \node [font=\huge, rotate around={0:(0,0)}] at (13.3,9) {2.33 m};
    \node [font=\huge, rotate around={90:(0,0)}] at (-1,5) {1 m};
    \node [font=\huge, rotate around={30:(0,0)}] at (4.2,11.5) {1 m}; 
    \node [font=\huge, rotate around={0:(0,0)}] at (13.3,2) {Flat Plate (No Slip) };
    \node [font=\huge, rotate around={30:(0,0)}] at (2,6) {Inflow};
    \node [font=\huge, rotate around={0:(0,0)}] at (13.3,11) {Slip};
    \node [font=\huge, rotate around={30:(0,0)}] at (26.3,6) {Outflow};
    \node [font=\huge, rotate around={50:(0,0)}] at (5,2) {Slip};
    \node [font=\huge,color = gray, rotate around={0:(0,0)}] at (16.3,7) {Slip};
    \node [font=\huge, rotate around={0:(0,0)}] at (12.3,7) {Slip};

    \draw [->, >=Stealth,line width=0.75mm] (-4.0,-1.0) -- (-4.0,1.0);
    \node [font=\huge, rotate around={0:(0,0)}] at (-4.3,1.0) {Y};
    
    \draw [->, >=Stealth,line width=0.75mm] (-4.0,-1.0) -- (-2.0,-1.0);
    \node [font=\huge, rotate around={0:(0,0)}] at (-2.1,-1.5) {X};

    \draw [->, >=Stealth,line width=0.75mm] (-4.0,-1.0) -- (-2.1, 0.0);
    \node [font=\huge, rotate around={0:(0,0)}] at (-2.3, 0.5) {Z};

    \draw [->, >=Stealth,line width=2.5mm, color = gray] (10.5,4.2) -- (14.5,4.2);
    \node [font=\LARGE, rotate around={0:(0,0)}] at (11.5,4.8) {Flow direction};
\end{circuitikz}
}%
\caption{Flow over a flat plate geometry setup, illustrating the dimensions of the rectangular channel and the placement of the flat plate. Includes inlet conditions such as specified velocity and pressure gradient, along with slip and no-slip boundary condition sections.}
\label{fig:FlatPlateayout}
\end{figure}

A slip boundary condition is enforced for the section preceding the flat plate, measuring $0.33 \, \text{m}$, while the flat plate itself is treated as a no-slip boundary. The inlet conditions specify a velocity of $U = 69.4 \, \text{m/s}$ and a pressure gradient of zero. The fluid properties remain the same as the previous case, using air with incompressible flow, with a density of $\rho = 1.0 \, \text{kg/m}^3$ and viscosity of $\nu = 1.388 \times 10^{-5} \, \text{m}^2/\text{s}$. The simulation continues until a steady state is achieved. The flow behavior at the wall is studied using the skin friction coefficient $C_{\mathrm{f}}$ computed with the help of  wall shear stress $\tau_{\mathrm{w}}$, 

\begin{align*}
    Re_x &= \frac{\rho_{\infty} U_{\infty} x}{\mu_{\infty}}\,, \quad &
    \tau_{\mathrm{w}} &= \mu_{\mathrm{w}}\left(\frac{\partial U}{\partial y}\right)_{\mathrm{w}}\,, \quad &
    C_{\mathrm{f}} &= \frac{\tau_{\mathrm{w}}}{\frac{1}{2} \rho_{\infty} U_{\infty}^2}\,,
\end{align*}
where $\rho_{\infty}$, $U_{\infty}$, $\mu_{\infty}$ are the bulk density, velocity, and viscosity values well above the boundary layer. Here, $x$ represents the distance along the flat plate where the respective values are computed. To compute the wall shear stress $\tau_{\mathrm{w}}$, the wall viscosity $\mu_{\mathrm{w}}$ and velocity gradient at the wall $\left(\frac{\partial U}{\partial y}\right)_{\mathrm{w}}$ are used. The computed skin friction coefficient $C_{\mathrm{f}}$ is validated against analytical expressions $C_{\mathrm{f, analytic}}=0.0592 \mathrm{Re}_x^{-1/5}$ \cite{granville_drag_1977, schlichting_boundary-layer_2000}.

We now evaluate the three proposed boundary treatments for wall function application: (i) the closest-neighbor (CN) method, (ii) the nearest-band neighbor (NBN) method, and (iii) the shifted-boundary method (SB) for the test case of turbulent flow over a smooth, zero-pressure-gradient flat plate. Instantaneous skin-friction  $C_{\mathrm{f}}(x)$ data without temporal averaging, extracted along the streamwise direction at the midspan line ($y = 0.5\,\mathrm{m}$) are validated against the analytical data. 
Unless stated otherwise, all simulations employ identical parameters, with the $k-\omega$ turbulence model and resolution $h = 0.0035$ m for the CN method and the NBN method. The resolution for the SB method is chosen to be a coarser resolution of $h = 0.005$ m. The resolution is determined via grid-convergence studies; a coarser resolution of $h = 0.005$ m is chosen for the SB method to maintain a similar $y^{+}$ value for both NBN method and SB method. 
Wall-adjacent points are constrained to lie within the logarithmic layer to ensure valid wall function enforcement. The exact $y^+$ band depends on the local resolution but remains within the log-layer in all cases. 

\subsubsection{Closest-neighbor (CN) Method Analysis}
\begin{figure}[!htb]
    \centering
    \includegraphics[width=0.6\textwidth]{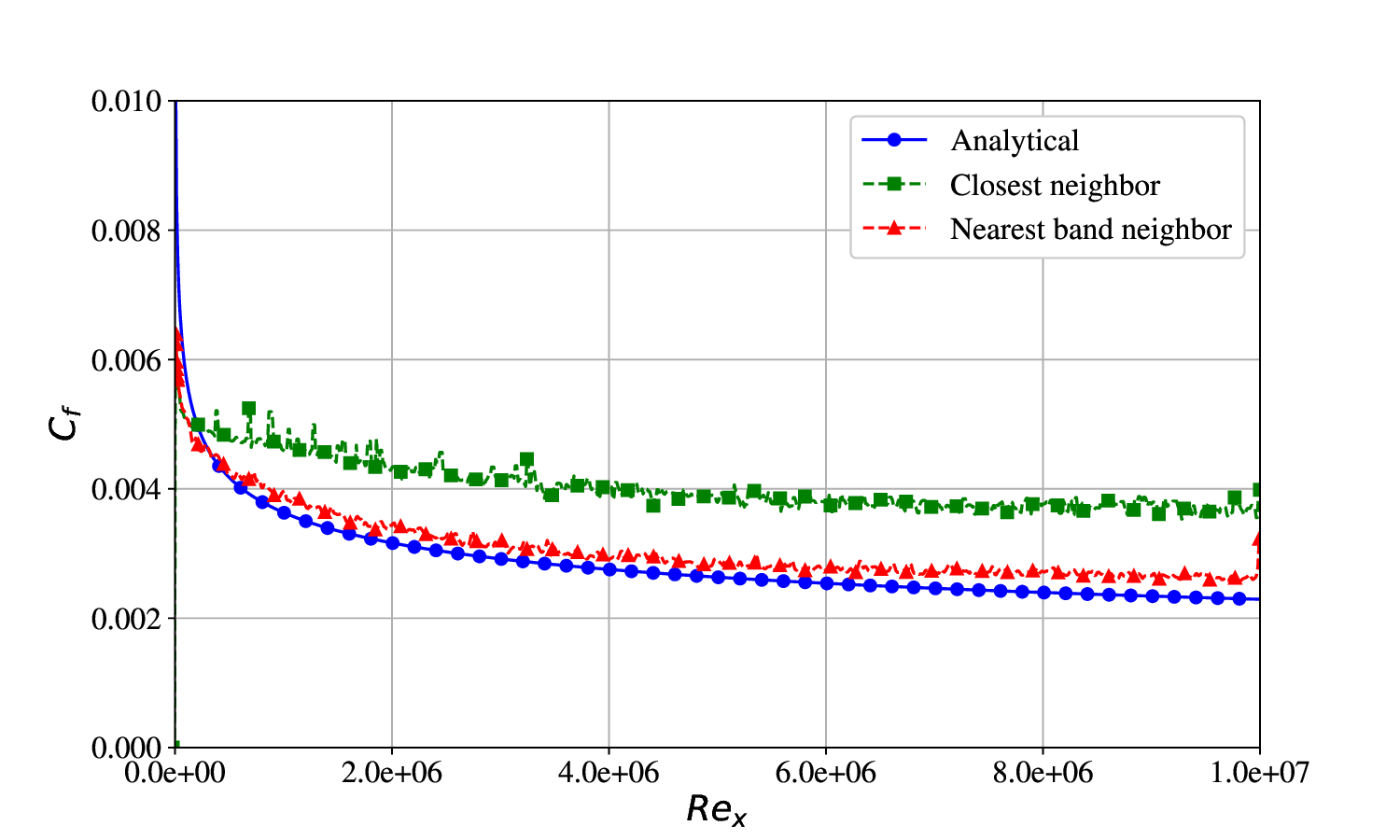}
    \caption{Skin friction coefficient $C_{\mathrm{f}}$ comparison of the closest neighbor (CN) method (green) with the Nearest-band neighbor (NBN) method (red) and the analytical expression (blue). Comparison of $C_{\mathrm{f}}$  shows a large deviation in the prediction of $C_{\mathrm{f}}$ for the CN method due to inadequate points selected to apply the wall function.}
    \label{fig:ClosestNeighborMethod}
\end{figure}

As described in Section~\ref{sec:ClosestNeighbor}, the closest-neighbor (CN) method selects only the single nearest interior point to each wall point to apply the wall function. This selection isn’t uniform along the wall, so some wall points get proper treatment while others do not. Because of this, the wall function is applied unevenly, and the wall resistance is underestimated. As a result, the skin friction $C_{\mathrm{f}}$ extracted along the flat plate (at $y=0.5$~m) is higher than expected, and the velocities are also higher, see Figure~\ref{fig:ClosestNeighborMethod}. We also observe large instantaneous fluctuations. As a result, the CN method is not used for further simulations.

\subsubsection{Nearest-band neighbor (NBN) Method Analysis}
With the NBN method, instead of selecting only the closest point, we select all interior points within a certain distance $\delta h$ from the wall (see Section~\ref{sec:NearestBandNeighbor}). Here, we use the selection height $\delta h$ so that there are always enough points selected near the wall to apply the wall function evenly. This results in a more uniform wall resistance and more accurate simulations. We also observe reduced oscillations when compared with the CN method, see Figure~\ref{fig:ClosestNeighborMethod}. 

In this section, we systematically evaluate the NBN method. The analysis begins with a grid independence study, followed by an assessment of time-step sensitivity. We further examine the effects of different wall function formulations, sensitivity to the selection height, and the overall behavior of various turbulence models.

\begin{figure}[!htb]
    \centering
    \begin{minipage}[t]{0.48\textwidth}
        \centering
        \includegraphics[width=\textwidth, height=0.25\textheight, keepaspectratio]{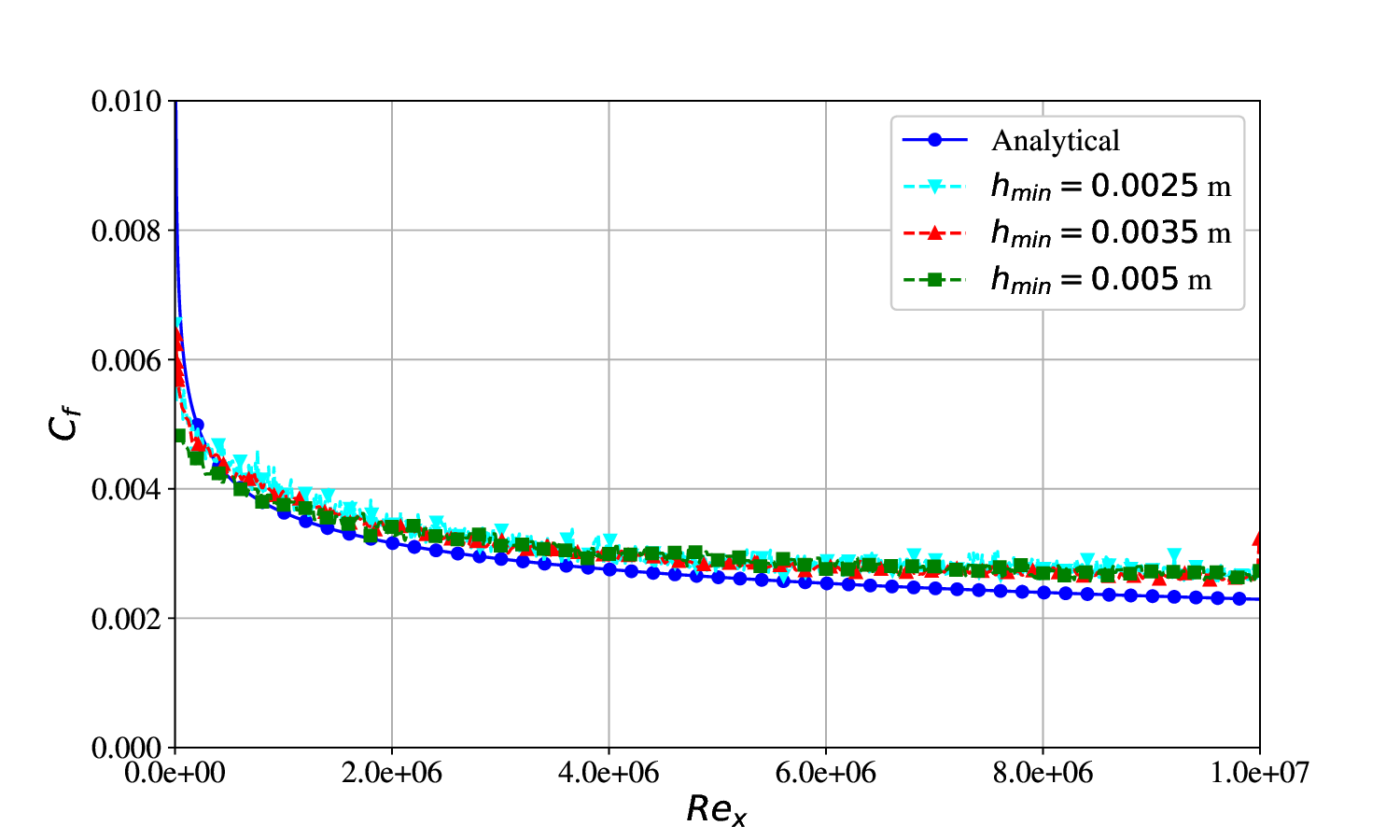}
        \subcaption{Comparison of skin friction coefficient $C_{\mathrm{f}}$ for various resolutions.}
        \label{fig:NearestSelectiongrid_cf_compare}
    \end{minipage}
    \hfill
    \begin{minipage}[t]{0.48\textwidth}
        \centering
        \includegraphics[width=\textwidth, height=0.25\textheight, keepaspectratio]{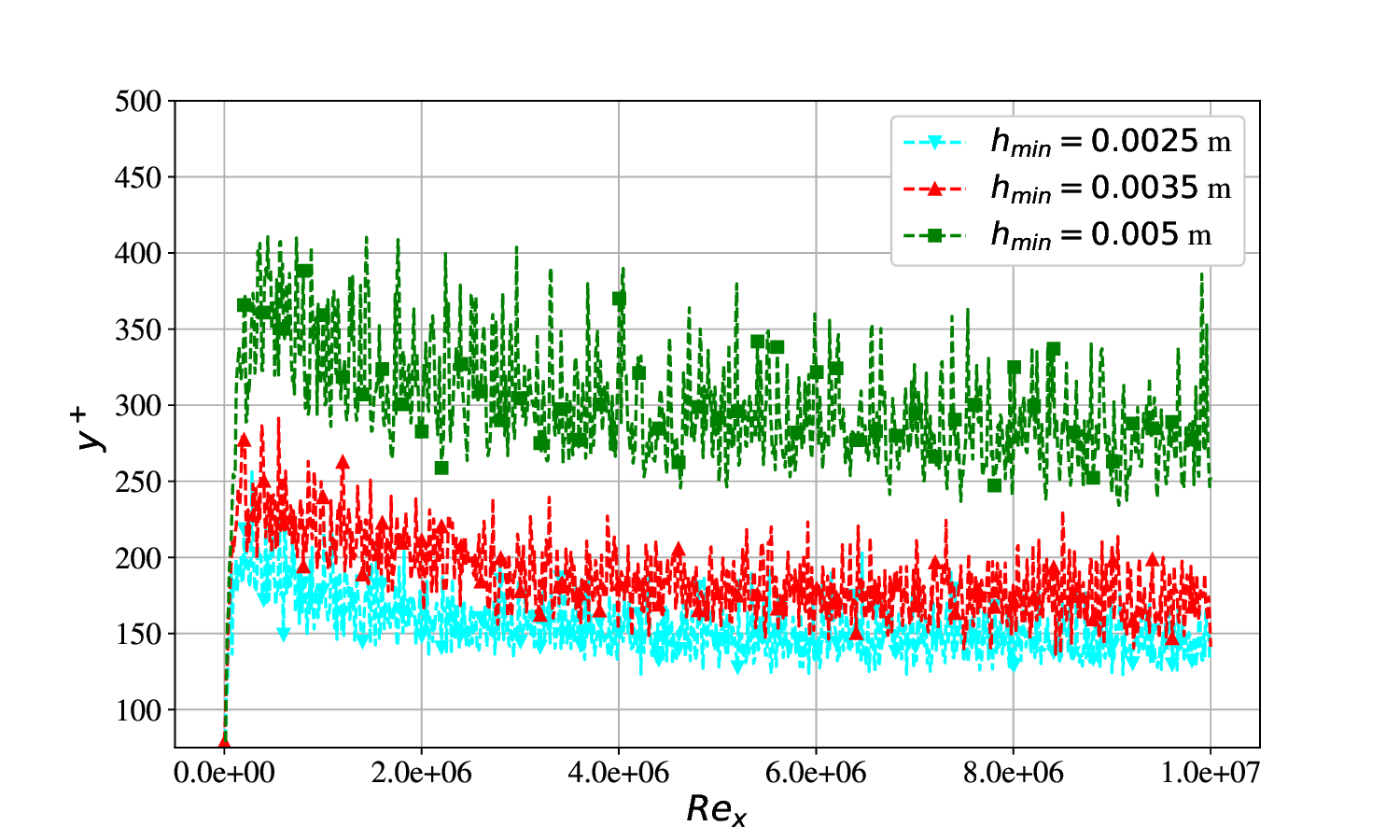}
        \subcaption{Comparison of $y^+$ along the flat plate for various resolutions.}
        \label{fig:NearestSelectionyplus_compare}
    \end{minipage}
    \caption{ The grid independent study is carried out by varying the resolution of the point cloud of the simulation domain to verify the dependency on the grid. The figure~\subref{fig:NearestSelectiongrid_cf_compare} shows the comparison of skin friction coefficient $C_{\mathrm{f}}$ with various resolutions $h_{\mathrm{min}}$ using the nearest band neighbor method. The plots show the resolution of the points close to the wall varying from $h_{\mathrm{min}} = 0.005 \text{ m}$ (green), $h_{\mathrm{min}} = 0.0035 \text{ m}$ (red) and $h_{\mathrm{min}} = 0.0025 \text{ m}$ (cyan) against the analytical expression (blue) where we can see that all the resolution provide the similar results except at the start of the flat plate. The figure~\subref{fig:NearestSelectionyplus_compare} shows the variation of non-dimensional wall distance $y^+$ for various resolutions $h_{\mathrm{min}}$. Due to high fluctuation the simulation results are down-sampled and smoothed. Even with the smoothing, due to the staggered position of the points, we notice a large variation in $y^+$ values. Even with the fluctuation, we clearly see that the decrease in $h_{\mathrm{min}}$ directly reduces the value of $y^+$. }
    \label{fig:NearestSelectionGridStudy}
\end{figure}

For grid independence, the NBN method is tested with different spatial resolutions. The point cloud resolution, $h$, is chosen such that it is minimum near the flat plate at $h_{\mathrm{min}}$ and gradually coarsened to a maximum resolution of $h_{\mathrm{max}}$ in the far field. For all cases in this study, the coarsest resolution is maintained at $h_{\mathrm{max}} = 0.007 \text{ m}$, while the near-wall resolution is varied between $h_{\mathrm{min}} = 0.0025 \text{ m}$ and $0.005 \text{ m}$. The selection height for the interior points is fixed at $\delta h = 0.5 h$. Figure~\ref{fig:NearestSelectiongrid_cf_compare} compares the computed skin friction coefficient, $C_{\mathrm{f}}$, against the analytical solution. The results demonstrate that varying the resolution yields no substantial deviation in the overall $C_{\mathrm{f}}$ profile. Minor variations observed are due to the positioning and distribution of the interior points. Notably, at the leading edge of the plate, the finer resolution produces a slightly higher $C_{\mathrm{f}}$ value that more closely aligns with the analytical solution. Figure~\ref{fig:NearestSelectionyplus_compare} illustrates the variation in the non-dimensional wall distance, $y^+$, for the different resolutions. For clarity of visualization, the $y^+$ data has been down-sampled and smoothed. As expected, a reduction in $h_{\mathrm{min}}$ (which corresponds to a finer point cloud) leads to an overall decrease in $y^+$, as the interior points are positioned closer to the boundary. The results show pronounced spatial fluctuations in $y^+$ along the length of the plate for all resolutions. The irregular point cloud and Lagrangian framework result in varying distances between interior points and the boundary, leading to these oscillations. The amplitude of these fluctuations is directly linked to the band of selected neighbors: interior points closest to the wall yield lower $y^+$ values, whereas those near the upper limit of the selection band generate higher values. Thus, the amplitude of the $y^+$ fluctuations decreases with finer resolutions. The approximate maximum $y^+$ values observed are $300$ for $h_{\mathrm{min}} = 0.005 \text{ m}$, $175$ for $h_{\mathrm{min}} = 0.0035 \text{ m}$, and $150$ for $h_{\mathrm{min}} = 0.0025 \text{ m}$. Resolutions coarser than $h_{\mathrm{min}} = 0.005 \text{ m}$ are excluded from this study, as the resulting neighbor band would capture points with $y^+$ values significantly exceeding $300$, which violates the recommended limits for standard wall functions \cite{launder_numerical_1974,kalitzin_near-wall_2005}.

\begin{figure}[!htb]
    \centering
    \includegraphics[width=0.6\textwidth]{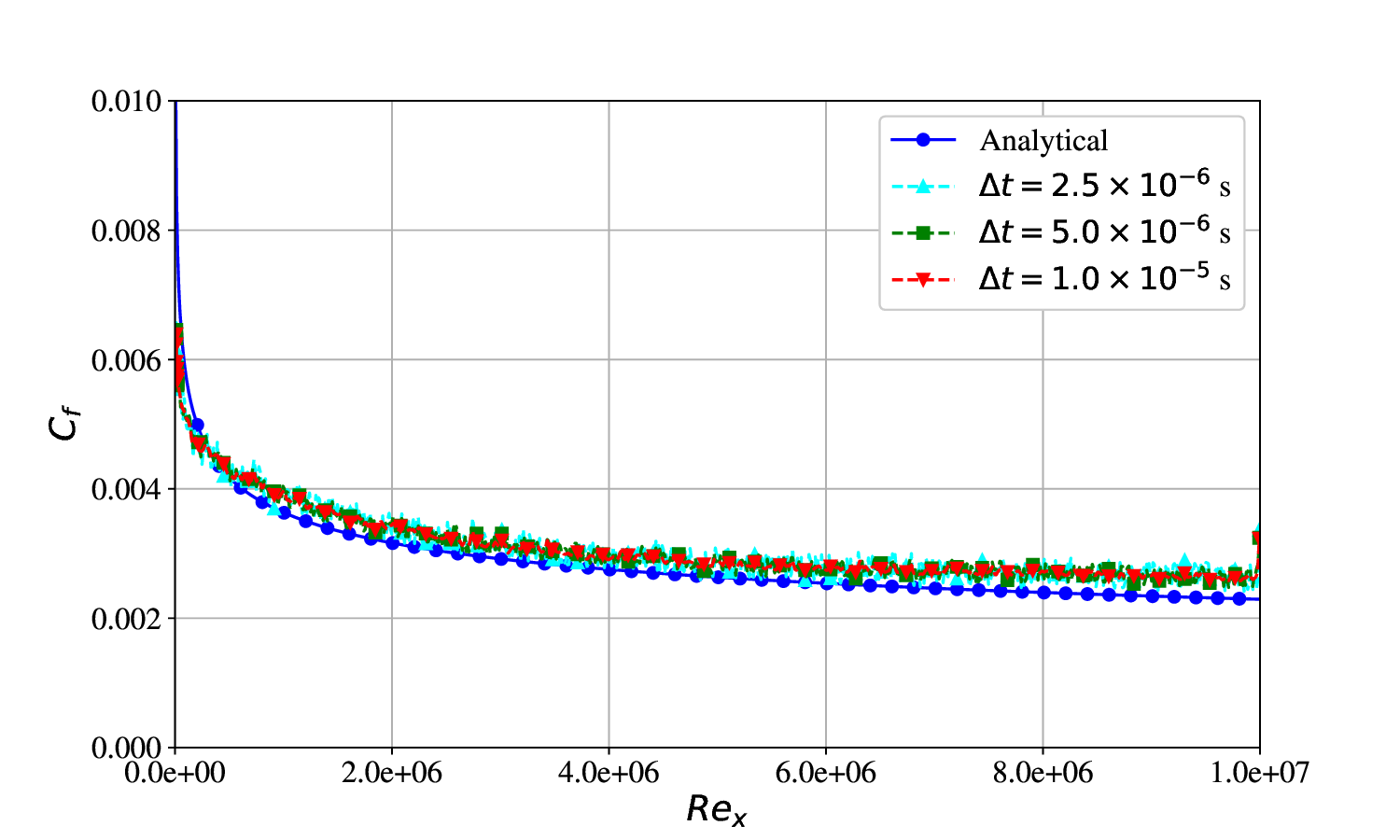}
    \caption{Comparison of Skin Friction Coefficient $C_{\mathrm{f}}$, of various timestep size ($\Delta t$), against the analytical expression (blue) using the nearest-band neighbor method in turbulent flow over a flat plate. The figure indicates that there is no significant difference in the $C_{\mathrm{f}}$ values between different time step sizes $\Delta t = 1.0 \times 10^{-5} \, \text{s}$, $5.0 \times 10^{-6} \, \text{s}$ and $2.5 \times 10^{-6} \, \text{s}$. }
    \label{fig:NearestTimeStepStudy}
\end{figure}

In addition to spatial resolution, the influence of the time-step size, $\Delta t$, is analyzed to ensure temporal convergence. The time steps are selected to maintain a Courant--Friedrichs--Lewy (CFL) number well below $1$. Three time-step sizes are evaluated: $\Delta t = 1.0 \times 10^{-5} \, \text{s}$, $5.0 \times 10^{-6} \, \text{s}$, and $2.5 \times 10^{-6} \, \text{s}$. The computed $C_{\mathrm{f}}$ values for these cases are compared against the analytical solution in Figure~\ref{fig:NearestTimeStepStudy}. The skin friction profiles for all three time steps overlap almost perfectly, with only negligible differences arising from the instantaneous interior point positions. Together, the grid independence and time-step sensitivity studies confirm the numerical stability and robustness of the proposed NBN method.

\begin{figure}[!htb]
    \centering
    \includegraphics[width=0.6\textwidth]{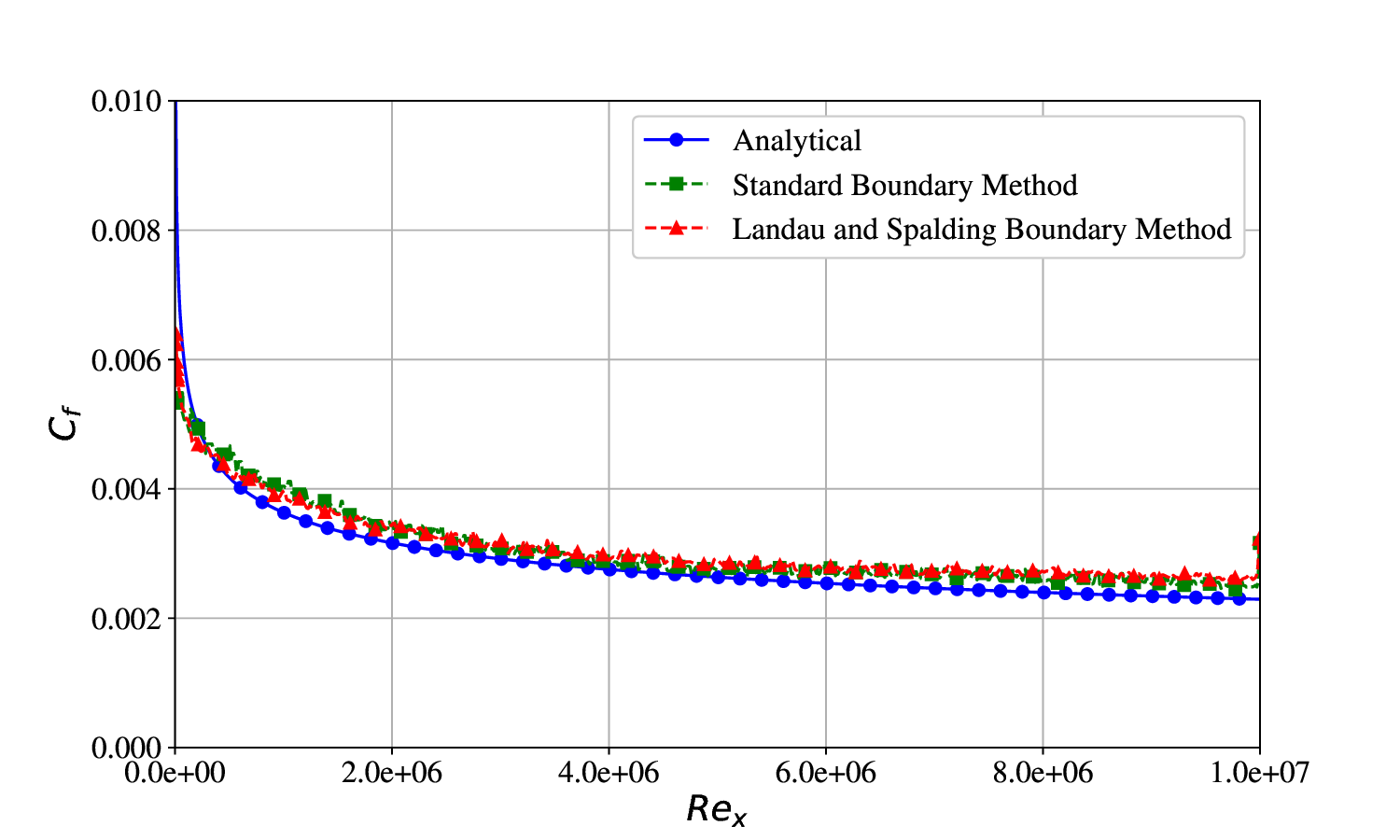}
    \caption{Comparison of Skin friction coefficient $C_{\mathrm{f}}$ for different wall function methods using the NBN method. The plot shows a comparison of the standard wall function (green) and Launder--Spalding wall function (red) against the analytical expression (blue) where we see no large difference between the two wall-function methods. }
    \label{fig:NearestWallFunctionCompare}
\end{figure}

We now compare the two strategies of the standard method and Launder--Spalding method discussed in Section \ref{sec:WallBoundedTurbulentFlow} applied to the wall-adjacent points when enforcing wall functions. For the zero-pressure-gradient flat-plate case, both methods yield comparable $C_{\mathrm{f}}(x)$ as seen in the Figure~\ref{fig:NearestWallFunctionCompare}. 
We therefore proceed with the Launder--Spalding method in subsequent simulations because it preserves transport consistency for $k$ while utilizing wall-stress information in production. 

The next step is to assess the sensitivity of the selection height $\delta h$. Selection height plays a crucial role in applying wall functions. If the selection height is inadequate, fewer or no points may be selected to apply the wall function, which will produce the same or worse results as the CN method. If the selection height is large, points outside of the boundary layer will be selected, which is not the correct approach. Hence, the selection height study is crucial to determine the appropriate height that needs to be selected to apply the wall function.

We consider several values of the parameter $\delta$, namely $\delta h \in \{0.3h,\,0.5h,\,0.7h,\,1.0h\}$. We choose $\delta h > 0.2h$ since $r_{\mathrm{min}}h = 0.2\,h$, which means that there are no points closer than $0.2h$ to the wall (see Figure~\ref{fig:DeletionofPointsClosestToTheBoundary}). Thus, setting $\delta h < 0.2h$ would make the NBN method similar to the CN method with too few points selected. On the other hand, setting $\delta h > h$ means picking points too far away, beyond the neighborhood of the wall, which would also reduce accuracy.

\begin{figure}[!htb]
    \centering
    \resizebox{0.4\textwidth}{!}{
        \begin{tikzpicture}
    \pgfmathsetseed{1218} 
    \draw[thick, name path=curve] (0,0) to[out=30,in=150] (4,0);
    \draw[dotted, thick, name path=shifted_curve] (-0.2,0.44) to[out=30,in=150] (4.2,0.44);
    \foreach \x in {0,1,...,9} {
        \pgfmathsetmacro{\t}{\x/9} 
        \path[name path=vert] ({\t*4}, -1) -- ({\t*4}, 1); 
        \draw [name intersections={of=curve and vert, by=p}, red, fill=red!50] (p) circle (2pt);
        \coordinate (R\x) at (p); 
        \ifnum\x=2
            \pgfmathsetmacro{\offset}{rand*0.2 + 0.8}
            \path[blue, fill=blue!50] (0.95, 1.1) circle  (2pt); 
            \coordinate (B\x) at (0.95, 1.1);

        \else\ifnum\x=5
            \pgfmathsetmacro{\offset}{rand*0.2 + 0.8}
            \path[blue, fill=blue!50] (2.1, 1.2) circle (2pt); 
            \coordinate (B\x) at (2.1, 1.2);

        \else
            \pgfmathsetmacro{\offset}{rand*0.2 + 0.8} 
            \pgfmathsetmacro{\myrand}{rand*0.05} 
            \path[blue, fill=blue!50] ($(p) + (\myrand, \offset)$)  circle (2pt);
            \coordinate (B\x) at ($(p) + (\myrand, \offset)$);

        \fi\fi

        \pgfmathsetmacro{\offset}{rand*0.22 + 1.3} 
        \path[blue, fill=blue!50] ($(p) + (rand*0.05, \offset)$) circle (2pt);

        \pgfmathsetmacro{\offset}{rand*0.27 + 1.8} 
        \path[blue, fill=blue!50] ($(p) + (rand*0.05, \offset)$) circle (2pt);

        \pgfmathsetmacro{\offset}{rand*0.21 + 2.3} 
        \path[blue, fill=blue!50] ($(p) + (rand*0.05, \offset)$) circle (2pt);

        }
        \draw [|<->|, >=latex,dotted] (4.15,0.47) -- (3.85,0.02);
    
        \node [font=\small, rotate around={-10:(0,0)}] at (4.6,0.17) {$r_{\mathrm{min}} h$};    
\end{tikzpicture}
    }
    \caption{Deletion of points if the points get closer than $r_{\mathrm{min}} h$ even between the wall (red) and fluid points (blue). Representation of the point cloud for a simulation case with curved boundary, where the fluid points are deleted if they move closer than the value $r_{\mathrm{min}} h$ to avoid clustering, which gives rise to a void region between the wall points and the fluid points.
    }
    \label{fig:DeletionofPointsClosestToTheBoundary}
\end{figure}

\begin{figure}[!htb]
    \centering
    \begin{minipage}[t]{0.48\textwidth}
        \centering
        \includegraphics[width=\textwidth, height=0.25\textheight, keepaspectratio]{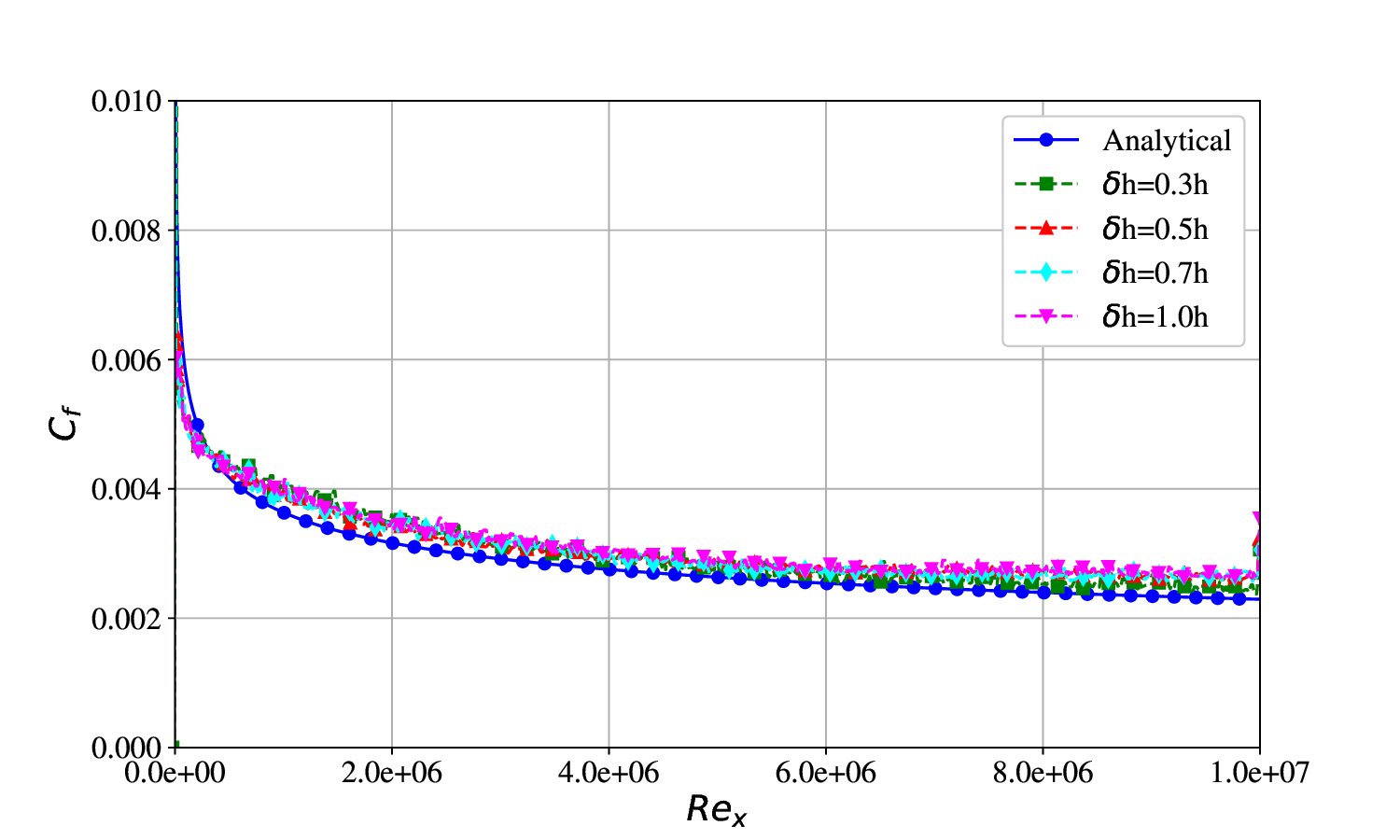}
        \subcaption{Comparison of skin friction coefficient $C_{\mathrm{f}}$ for complete length of the flat plate.}
        \label{fig:NearestSelectionHeight_complete}
    \end{minipage}
    \hfill
    \begin{minipage}[t]{0.48\textwidth}
        \centering
        \includegraphics[width=\textwidth, height=0.25\textheight, keepaspectratio]{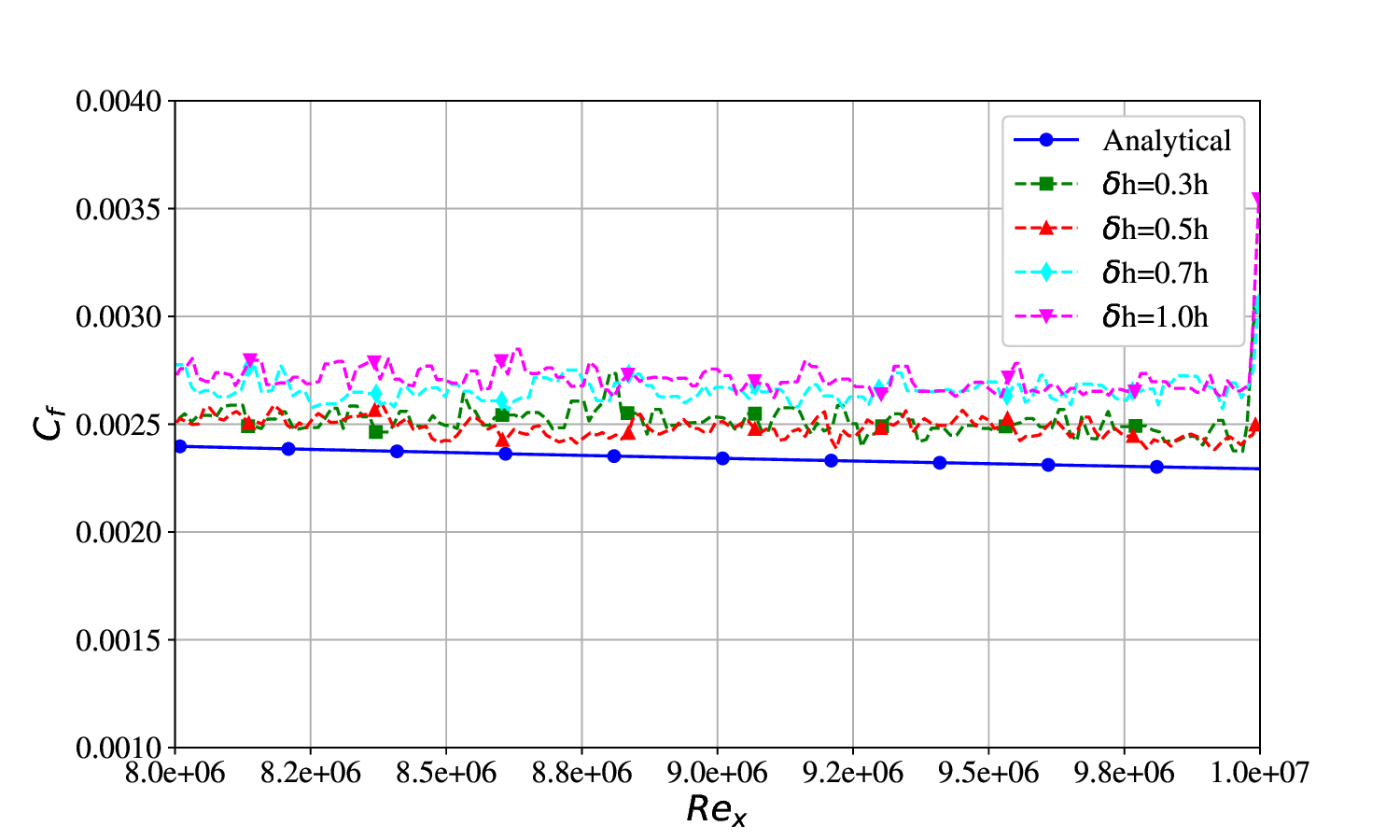}
        \subcaption{Zoomed view of the final section (between $Re_{\mathrm{x}} = 8.0 \times 10^{6}$ to $1.0 \times 10^{7}$) for detailed visualization.}
        \label{fig:NearestSelectionHeight_zoom}
    \end{minipage}
    \caption{Comparison of skin friction coefficient $C_{\mathrm{f}}$ with various selection heights $\delta h$ using the nearest band neighbor method. The plots show selection heights of $\delta h = 0.3h$ (green), $\delta h = 0.5h$ (red), $\delta h = 0.7h$ (cyan), $\delta h = 1h$ (pink) against the analytical expression (blue). The zoomed view reveals that $\delta h = 0.5h$ provides the best match with the analytical solution, followed by $\delta h = 0.3h$, while $\delta h = 0.7h$ and $\delta h = 1h$ show slight deviations due to the selection of a larger number of points for applying the wall function.}
    \label{fig:NearestSelectionHeightComparison}
\end{figure}

Skin friction results for different $\delta$ values are shown in Figure~\ref{fig:NearestSelectionHeightComparison}. All these cases generally follow the analytical trend, but there are some fluctuations caused by random point positions and varying wall-normal distances. If we focus on the well-developed region near the end of the plate, $\delta h = 0.3h$ and $\delta h = 0.5h$ agree best with the analytical values, while $\delta h =1.0h$ is less accurate because it includes points far from the wall corresponding to higher $y^+$ values, which increases errors and computational cost. Based on these results, we use $\delta h = 0.5h$ for further NBN simulations.

\begin{figure}[!htb]
    \centering
    \includegraphics[width=0.6\textwidth]{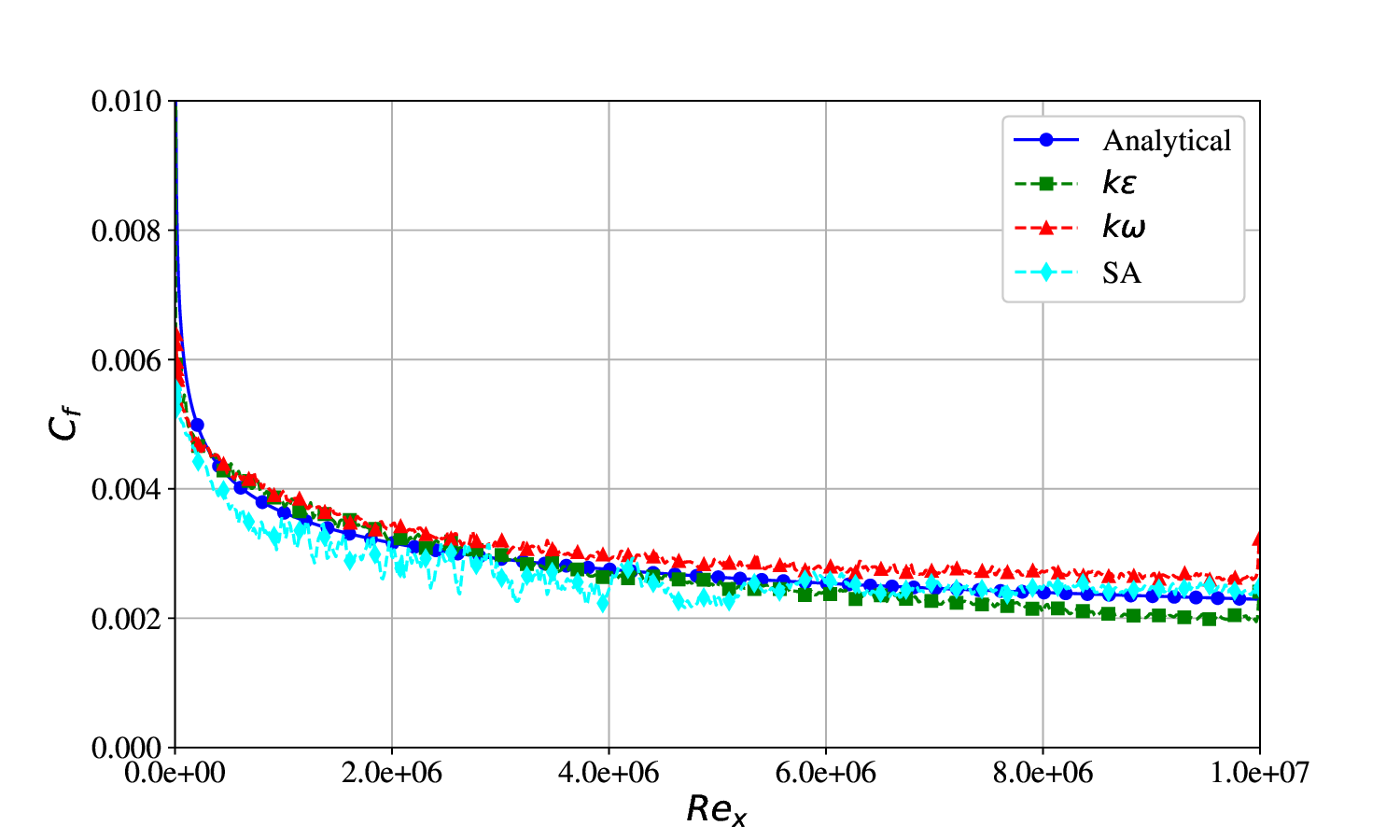}
    \caption{Comparison of the skin friction coefficient, $C_{\mathrm{f}}$, for various turbulence models---Spalart--Allmaras (SA, green line), $k-\varepsilon$ (red line), and $k-\omega$ (cyan line)---against the analytical expression. The simulations employ the nearest-band neighbor method for turbulent flow over a flat plate. The results indicate that the $k-\omega$ model provides superior accuracy near the wall compared to the $k-\varepsilon$ and SA models.}
    \label{fig:NearestTurbulenceComparison}
\end{figure}

Using the NBN method with $\delta h = 0.5h$, we evaluate the performance of three turbulence models: Spalart--Allmaras (SA), standard $k-\varepsilon$, and $k-\omega$ (Wilcox, 2006), as shown in Figure~\ref{fig:NearestTurbulenceComparison}. For the SA model, a Neumann wall boundary condition is applied, and the wall-adjacent eddy viscosity, $\nu_{\mathrm{t}}$, is modified according to the wall function. The SA model exhibits fluctuations in the skin friction coefficient, $C_{\mathrm{f}}$, near the leading edge of the plate. This behavior is due to the model's single-equation formulation, which explicitly relies on the wall distance to compute the destruction of eddy viscosity. The sensitivity to distance, combined with the variations in wall stress and the staggered point distribution, directly induces these fluctuations \cite{spalart_one-equation_1992,huang_skin_1993}. As the boundary layer thickens further downstream, the relative impact of the varying point distances decreases, and the fluctuations are dampened. 

\begin{figure}[!htb]
    \centering
    \includegraphics[width=0.6\textwidth]{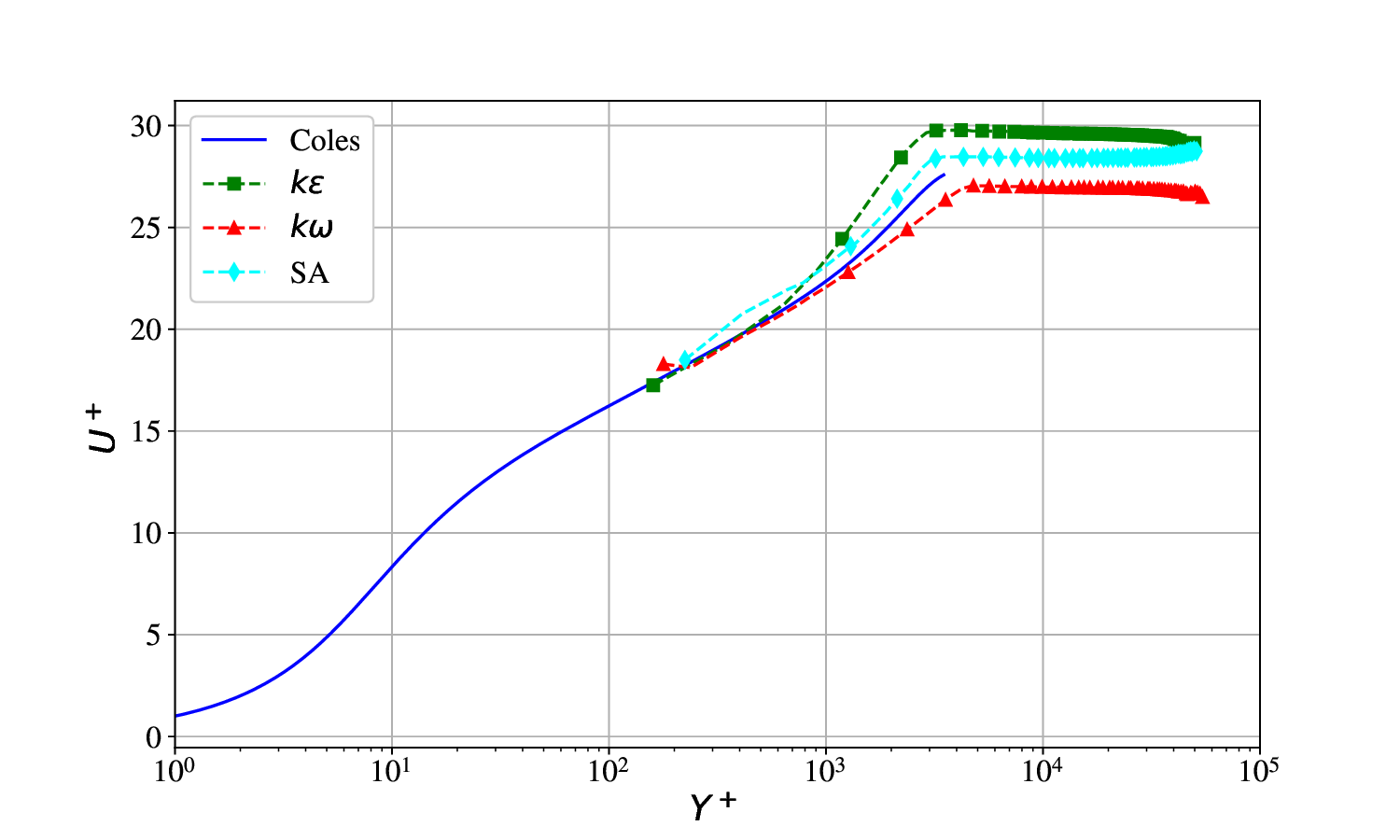}
    \caption{Comparison of the non-dimensional velocity profiles ($u^+$ versus $y^+$) along the wall-normal direction for various turbulence models---SA (green line), $k-\varepsilon$ (red line), and $k-\omega$ (cyan line)---against the analytical expression derived from Coles' law of the wake \cite{coles_law_1956} (blue). The simulations employ the nearest-band neighbor method for turbulent flow over a flat plate. The SA model demonstrates a better overall fit with the analytical velocity profile. In contrast, the $k-\omega$ model underpredicts the velocity near the bulk flow, while the $k-\varepsilon$ model overpredicts it.}
    \label{fig:NearestTurbulenceComparisonUplusYPlus}
\end{figure}

In contrast, the $k-\varepsilon$ and $k-\omega$ formulations are two-equation models that are more numerically diffusive than the SA model. The $k-\varepsilon$ model, in particular, is highly diffusive and heavily depends on the wall function \cite{menter_two-equation_1994,wilcox_reassessment_1988}. Consequently, the fluctuating wall distances result in noisy wall stress approximations, which smear the velocity gradients and lead to a slight underprediction of $C_{\mathrm{f}}$ further along the plate. On the other hand, the $k-\omega$ model is designed to be highly responsive to near-wall gradients \cite{menter_zonal_1993,huang_skin_1993}. In this framework, the high fluctuation of the near-wall stresses causes a gradual overprediction of $C_{\mathrm{f}}$ along the flat plate. Finally, the non-dimensional velocity profiles ($u^+ = u/u_\tau$) extracted at $x = 1.0\,\mathrm{m}$ and $y = 0.5\,\mathrm{m}$ demonstrate that the accuracy of the wall shear stress prediction directly governs the velocity profile, as illustrated in Figure~\ref{fig:NearestTurbulenceComparisonUplusYPlus}.

\subsubsection{Shifted-boundary (SB) Method Analysis}
In a Lagrangian framework, point positions evolve with the flow velocity, and points are dynamically added or removed to maintain spatial uniformity. As a result, the non-dimensional wall distance, $y^+$, at wall-adjacent points cannot be kept constant. This motivates the adoption of the shifted boundary (SB) method to ensure consistent near-wall point distances. In this approach, the wall function is enforced at virtually shifted points, all located at a precisely fixed distance from the wall, $\alpha h$. The diffusion operator utilizes a momentum height, $\beta h$ (as detailed in Section~\ref{sec:ShiftedBoundary}). By fixing this evaluation distance, the SB method guarantees controlled and uniform wall function application, regardless of the underlying point motion. Similar to the NBN evaluation, we systematically analyze the SB method by conducting a grid independence study before proceeding to evaluate its sensitivity to the shift height and its performance across various turbulence models.

\begin{figure}[!htb]
    \centering
    \begin{minipage}[t]{0.48\textwidth}
        \centering
        \includegraphics[width=\textwidth, height=0.25\textheight, keepaspectratio]{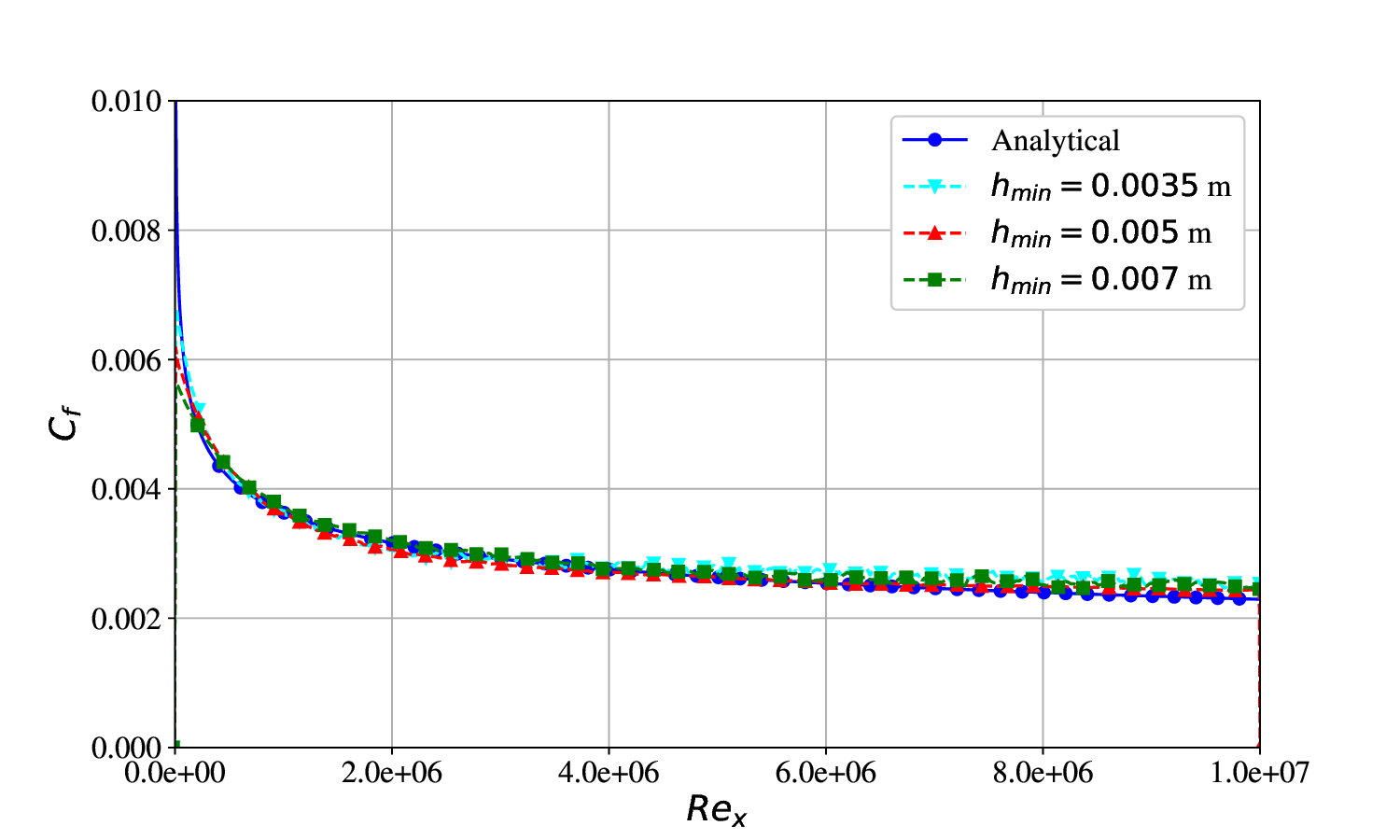}
        \subcaption{Comparison of skin friction coefficient $C_{\mathrm{f}}$ for various resolutions.}
        \label{fig:SBFlatPlateGridStudyCF}
    \end{minipage}
    \hfill
    \begin{minipage}[t]{0.48\textwidth}
        \centering
        \includegraphics[width=\textwidth, height=0.25\textheight, keepaspectratio]{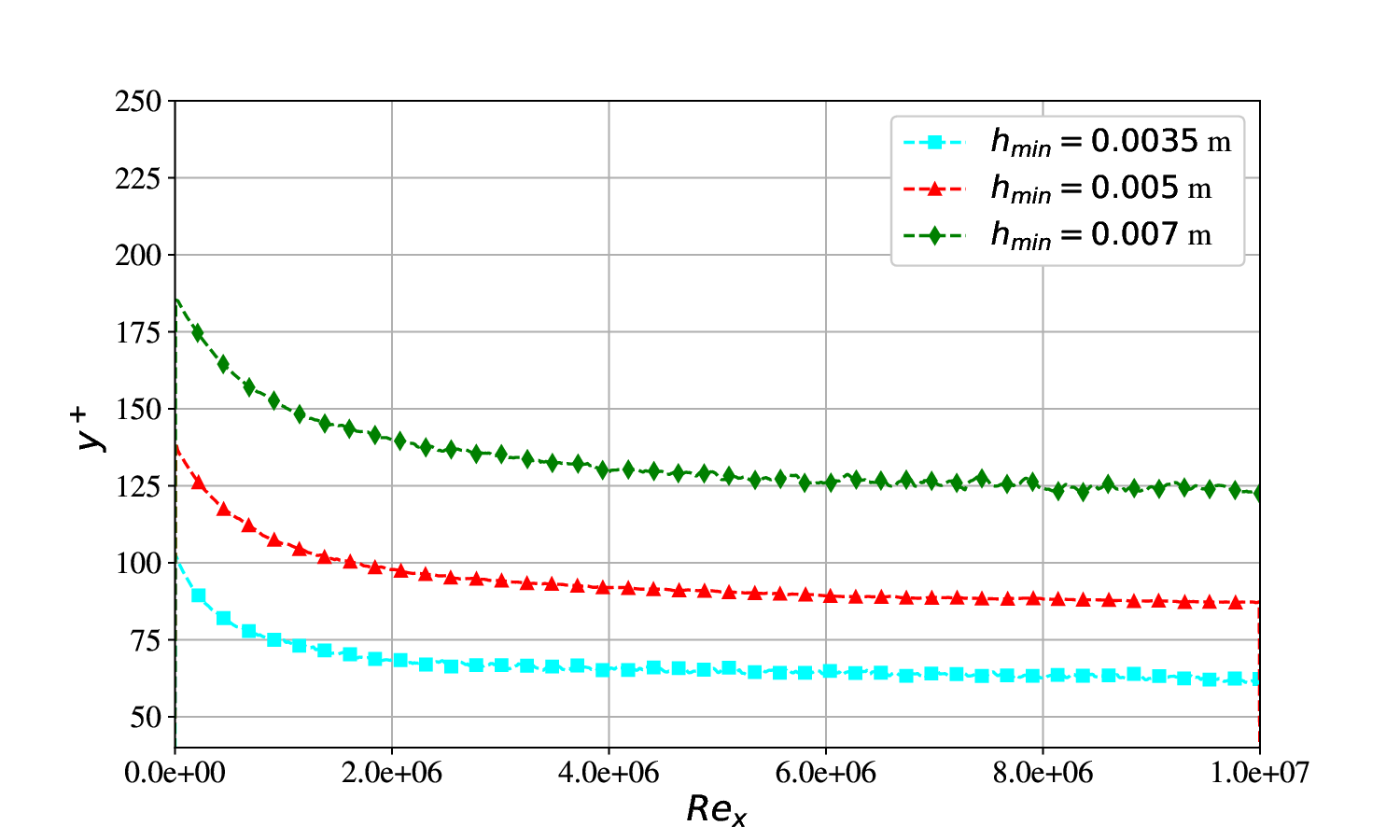}
        \subcaption{Comparison of $y^+$ along the flat plate for various resolutions.}
        \label{fig:SBFlatPlateGridStudyYplus}
    \end{minipage}
    \caption{Grid independence study verifying the spatial resolution dependency of the shifted boundary (SB) method for turbulent flow over a flat plate. Figure~\subref{fig:SBFlatPlateGridStudyCF} compares the skin friction coefficient, $C_{\mathrm{f}}$, computed with varying near-wall resolutions---$h_{\mathrm{min}} = 0.007\,\text{m}$ (green), $0.005\,\text{m}$ (red), and $0.0035\,\text{m}$ (cyan)---against the analytical expression (blue). All resolutions provide similar results, with slight divergence only at the leading edge. Figure~\subref{fig:SBFlatPlateGridStudyYplus} displays the variation of the non-dimensional wall distance, $y^+$, for the same resolutions. A decrease in $h_{\mathrm{min}}$ directly reduces the value of $y^+$. Notably, the fixed shift distance of $\alpha h = 0.1 h$ ensures a perfectly smooth $y^+$ distribution without the spatial oscillations present in standard collocation methods.}
    \label{fig:ShiftedGridStudyComparison}
\end{figure}

To verify grid independence, simulations were performed using three near-wall spatial resolutions: $h_{\mathrm{min}} = 0.007\,\text{m}$, $0.005\,\text{m}$, and $0.0035\,\text{m}$. For these tests, the shift height was fixed at $\alpha h = 0.1 h$. Figure~\ref{fig:SBFlatPlateGridStudyCF} compares the computed skin friction coefficient, $C_{\mathrm{f}}$, against the analytical solution for these grids. The results demonstrate that all three resolutions yield consistent $C_{\mathrm{f}}$ profiles along the length of the plate, with only negligible fluctuations. A minor deviation between the resolutions is observed at the leading edge, where the closer proximity of the finer resolution's shifted points to the wall induces higher velocity gradients and, correspondingly, slightly higher skin friction values.

Figure~\ref{fig:SBFlatPlateGridStudyYplus} illustrates the $y^+$ distribution along the flat plate for the tested resolutions. In contrast to the NBN method, the SB approach produces a smooth and stable $y^+$ profile free of large spatial oscillations. This stability is directly related to the constant shift height. Because the wall-normal distance for all boundary points is uniform, the noisy wall stress approximations seen in the NBN method are eliminated. This translates into the smooth skin friction profiles observed previously. As expected, a reduction in $h_{\mathrm{min}}$ naturally decreases the wall distance, which corresponds to lower $y^+$ values. Because the SB method permits smaller effective near-wall distances than the NBN method while maintaining $y^+$ within a stable range, resolution of $h_{\mathrm{min}} = 0.005\,\text{m}$ provides a balance between accuracy and computational efficiency. This resolution is adopted for all subsequent SB simulations.

\begin{figure}[!htb]
    \centering 
    \includegraphics[width=0.6\textwidth]{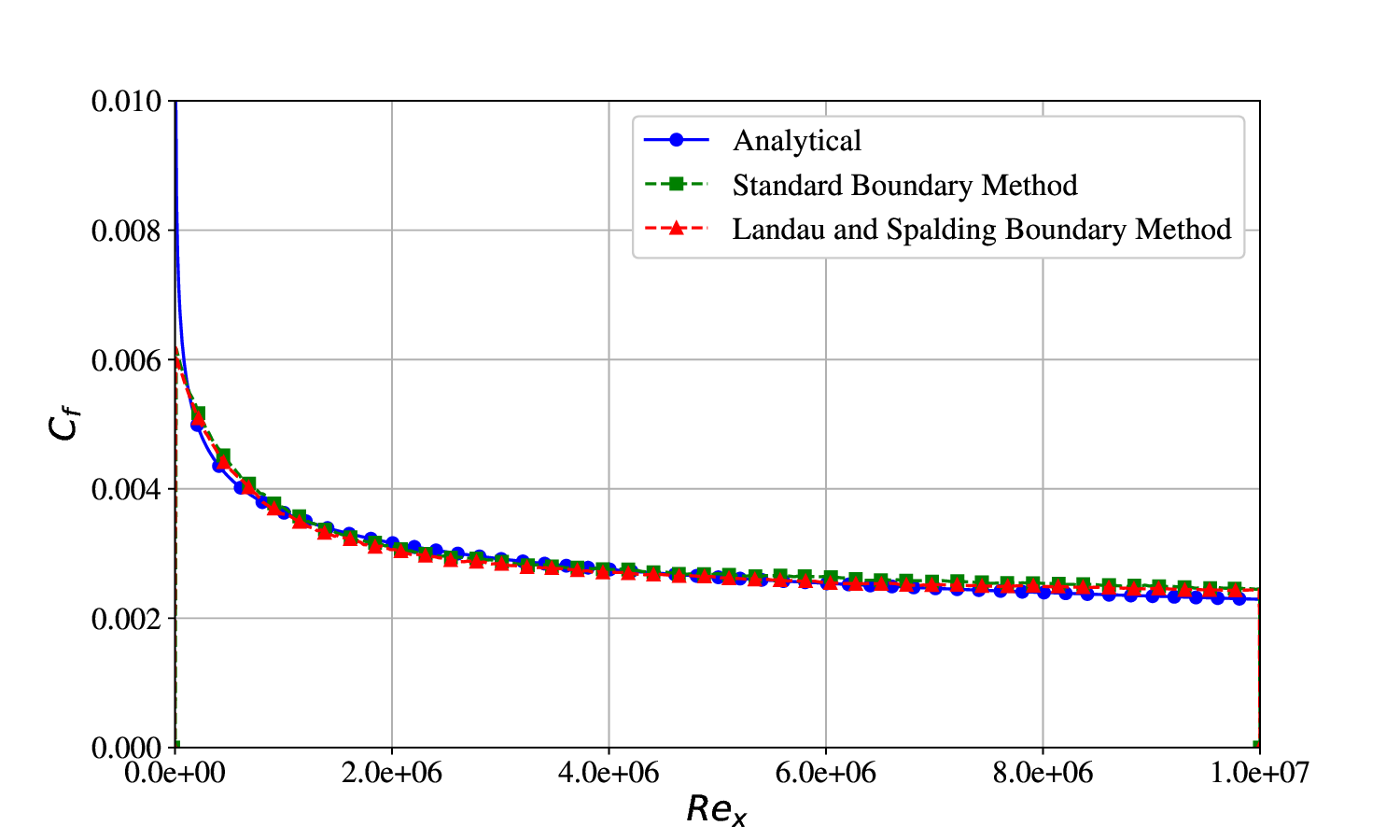}
    \caption{Comparison of Skin Friction Coefficient $C_{\mathrm{f}}$ for shifted boundary method with Standard wall-function method (green) and Launder--Spalding method (red) against the analytical solution (blue) (von Karman, Theodore). The simulation is carried out using the $k-\omega$ turbulence model with resolution $h$ of 0.005, shifted height $\alpha h = 0.1 h$, and momentum thickness of $\beta h = 0.2h$.}
    \label{fig:ShiftedBoundaryWallMethods}
\end{figure}

Initial comparisons between standard and Launder--Spalding wall functions employ shifted height $\alpha h = 0.1h$ and momentum thickness $\beta h = 2.0\alpha h$ with the $k-\omega$ turbulence model. Both methods yield comparable results due to the simple flat plate configuration lacking high turbulent fluctuations or pressure gradients, see Figure~\ref{fig:ShiftedBoundaryWallMethods}. The Launder--Spalding formulation is selected for subsequent analyses.

\begin{figure}[!htb]
    \centering
    \includegraphics[width=0.6\textwidth]{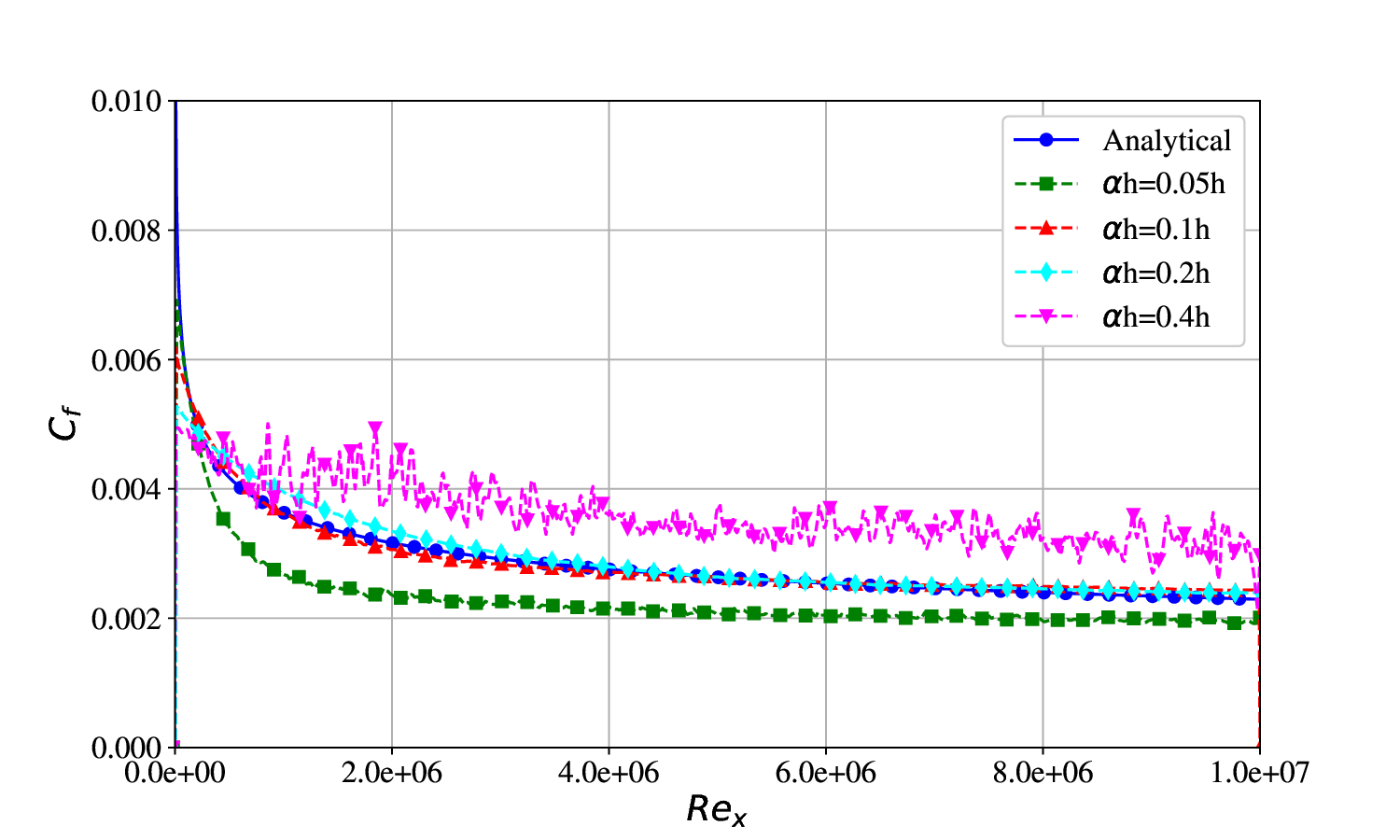}
    \caption{Comparison of Skin Friction Coefficient $C_{\mathrm{f}}$ with different shifted height {$\alpha h$} with shifted boundary method for shifted height of $\alpha h = 0.05h$ (green), $\alpha h = 0.1h$ (red), $\alpha h = 0.2h$ (cyan), $\alpha h = 0.4h$ (pink). The comparison shows that the shifted height of $\alpha h = 0.1h$ to $0.2h$ give the best match to the analytical solution, with $0.2h$ and $0.1h$ giving a larger deviation with respect to the analytical expression.}
    \label{fig:ShiftedBoundaryAlphaHVariation}
\end{figure}

As the wall functions are selected, the next step is to study the effect of the virtual shift height $\alpha h$. For this, shifted heights of $\alpha h = 0.05h$, $0.1h$, $0.2h$, and $0.4h$ are tested, with $\beta h = 2\alpha h$ (see Figure~\ref{fig:ShiftedBoundaryAlphaHVariation}). At $\alpha h = 0.05h$, the shifted and momentum points lie in regions with very few neighbors, which biases the wall function and leads to underprediction of $C_{\mathrm{f}}$. For $\alpha h = 0.1h$ and $0.2h$, the shifted point is located within the region where no other points are present due to the point deletion criteria $r_{\mathrm{min}}h$, and $\beta h$ sits at the interface with the interior, ensuring accurate results. When $\alpha h = 0.4h$, the shifted point is too far from the wall, even resulting in interior points situated between the boundary and shifted points to which the wall function is not applied. This leads to inconsistency in how the wall functions are imposed, leading to large errors similar to those of the CN method. The range $0.1h$ to $0.2h$ (with $\beta h = 2\alpha h$) yields the best performance. For all further SB cases, $\alpha h = 0.1h$ is used. 

\begin{figure}[!htb]
    \centering
    \includegraphics[width=0.6\textwidth]{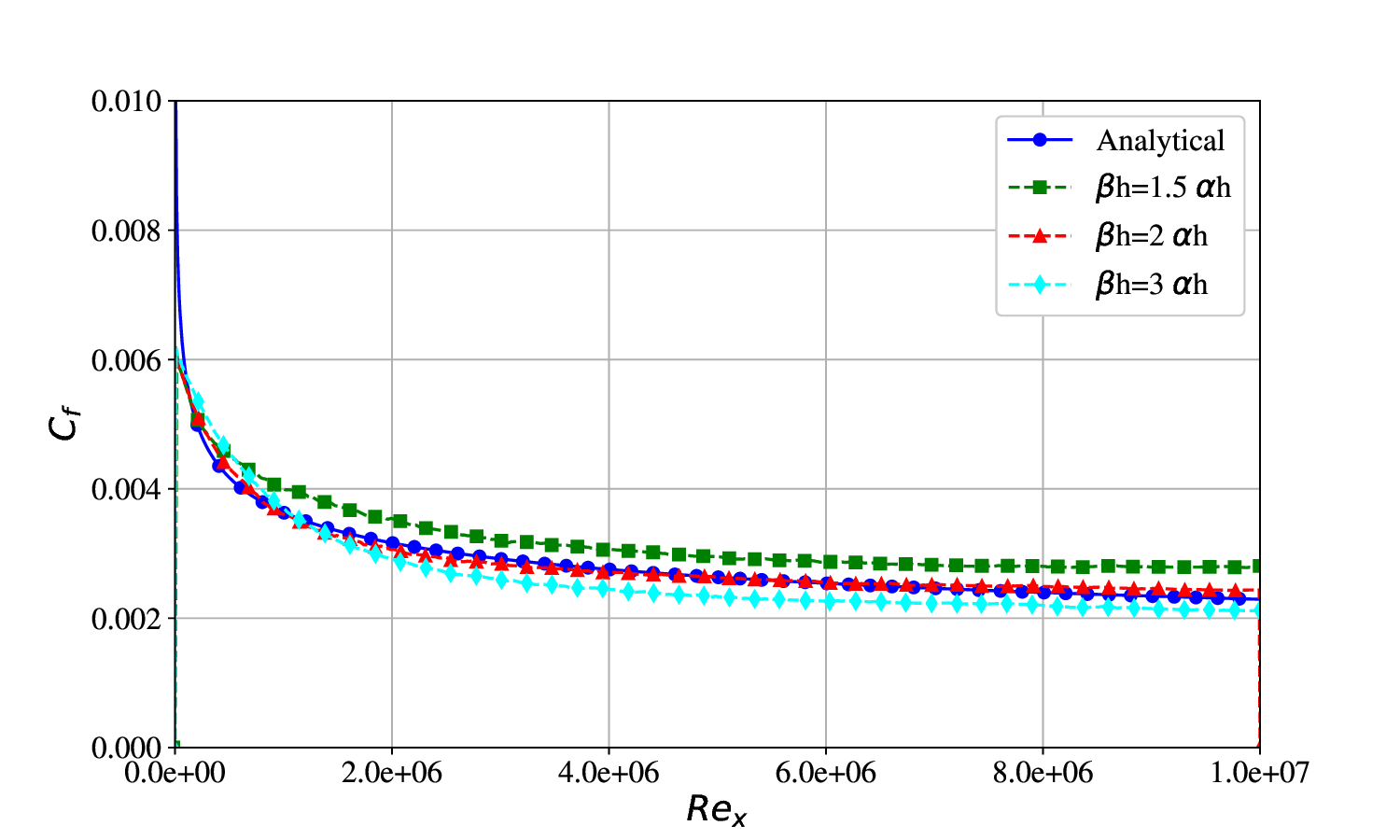}
    \caption{Comparison of the skin friction coefficient, $C_{\mathrm{f}}$, for various momentum heights---$\beta h = 1.5 \alpha h$ (green), $\beta h = 2.0 \alpha h$ (red), and $\beta h = 3.0 \alpha h$ (cyan)---using the shifted boundary method. The $C_{\mathrm{f}}$ comparison demonstrates that maintaining an equidistant spacing between $\alpha h$ and $\beta h$ (i.e., $\beta h = 2.0 \alpha h$) yields superior accuracy.}
    \label{fig:ShiftedBoundaryBetaHVariation}
\end{figure}

Next, we investigate the influence of the momentum height, $\beta h$, while keeping the shifted height, $\alpha h$, constant. Fixing $\alpha h = 0.1h$, we evaluate the accuracy of the diffusion operator using $\beta h$ values of $1.5\alpha h$, $2.0\alpha h$, and $3.0\alpha h$, as shown in Figure~\ref{fig:ShiftedBoundaryBetaHVariation}. The configuration with $\beta h = 2.0\alpha h$ yields the most accurate results, as it is equidistant spacing between the physical wall, the shifted evaluation point, and the momentum height. Setting $\beta h$ too small (placing the stencil too close to the wall) or too large (extending too deeply into the interior) introduces geometric asymmetry and numerical errors. Maintaining $\beta h \geq r_{\mathrm{min}}h$ alongside this near-equidistant spacing ensures optimal accuracy.

We now examine the relationship between the shifted height ($\alpha h$), the momentum height ($\beta h$), and the underlying grid resolution. Because $\alpha h$ and $\beta h$ are defined as functions of the local resolution $h$, their absolute physical distances will vary if the resolution changes. While this scaling is straightforward for flat surfaces, special considerations are necessary for complex geometries with significant curvature. When shifting points along the normal vector of a curved (convex or concave) surface, the physical distance between adjacent shifted points can either expand or contract relative to their original spacing on the boundary. This can lead to points diverging too far apart or overlapping. Furthermore, we decompose the gradient ($\nabla$) and Laplacian ($\Delta$) operators into tangential and normal directions, which assumes a locally flat boundary. Extending the SB method to geometries with strong curvature will require careful treatment of these normal vectors and overlapping regions to maintain simulation integrity.

\begin{figure}[!htb]
    \centering
    \includegraphics[width=0.6\textwidth]{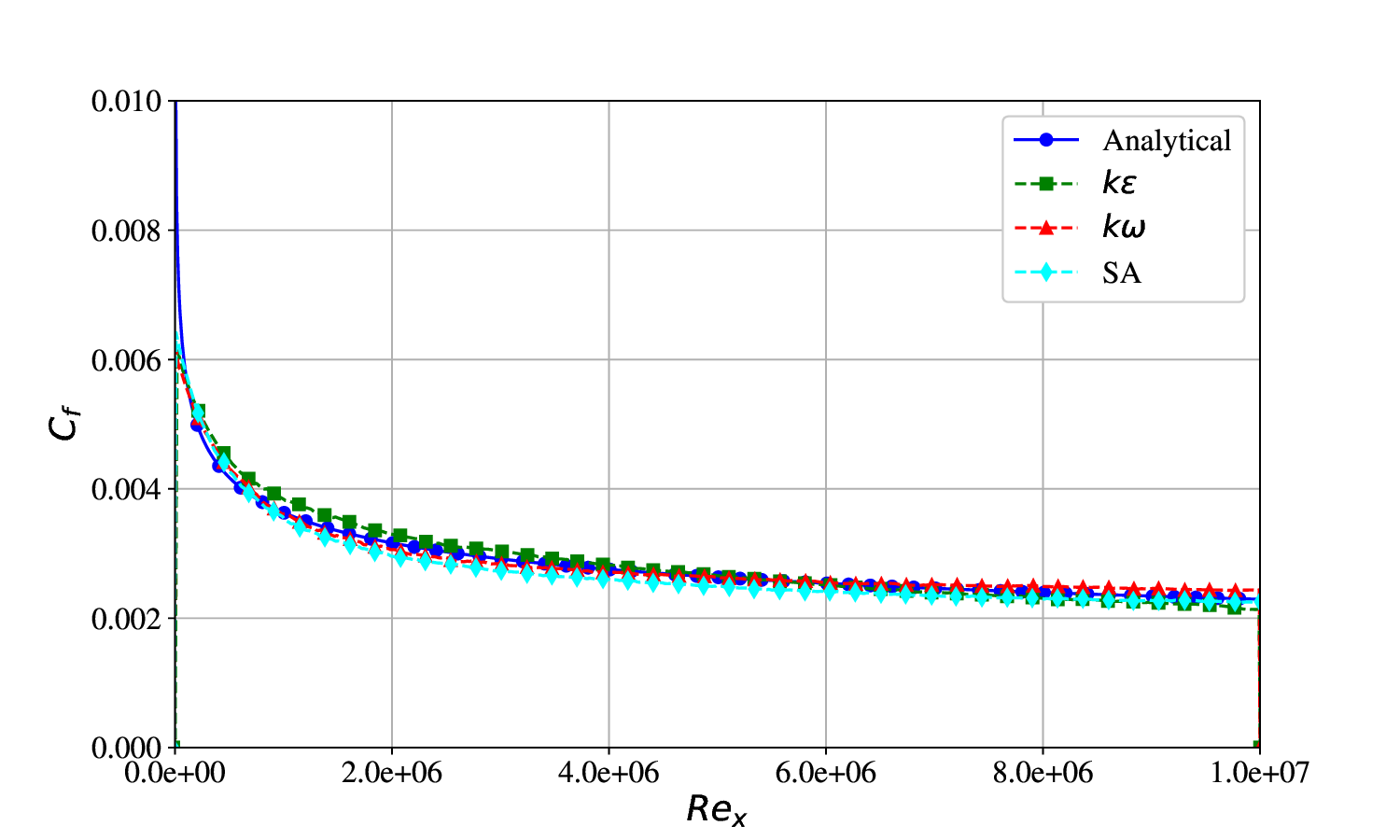}
    \caption{Comparison of Skin Friction Coefficient $C_{\mathrm{f}}$ for various turbulence models (SA (Green line),  $k-\varepsilon$ (Red line), and $k-\omega$ (Cyan line)) against the analytical expression (blue) using the shifted boundary method in turbulent flow Over a flat plate: similar trend can be observed between all the turbulence models and match well with respect to analytical expression with slight variation at the end of the flat plate.}
    \label{fig:ShiftedBoundaryTurbulenceModels}
\end{figure}

Using the optimal parameters $\alpha h = 0.1h$ and $\beta h = 0.2h$, we evaluate the performance of the three turbulence models: SA, $k-\varepsilon$, and $k-\omega$. As shown in Figure~\ref{fig:ShiftedBoundaryTurbulenceModels}, the models produce nearly overlapping $C_{\mathrm{f}}$ curves, with only minor deviations emerging toward the trailing edge of the flat plate. Because the shift heights remain strictly constant in the SB method, the wall stress computations are highly stable, drastically reducing the fluctuations previously observed with the NBN method. The slight remaining differences are due to the formulations of the turbulence models themselves. As noted earlier, the highly diffusive nature of the $k-\varepsilon$ model leads to a slight underprediction of $C_{\mathrm{f}}$. where as, the $k-\omega$ model's sensitivity to near-wall velocity gradients results in increased turbulent production, causing a slight overprediction of $C_{\mathrm{f}}$. Despite these minor variations, the fixed point distances in the SB method ensure a significantly more stable response across all models compared to the NBN method. 

\begin{figure}[!htb]
    \centering
    \includegraphics[width=0.6\textwidth]{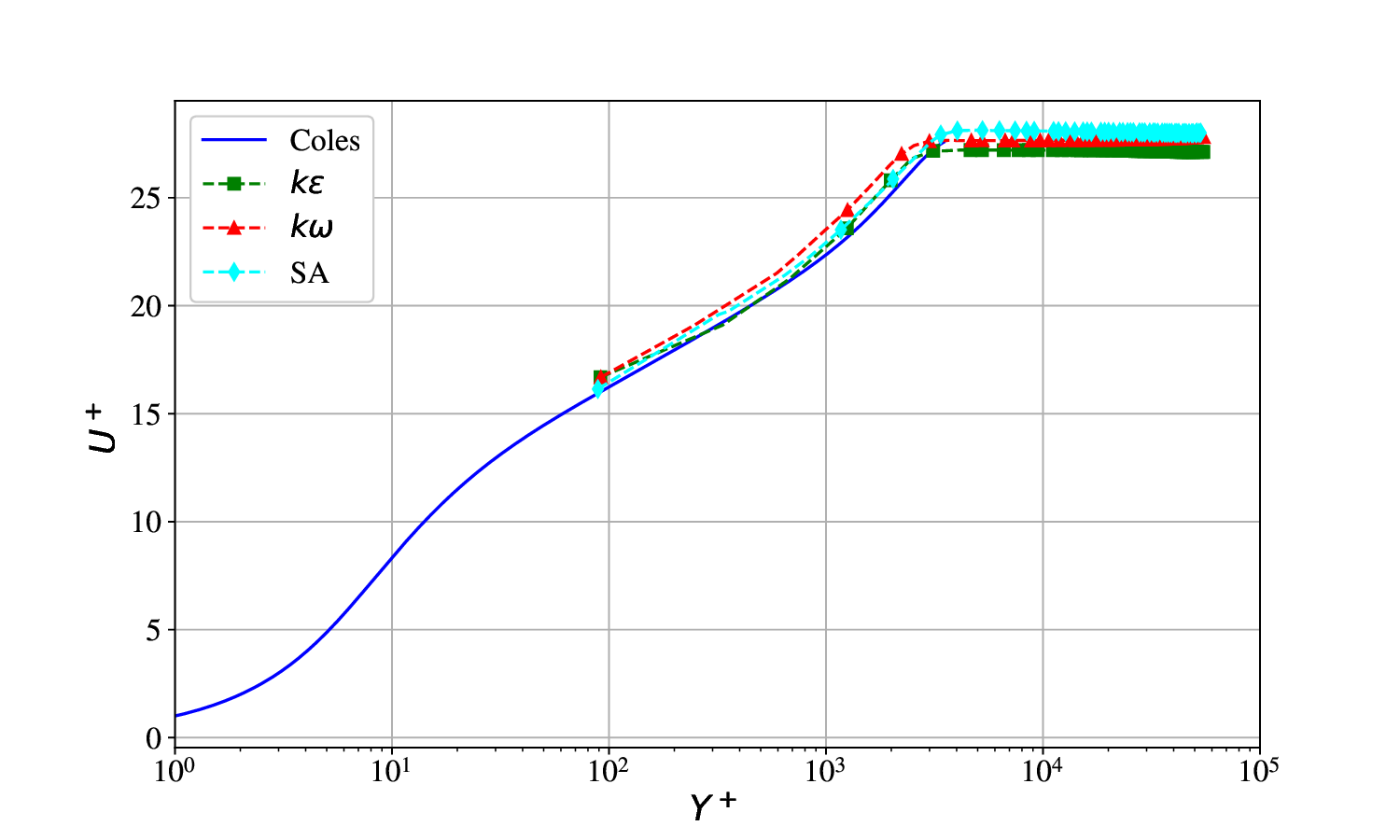}
    \caption{Comparison of velocity profile at the wall normal direction with respect to $U^{+}$ , $y^{+}$ for various turbulence models (SA (Green line),  $k-\varepsilon$ (Red line), and $k-\omega$ (Cyan line)) against Cole's theory \cite{coles_law_1956} (blue) using the shifted boundary method in turbulent flow over a flat plate. The results show a similar profile with minor variation towards the bulk flow region.}
    \label{fig:ShiftedBoundaryWallYplusUplus}
\end{figure}

Furthermore, the non-dimensional velocity profiles ($u^+$ versus $y^+$) for all three models demonstrate excellent agreement with the analytical solution near the wall, with only minor variations emerging in the outer wake region towards the bulk flow as seen in Figure~\ref{fig:ShiftedBoundaryWallYplusUplus}. Compared to the NBN approach, the SB method maintains a fixed wall-normal sampling distance, enabling it to achieve lower, more consistent $y^+$ values for a given spatial resolution. This stability not only helps in the accuracy and robustness of the simulation but also permits the use of coarser grids while keeping $y^+$ within the recommended valid range.

\subsubsection{Computation cost and performance analysis}

While selecting the boundary treatment method, in addition to accuracy, the computation cost required for each method is also crucial. We compare the computation cost for each method to carry out the different steps in the algorithm and the total cost. The CN method is excluded due to poor performance. We compare the remaining NBN and SB methods to reveal a trade-off between the computational overhead required for the boundary treatment and the cost savings gained from a coarser point cloud resolution.

The Shifted Boundary Method (SB) successfully achieves a comparable near-wall resolution (i.e., the same $\boldsymbol{y}^{+}$ value) to the Nearest Band Neighbor Method (NBN), even when employing a resolution 30\% larger (coarser) than that required for the NBN. Crucially, this yields the same skin friction $C_{\mathrm{f}}$ results comparable to the NBN, particularly for the flat plate benchmark case.

Whereas for NBN, achieving this identical $\boldsymbol{y}^{+}$ value with increased resolution causes the point cloud density to increase in all three directions, leading to an overall increase of approximately 50\% to 60\% in the total number of simulation points. With a 50\% increase in the number of points, the total computation time increases by roughly 50\%, raising the total time per time step from $92.5$ seconds to approximately $152.6$ seconds for the NBN method. This demonstrates a clear net advantage for the SB in large simulation cases where minimizing the total number of points is critical for efficiency.

The computational cost of the NBN and SB boundary treatment operations was compared based on a simulation setup comprising approximately $\approx 5 \times 10^5$ points. All simulations were carried out using MPI parallelization on a single compute node equipped with Intel Xeon Gold 6240R processors (48 CPU cores in total) and 384 GB of RAM. To ensure reliable performance data, all reported CPU times are averaged over one thousand time steps.

The NBN has two primary steps contributing to its computational cost: (i) Selection of interior neighbors next to the wall based on the distance ${\delta}h$. (ii) Computation of the wall stress using the wall function for all selected points for both the momentum equation and turbulence equation. The time taken for these two setups are shown in the table~\ref{Table:ComputationCostNBN}

\begin{table}[!htb]
    \centering
    \caption{Computational time for nearest-band neighbor (NBN) method}
    \begin{tabular}{lc}
        \toprule
        \textbf{Operation} & \textbf{Time Taken (seconds)} \\
        \midrule
        Nearest neighbors search (Step 1) & $0.07$ \\
        Applying the wall function (Step 2) & $0.05$ \\
        \bottomrule
    \end{tabular}
    \label{Table:ComputationCostNBN}
\end{table}

The main computational costs for the SB are associated with the geometry and derivative calculations at the boundary. The first is the operation of the virtual shift of the boundary points to a distance $\alpha h$ in the normal direction. The second is the computation of differential operators with respect to the normal and tangential directions.
The differential operator split computation is required for the momentum equation and may also be necessary for turbulence equations (e.g., the kinetic energy ($k$) equation, especially when using the Launder--Spalding wall function). The computation time taken for each of these steps is shown in Table~\ref{Table:ComputationCostSB}

\begin{table}[!htb]
    \centering
    \caption{Computational time for shifted boundary (SB) method}
    \begin{tabular}{lc}
        \toprule
        \textbf{Operation} & \textbf{Time Taken (seconds)} \\
        \midrule
        Virtual shift of points & $0.04$ \\
        Computing normal and tangential components (Momentum Equation) & $1.30$ \\
        Computing gradients/Source Terms ($k$ Equation) & $0.90$ \\
        Computing wall stresses (Turbulence Equation) & $0.07$ \\
        \bottomrule
    \end{tabular}
    \label{Table:ComputationCostSB}
\end{table}

The total time required for the SB boundary treatment operations is significantly higher than the NBN, with the NBN being approximately 19.8 times faster for the localized boundary treatment steps alone.

However, the complete time taken to compute all primary flow equations (momentum, turbulence, and pressure) for all points is approximately $92.5$ seconds per time step. When benchmarked against this total simulation time, the additional cost of the SB is comparatively low. Therefore, the SB offers a significant net advantage by preventing the large $\sim 50\%$ increase in total simulation time that results from the finer point cloud resolution required to make the NBN comparable in near-wall resolution.

\subsection{Flow around a NACA12 wing }

The next simulation test case considers the flow around a curved boundary. For this, we chose a 3D flow simulation around a wing with a \textbf{NACA 0012} airfoil profile. The wing is placed in a rectangular channel, with a chord length of \SI{1}{\meter} and a width of \SI{1}{\meter}. The channel itself is \SI{10}{\meter} long and \SI{5}{\meter} high, with the channel width matching the wing's width. The wing is positioned \SI{2}{\meter} from the inflow and \SI{2.5}{\meter} above the bottom wall, as shown in Figure \ref{fig:NACA0012Layout}. The side, top, and bottom walls of the channel have slip boundary conditions. 

\begin{figure}[!htb]
\centering
\resizebox{0.7\textwidth}{!}{
\begin{circuitikz}
    \draw (0,-1) rectangle (10,3);

    \draw[gray] (1,-0.5) rectangle (11,3.5);

    \draw[thick] plot[domain=0:1, samples=100] 
        ({2 + \x}, {1 + 1.0 * (0.6 * (0.2969 * sqrt(\x) - 0.1260 * \x - 0.3516 * \x^2 + 0.2843 * \x^3 - 0.1036 * \x^4))});
    \draw[thick] plot[domain=0:1, samples=100] 
        ({2 + \x}, {1 + 1.0 * (-0.6 * (0.2969 * sqrt(\x) - 0.1260 * \x - 0.3516 * \x^2 + 0.2843 * \x^3 - 0.1036 * \x^4))});

    \draw[thick, gray] plot[domain=0:1, samples=100] 
        ({3 + \x}, {1.5 + 1.0 * (0.6 * (0.2969 * sqrt(\x) - 0.1260 * \x - 0.3516 * \x^2 + 0.2843 * \x^3 - 0.1036 * \x^4))});
    \draw[thick, gray] plot[domain=0:1, samples=100] 
        ({3 + \x}, {1.5 + 1.0 * (-0.6 * (0.2969 * sqrt(\x) - 0.1260 * \x - 0.3516 * \x^2 + 0.2843 * \x^3 - 0.1036 * \x^4))});

    \draw[gray] (0,-1) -- (1,-0.5);
    \draw[gray] (10,-1) -- (11,-0.5);
    \draw[gray] (0,3) -- (1,3.5);
    \draw[gray] (10,3) -- (11,3.5);

    \draw[gray] 
        ({2 + 0}, {1 + 1.0 * (0.6 * (0.2969 * sqrt(0) - 0.1260 * 0 - 0.3516 * 0^2 + 0.2843 * 0^3 - 0.1036 * 0^4))}) -- 
        ({3 + 0}, {1.5 + 1.0 * (0.6 * (0.2969 * sqrt(0) - 0.1260 * 0 - 0.3516 * 0^2 + 0.2843 * 0^3 - 0.1036 * 0^4))});
    \draw[gray] 
        ({2 + 1}, {1 + 1.0 * (0.6 * (0.2969 * sqrt(1) - 0.1260 * 1 - 0.3516 * 1^2 + 0.2843 * 1^3 - 0.1036 * 1^4))}) -- 
        ({3 + 1}, {1.5 + 1.0 * (0.6 * (0.2969 * sqrt(1) - 0.1260 * 1 - 0.3516 * 1^2 + 0.2843 * 1^3 - 0.1036 * 1^4))});
    \draw[gray] 
        ({2 + 0}, {1 + 1.0 * (-0.6 * (0.2969 * sqrt(0) - 0.1260 * 0 - 0.3516 * 0^2 + 0.2843 * 0^3 - 0.1036 * 0^4))}) -- 
        ({3 + 0}, {1.5 + 1.0 * (-0.6 * (0.2969 * sqrt(0) - 0.1260 * 0 - 0.3516 * 0^2 + 0.2843 * 0^3 - 0.1036 * 0^4))});
    \draw[gray] 
        ({2 + 1}, {1 + 1.0 * (-0.6 * (0.2969 * sqrt(1) - 0.1260 * 1 - 0.3516 * 1^2 + 0.2843 * 1^3 - 0.1036 * 1^4))}) -- 
        ({3 + 1}, {1.5 + 1.0 * (-0.6 * (0.2969 * sqrt(1) - 0.1260 * 1 - 0.3516 * 1^2 + 0.2843 * 1^3 - 0.1036 * 1^4))});

    \node [font=\scriptsize, color = gray, rotate around={20:(0,0)}] at (0.5,1.00) {Inflow};
    \node [font=\scriptsize, rotate around={20:(0,0)}] at (10.5,1.60) {Outflow};
    \node [font=\scriptsize, rotate around={0:(0,0)}] at (2.9,1.2) {Wing};
    \node [font=\scriptsize, rotate around={0:(0,0)}] at (5,3.2) {Slip};
    \node [font=\scriptsize, rotate around={0:(0,0)}] at (4.6,1.6) {Slip};
    \node [font=\scriptsize, color = gray, rotate around={0:(0,0)}] at (5.8,-0.8) {Slip};
    \node [font=\scriptsize, color = gray,  rotate around={0:(0,0)}] at (7,1.8) {Slip};

    \draw [<->, >=Stealth] (0,-1.1) -- (10,-1.1);
    \draw [<->, >=Stealth] (-0.1,-1.0) -- (-0.1,3.0);
    \draw [<->, >=Stealth] (0.2,3.0) -- (1.15,3.5);
    \draw [<->, >=Stealth] (3.0,1.6) -- (4.0,1.6);
    \draw [<->, >=Stealth] (2.5,-1.0) -- (2.5,0.95);
    \draw [<->, >=Stealth] (3.0,1.0) -- (10.0,1.0); 

    \node [font=\scriptsize, rotate around={0:(0,0)}] at (5.0,-1.25) {10 m};
    \node [font=\scriptsize, rotate around={90:(0,0)}] at (-0.25,1.0) {5 m};
    \node [font=\scriptsize, rotate around={30:(0,0)}] at (0.8,3.2) {1 m};
    \node [font=\scriptsize, rotate around={90:(0,0)}] at (2.35,0.2) {2.5 m};
    \node [font=\scriptsize, rotate around={0:(0,0)}] at (3.5,1.7) {1 m};
    \node [font=\scriptsize, rotate around={0:(0,0)}] at (6.0,1.15) {7 m};

    \draw [->, >=Stealth] (-1.5,-1.25) -- (-1.5,-0.5);
    \node [font=\scriptsize, rotate around={0:(0,0)}] at (-1.7,-0.5) {Y};
    
    \draw [->, >=Stealth] (-1.5,-1.25) -- (-0.75,-1.25);
    \node [font=\scriptsize, rotate around={0:(0,0)}] at (-0.75,-1.5) {X};

    \draw [->, >=Stealth] (-1.5,-1.25) -- (-0.85, -0.8);
    \node [font=\scriptsize, rotate around={0:(0,0)}] at (-0.9, -0.65) {Z};

    
\end{circuitikz}
}
\caption{Flow over a 3D wing with NACA-0012 profile, illustrating the dimensions of the rectangular channel and the placement of the wing. Includes inlet conditions such as specified velocity and pressure gradient, along with slip and no-slip boundary condition sections.}
\label{fig:NACA0012Layout}
\end{figure}

For the simulation of flow over the 3D NACA 0012 wing, the fluid is modeled as air with a density of $\rho = \SI{1}{\kilogram\per\meter\cubed}$ and a kinematic viscosity of $\nu = \SI{8.544e-6}{\meter\squared\per\second}$. The freestream inflow velocity is set to $V_{\infty} = \SI{51.2}{\meter\per\second}$ at a $0^\circ$ angle of attack. This yields a Mach number of approximately $M = 0.15$, allowing the flow to be treated as purely incompressible. Based on the wing's chord length, the resulting Reynolds number is $Re = 6 \times 10^6$. This specific Reynolds number was carefully selected to ensure the flow becomes fully turbulent almost immediately. While lower values, such as $Re = 10^5$, technically fall within the turbulent regime, they exhibit a prolonged laminar-to-turbulent transition. This delayed transition induces variations in the boundary layer thickness, complicating direct comparisons with established benchmark data for skin friction and pressure coefficients. Therefore, the $Re = 6 \times 10^6$ case ensures a more robust and reliable validation of the fully turbulent boundary treatments.

\begin{figure}[!htb]
    \centering
    \begin{tikzpicture}
        \node[anchor=south west, inner sep=0] (image) at (0,0) 
            {\includegraphics[width=0.6\textwidth]{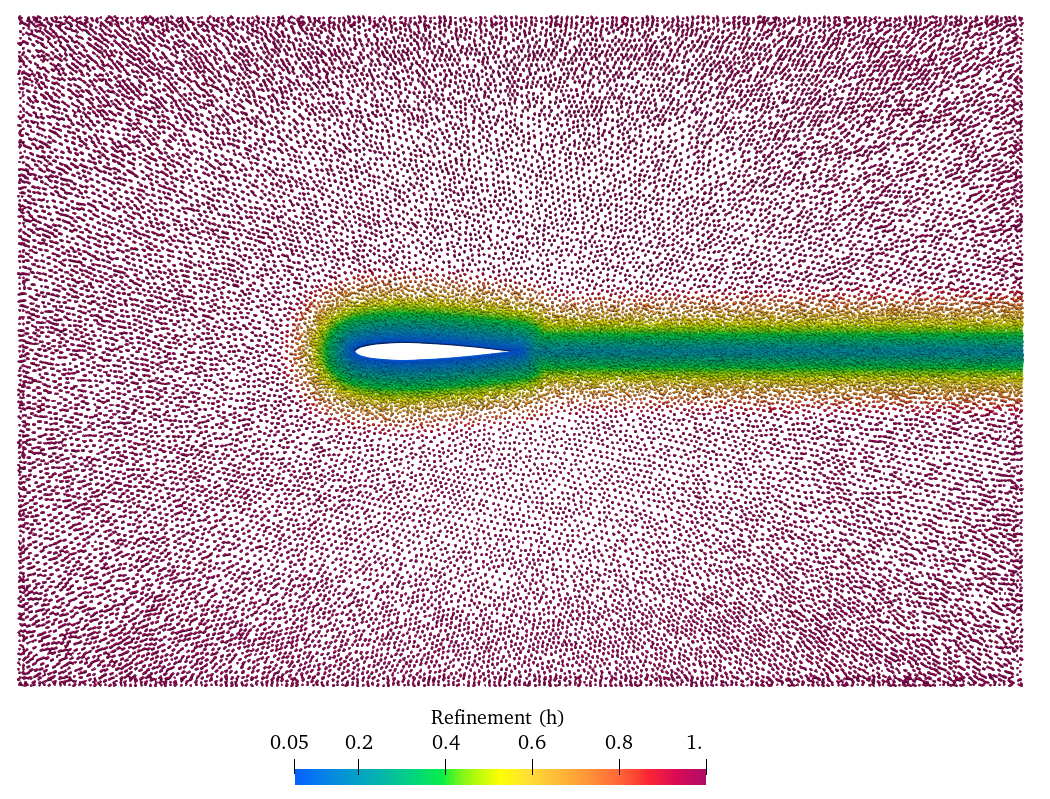}};
        
        \begin{scope}[x={(image.south east)}, y={(image.north west)}]
        \end{scope}
        
        \draw [->, >=Stealth, thick] (-0.8,0.5) -- (-0.8,1.3) node[left, font=\scriptsize] {Y};
        \draw [->, >=Stealth, thick] (-0.8,0.5) -- (-0.1,0.5) node[below, font=\scriptsize] {X};
    \end{tikzpicture}
    \caption{Refinement strategy for the flow around a NACA 0012 profile. The normalized refinement size $h$ varies from $0.05$ (blue) near the airfoil surface to $1.0$ (magenta) at the far-field boundary. All $h$ values are normalized, and this visualization serves to illustrate the gradient variation across the domain.}
    \label{fig:NACA0012Refinement}
\end{figure}

The spatial discretization for the meshfree method (see \cite{suchde2023point}) is generated such that the point cloud resolution, $h$, is highly refined near the wing surface and gradually coarsens with increasing distance into the far-field, as shown in Figure~\ref{fig:NACA0012Refinement}. Using this discretization strategy, we evaluate the two proposed boundary treatment methods: the nearest-band neighbor (NBN) method and the shifted boundary (SB) method. 

\begin{figure}[!htb]
    \centering
    \includegraphics[width=0.6\textwidth]{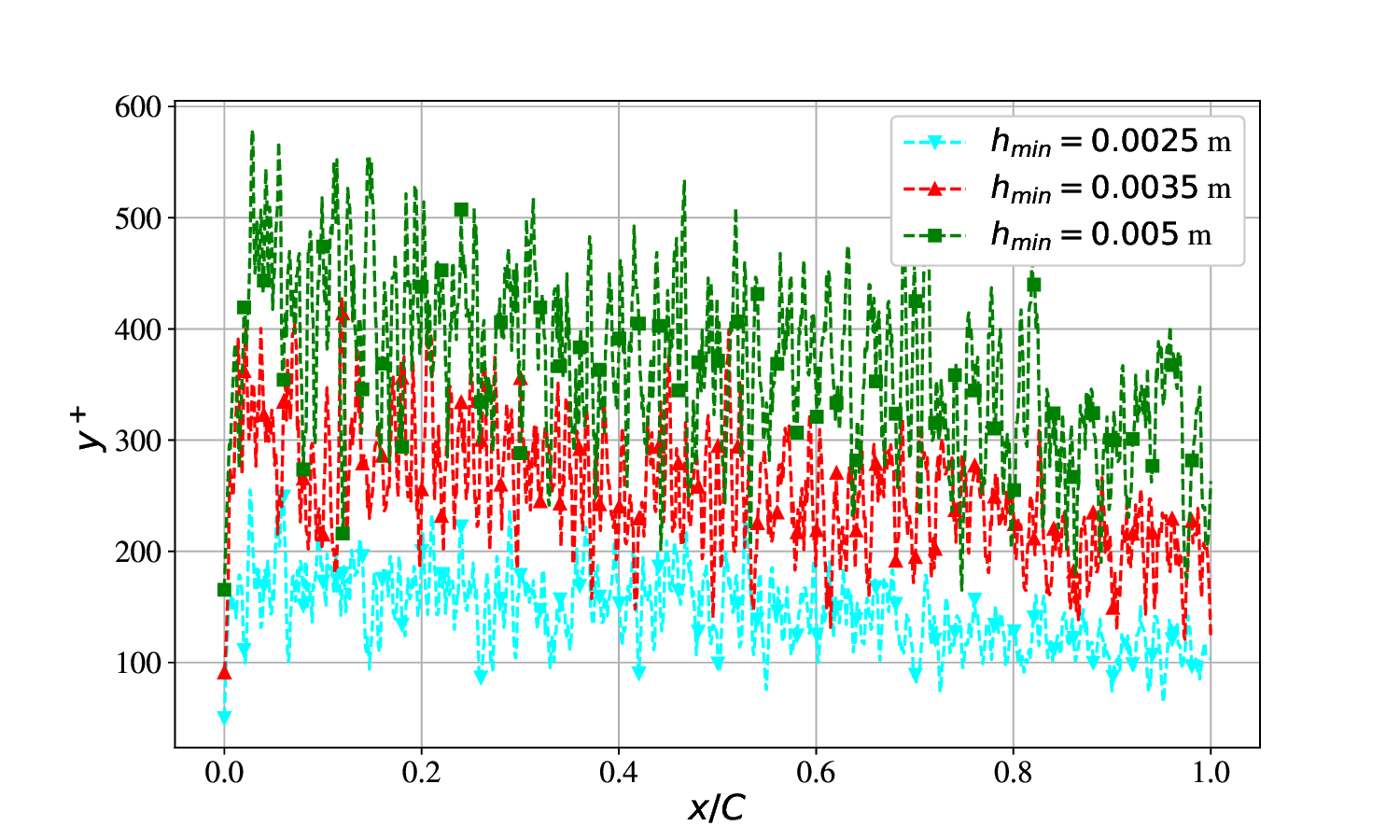}
    \caption{A grid independence study verifying the spatial resolution dependency of the nearest-band neighbor (NBN) method for turbulent flow over a 3D NACA 0012 wing. The figure displays the variation of the non-dimensional wall distance, $y^+$, with respect to the near-wall resolution, $h_{\mathrm{min}}$. Due to high-frequency spatial fluctuations caused by the staggered positions of the interior points, the simulation results have been down-sampled and smoothed. Even after smoothing, notable variations in the local $y^+$ values are visible. Across the tested resolutions, a decrease in $h_{\mathrm{min}}$ consistently reduces the overall magnitude of $y^+$.}
    \label{fig:NACA0012NBNYplusCompare}
\end{figure}

\begin{figure}[!htb]
    \centering
    \begin{minipage}[t]{0.48\textwidth}
        \centering
        \includegraphics[width=\textwidth, height=0.25\textheight, keepaspectratio]{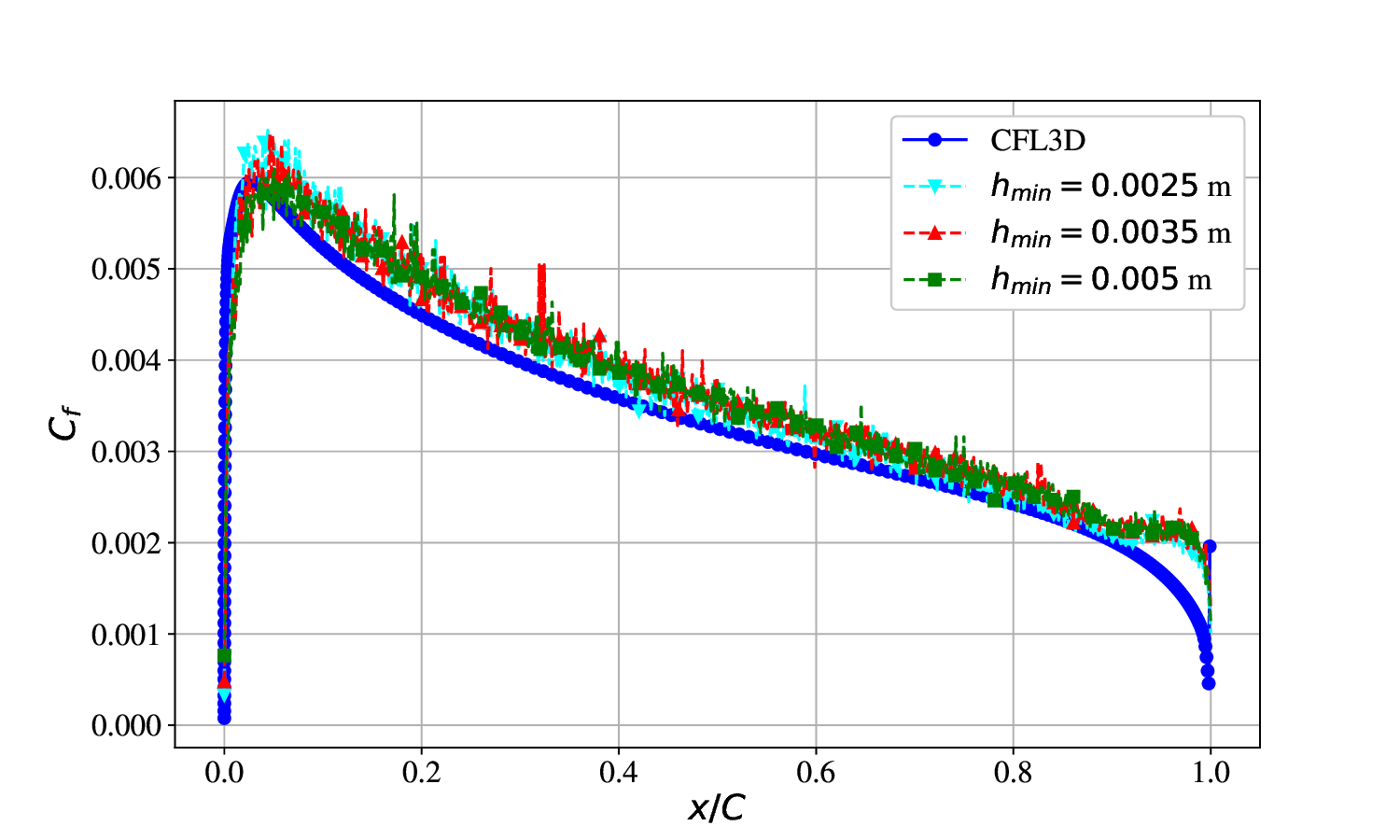}
        \subcaption{Comparison of skin friction coefficient $C_{\mathrm{f}}$ for various resolutions.}
        \label{fig:NBN_NACA_GridStudy_Cf}
    \end{minipage}
    \hfill
    \begin{minipage}[t]{0.48\textwidth}
        \centering
        \includegraphics[width=\textwidth, height=0.25\textheight, keepaspectratio]{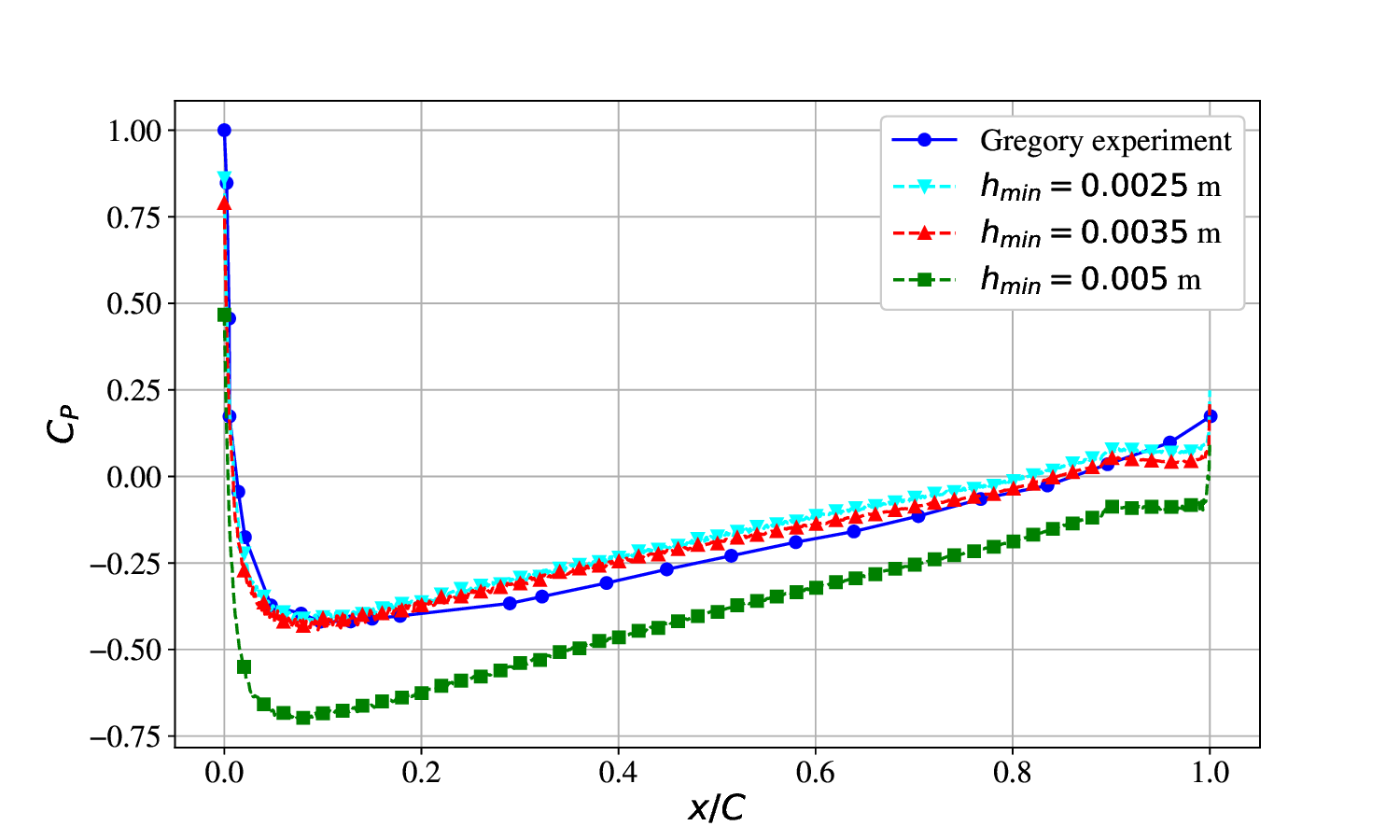}
        \subcaption{Comparison of pressure coefficient $C_{\mathrm{p}}$ for various resolutions.}
        \label{fig:NBN_NACA_GridStudy_Cp}
    \end{minipage}
    \caption{Grid independence study for the 3D NACA 0012 wing using the NBN method. Figure~\subref{fig:NBN_NACA_GridStudy_Cf} compares the skin friction coefficient, $C_{\mathrm{f}}$, computed with near-wall resolutions of $h_{\mathrm{min}} = \SI{0.005}{\meter}$ (green), $h_{\mathrm{min}} = \SI{0.0035}{\meter}$ (red), and $h_{\mathrm{min}} = \SI{0.0025}{\meter}$ (cyan) against reference CFD results from a 3D CFL solver (blue). All resolutions provide similar results, except at the trailing edge due to the numerical smoothing of the sharp wing tip. Figure~\subref{fig:NBN_NACA_GridStudy_Cp} shows the variation of the pressure coefficient, $C_{\mathrm{p}}$, for the same resolutions. The finer resolutions of $h_{\mathrm{min}} = \SI{0.0025}{\meter}$ and $h_{\mathrm{min}} = \SI{0.0035}{\meter}$ produce nearly identical results that closely match the experimental data \cite{gregory_low-speed_1970}, deviating only at the smoothed wing tip. The coarsest resolution, $h_{\mathrm{min}} = \SI{0.005}{\meter}$, noticeably underpredicts $C_{\mathrm{p}}$ due to an insufficient number of near-wall points and correspondingly larger $y^+$ values.}
    \label{fig:NACA_GridStudyComparison_cf_cp}
\end{figure}

To evaluate the NBN method on a complex geometry, a grid independence study was conducted for the flow around the NACA 0012 profile. Three near-wall resolutions were considered: $h_{\mathrm{min}} = 0.0025\,\text{m}$, $0.0035\,\text{m}$, and $0.005\,\text{m}$. The variation of $y^+$ along the wing can be seen from the figure~\ref{fig:NACA0012NBNYplusCompare}, where the $y^+$ values are down-sampled and a smoothed distribution is shown for better visualization. We observe spatial fluctuations that are larger than those seen in the flat plate case. This amplification is driven by the surface curvature and the multi-directional movement of the Lagrangian points relative to the local boundary tangent. Nevertheless, decreasing the resolution consistently reduces both the overall magnitude and the fluctuation band of the $y^+$ values. Figure~\ref{fig:NACA_GridStudyComparison_cf_cp} shows the variation of the skin friction and pressure coefficients for the different resolutions considered. The variation of skin friction coefficient, $C_{\mathrm{f}}$, remains stable across all three resolutions, even when the local $y^+$ values for the coarser grid extend beyond the ideal bounds of the boundary layer. A localized deviation in $C_{\mathrm{f}}$ is observed at the trailing edge. This is a numerical artifact caused by the geometric smoothing of the sharp wing tip, which induces a localized pressure drop and artificially elevates the skin friction.

Analyzing the pressure coefficient, $C_{\mathrm{p}}$, the finer resolutions ($h_{\mathrm{min}} = 0.0025\,\text{m}$ and $0.0035\,\text{m}$) align quite well with the experimental results, aside from minor discrepancies at the mid-chord and the smoothed trailing edge. In contrast, the coarsest resolution ($h_{\mathrm{min}} = 0.005\,\text{m}$) exhibits a pronounced underprediction of $C_{\mathrm{p}}$. This loss of accuracy is due to the insufficient density of near-wall points, which causes the $y^+$ value to exceed $500$ near the leading edge, leading to poor boundary layer approximations.

Table~\ref{tab:naca0012_coefficients_GridStudy} presents the aerodynamic coefficients for the grid study. For a symmetric airfoil at a $0^\circ$ angle of attack, the theoretical lift coefficient is $C_{\mathrm{L}} = 0$. While the finer resolutions produce $C_{\mathrm{L}}$ values closer to zero, a slight lift is still recorded, and the coarser resolution deviates further. Additionally, the drag coefficients ($C_{\mathrm{D}}$) are slightly elevated. This is due to the trailing edge curvature modifications and the numerical error due to staggered points and wall distances. The non-zero $C_{\mathrm{L}}$ is a known characteristic of Lagrangian meshfree methods: unlike structured finite volume meshes, the scattered point cloud is inherently asymmetric. This non-uniform distribution introduces minor asymmetric approximation errors within the differential operators, leading to slight deviations in the integrated forces \cite{benito_influence_2001}.

\begin{table}[!htb]
\centering
\caption{Aerodynamic Coefficients for NACA 0012 Airfoil at $0^\circ$ Angle of Attack for various resolutions using the NBN method.}
\label{tab:naca0012_coefficients_GridStudy}
\begin{tabular}{cccc}
\hline
\textbf{Resolution} & \textbf{Lift Coefficient ($C_{\mathrm{L}}$)} & \textbf{Drag Coefficient ($C_{\mathrm{D}}$)} \\
\hline
$h_{\mathrm{min}} = 0.0025 \text{ m}$ & 0.0084 & 0.0109 \\
$h_{\mathrm{min}} = 0.0035 \text{ m}$ & 0.0085 & 0.0157 \\
$h_{\mathrm{min}} = 0.005 \text{ m}$  & 0.0120 & 0.0038 \\
\hline
\end{tabular}
\end{table}

\begin{figure}[!htb]
    \centering
    \begin{minipage}[t]{0.48\textwidth}
        \centering
        \includegraphics[width=\textwidth, height=0.25\textheight, keepaspectratio]{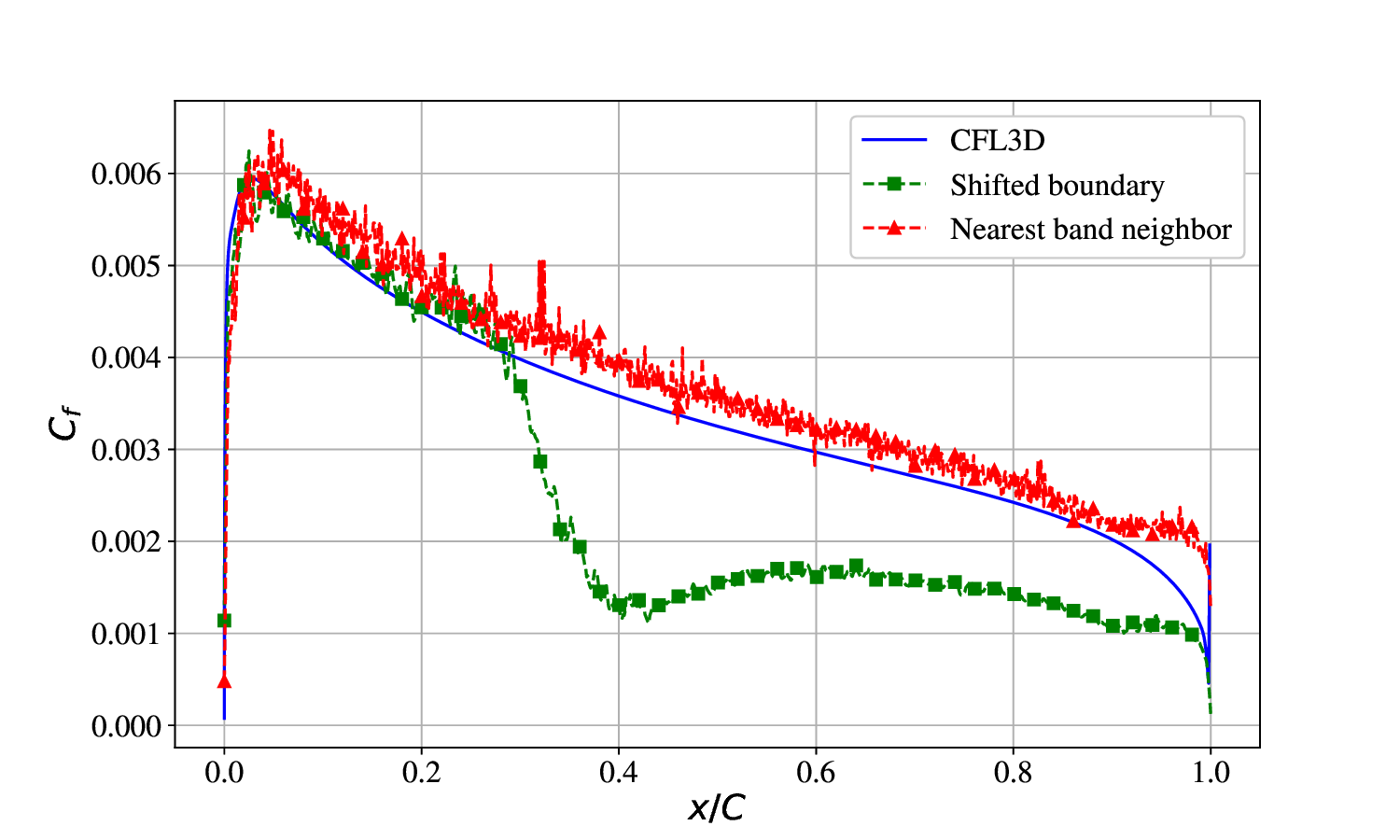}
        \subcaption{Comparison of skin friction coefficient $C_{\mathrm{f}}$ for SB and NBN methods.}
        \label{fig:SB_VS_NBN_NACA_Cf}
    \end{minipage}
    \hfill
    \begin{minipage}[t]{0.48\textwidth}
        \centering
        \includegraphics[width=\textwidth, height=0.25\textheight, keepaspectratio]{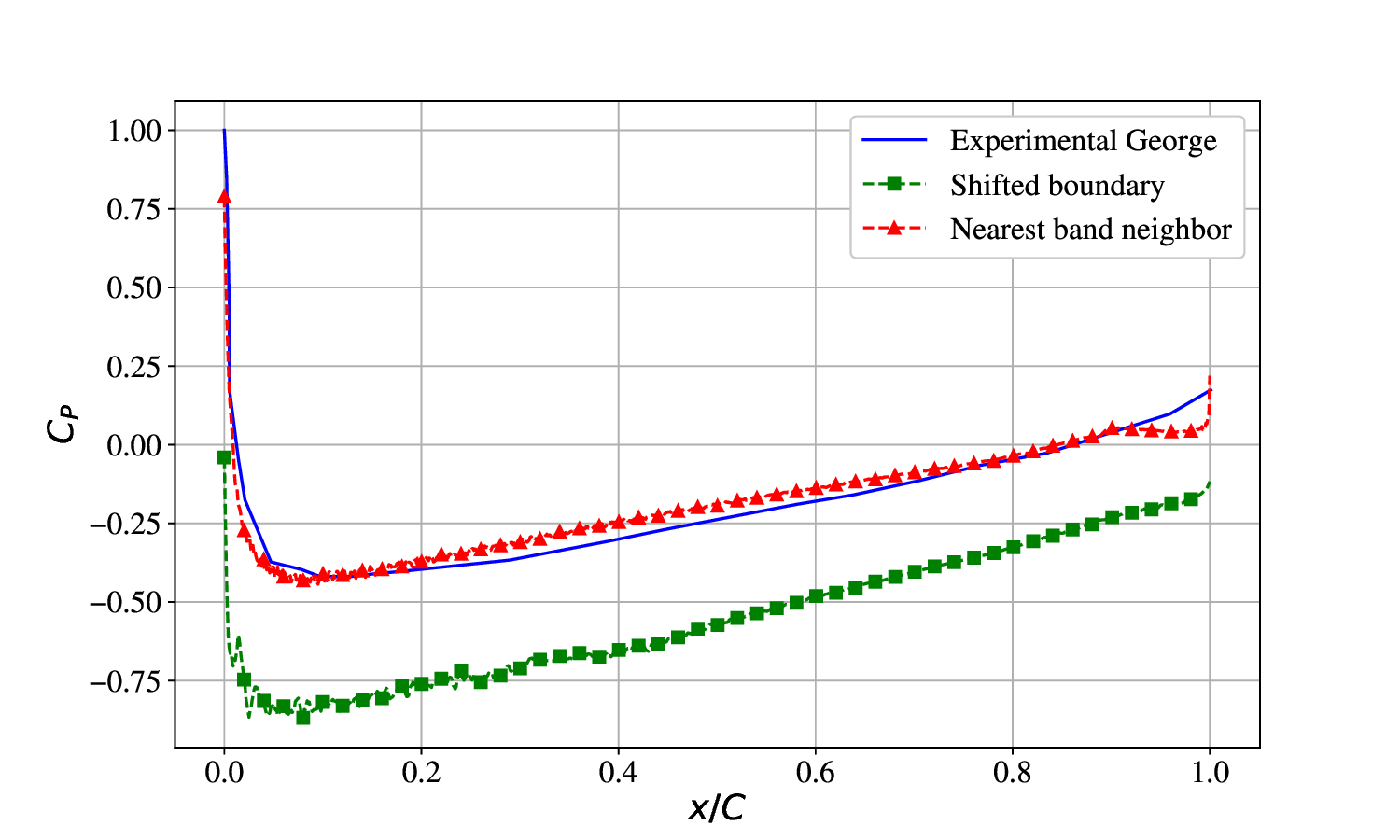}
        \subcaption{Comparison of pressure coefficient $C_{\mathrm{p}}$ for SB and NBN methods.}
        \label{fig:SB_VS_NBN_NACA_Cp}
    \end{minipage}
    \caption{Comparison of the skin friction coefficient ($C_{\mathrm{f}}$) and pressure coefficient ($C_{\mathrm{p}}$) along the length of the wing for the shifted boundary (SB, green) and nearest-band neighbor (NBN, red) methods against reference results from CFL3D \cite{freeman_verification_2014}. Simulations were performed using a near-wall resolution of $h = 0.0035\,\mathrm{m}$ and the $k-\omega$ turbulence model. Figure~\subref{fig:SB_VS_NBN_NACA_Cf} shows that the NBN method agrees well with the reference $C_{\mathrm{f}}$ data, except at the smoothed trailing edge. The SB method, however, deviates significantly past the thickest part of the wing due to uncorrected curvature effects. Figure~\subref{fig:SB_VS_NBN_NACA_Cp} demonstrates that the $C_{\mathrm{p}}$ is severely underpredicted by the SB method due to flow detachment caused by the lack of a stabilized pressure correction step in adverse gradients.}
    \label{fig:NACA0012CfCompare}
\end{figure}

Following the grid analysis for NBN, both the NBN and SB boundary treatment methods were evaluated using the Launder--Spalding wall function coupled with the standard $k-\omega$ turbulence model. For the NBN method, a near-wall resolution of $h = \SI{0.0035}{\meter}$ is used alongside a selection height of $\delta h = 0.4h$. For the SB method, the near-wall resolution is set to $h = \SI{0.0035}{\meter}$, with a shift height of $\alpha h = 0.1h$ and a momentum height of $\beta h = 0.2h$. We compare the resulting $C_{\mathrm{f}}$ and $C_{\mathrm{p}}$ distributions against reference data generated by the structured finite volume CFD solver, CFL3D \cite{freeman_verification_2014}, as shown in the figure~\ref{fig:NACA0012CfCompare}. The NBN method predicts the $C_{\mathrm{f}}$ well, closely tracking the CFL3D reference values as seen earlier from the grid study.

In contrast, the SB method exhibits significant deviations on this complex geometry. While it matches the reference data up to the thickest part of the wing ($x \approx 0.3\,\text{m}$), the $C_{\mathrm{f}}$ and $C_{\mathrm{p}}$ values drop suddenly/sharply thereafter. This behavior highlights a fundamental limitation in the current formulation of the SB method when applied to geometries with strong curvature and adverse pressure gradients. As established previously, the differential operator splitting in our current framework assumes a locally flat boundary plane. On a curved profile, shifting points along the normal vector alters their relative spatial distances compared to their distribution on the physical boundary. Without an explicit curvature correction applied to the Laplacian, these geometric distortions introduce severe numerical errors. Furthermore, the pressure correction step currently lacks the specialized stabilization required for the shifted points to keep the flow attached under adverse pressure gradients. Consequently, the simulated flow prematurely detaches past the maximum thickness of the wing, driving the observed drop in $C_{\mathrm{f}}$ and $C_{\mathrm{p}}$. The SB method performed exceptionally well for the flat plate precisely because that configuration lacked both surface curvature and adverse pressure gradients. Since extending the SB method to complex aerodynamic bodies requires fundamental modifications to the underlying meshfree operators (specifically, incorporating rigorous curvature corrections and an enhanced pressure correction step for the shifted points) along with methodological developments, this topic is deferred to our future work.

Finally, the overall lift and drag coefficients for both methods are summarized in Table~\ref{tab:naca0012_coefficients}. Experimental data \cite{ladson_effects_1988} for this configuration ($\text{Re} \approx 6 \times 10^6$, $0^\circ$ angle of attack) dictates a lift coefficient of $C_{\mathrm{L}} = 0$ and a drag coefficient between $C_{\mathrm{D}} = 0.006$ and $0.009$. Both methods yield a slightly non-zero lift coefficient due to the inherent asymmetry of the meshfree point cloud distribution discussed earlier. Despite the localized near-wall flow detachment, the SB method predicts $C_{\mathrm{D}}$ within the experimental range, since SB is unable to predict $C_{\mathrm{f}}$ and $C_{\mathrm{p}}$ the values of $C_{\mathrm{D}}$ and $C_{\mathrm{L}}$ cannot be relied on. The NBN method slightly overpredicts the drag ($C_{\mathrm{D}} = 0.0157$), which can be largely attributed to the geometric modifications at the trailing edge. Given the structural limitations of the SB method on curved profiles, we conclude that the NBN method currently provides a much more robust and reliable wall-treatment strategy for practical, complex geometries.

\begin{table}[!htb]
\centering
\caption{Comparison of Aerodynamic Coefficients for the NACA 0012 Airfoil at a $0^\circ$ Angle of Attack.}
\label{tab:naca0012_coefficients}
\begin{tabular}{lcc}
\hline
\textbf{Boundary Treatment} & \textbf{Lift Coefficient ($C_{\mathrm{L}}$)} & \textbf{Drag Coefficient ($C_{\mathrm{D}}$)} \\
\hline
Shifted boundary & 0.0102 & 0.0072 \\
Nearest-band neighbor & 0.0085 & 0.0157  \\
\hline
\end{tabular}
\end{table}

The evaluation of these wall-treatment strategies highlights for flat-surface applications, both the nearest-band neighbor (NBN) and shifted boundary (SB) methods successfully enforce wall functions. Though the SB method provides superior mathematical stability by maintaining a strictly constant evaluation distance. However, the transition to the 3D NACA 0012 airfoil revealed that the NBN method is significantly more robust and geometrically flexible for complex topologies. Conversely, the SB method struggled with the curved profile, exposing a critical vulnerability to both uncorrected geometric distortions and adverse pressure gradients. Shifting points along normal vectors without explicit curvature corrections introduces severe numerical diffusion, which, combined with the lack of a pressure correction, caused premature flow separation and a severe underprediction of aerodynamic forces. Extending the highly stable SB method to curved geometries will require fundamental advancements in curvature-aware differential operators, which remain a primary focus for future research.

\section{Conclusion}
In this paper, we addressed the challenges associated with implementing near-wall treatments in meshfree collocation methods. Specifically, we proposed two novel strategies for enforcing wall functions within a Lagrangian meshfree framework for URANS simulations. These techniques were developed, implemented, and systematically evaluated to establish a robust foundation for simulating high-Reynolds-number turbulent flows without the constraints of traditional grid generation.

The comparative analysis reveals clear performance differences between the evaluated approaches. The baseline closest neighbor method, while straightforward to implement, suffers from fundamental limitations due to non-uniform point selection. This inconsistency leads to gaps in wall function coverage, numerical instabilities, consistent overprediction of skin friction, and large instantaneous fluctuations, rendering it unsuitable for reliable wall treatment.

The nearest-band neighbor (NBN) method demonstrates substantial improvements over the closest neighbor approach. By selecting all interior points within a specified distance $\delta h$ from the wall, the NBN method ensures uniform and consistent wall function coverage. Numerical results indicate that a selection height of $\delta h = 0.5h$ provides an optimal balance between accuracy and computational efficiency. This Lagrangian approach necessitates dynamic point identification at each time step and introduces minor spatial fluctuations due to varying boundary distances. It proved to be highly robust and geometrically flexible. It successfully maintained attached flow and accurately predicted aerodynamic forces even on the complex, curved surface of the 3D NACA 0012 airfoil.

The shifted boundary (SB) method offers an alternative solution of virtually shifting boundary evaluation points by a fixed distance $\alpha h$ along the boundary normal vector toward the interior. This eliminates the need for dynamic point selection while enforcing strictly constant wall-normal distances. For flat surfaces, the optimal parameters ($\alpha h = 0.1h$ and $\beta h = 0.2h$) yielded smooth and stable results. Because the shift distance $\alpha h$ can be smaller than the selection distance $\delta h$ of the NBN method, the SB approach achieves valid $y^+$ ranges using significantly coarser point clouds. Although the SB boundary treatment operations are computationally more intensive per point, the ability to reduce the total simulation point count by approximately 50\% translates to significant net computational savings for large-scale, flat-surface simulations.

Regarding turbulence closures, the Spalart--Allmaras model provided the highest near-wall accuracy across the evaluated methods, followed by the $k-\omega$ and $k-\varepsilon$ models. Notably, the strictly fixed evaluation distances of the SB method enforced remarkable consistency across all three turbulence models, yielding nearly identical skin friction and velocity profiles on canonical flat plates.

However, validation against the 3D NACA 0012 profile revealed critical geometric constraints within the current SB formulation. While the NBN method successfully adapted to the airfoil's curvature, the SB method severely underpredicted skin friction and pressure coefficients past the thickest section of the wing. This divergence is driven by the lack of explicit curvature corrections when shifting points along normal vectors, which introduces severe numerical diffusion. In addition, the absence of a special treatment for correction pressure for shifted points at pressure-velocity coupling fails to withstand strong adverse pressure gradients.

Overall, this work establishes meshfree collocation methods as viable, highly adaptive alternatives to traditional mesh-based approaches for high-Reynolds-number turbulent flows. The NBN method currently stands as the most robust and geometrically flexible wall-treatment strategy for practical aerodynamic bodies. Future research will focus on extending the highly stable and efficient SB method to complex geometries. This will require the development of curvature-aware differential operators to correct normal-vector shifting distortions, as well as enhanced pressure correction algorithms to accurately capture flow separation and adverse pressure gradients.

\bibliographystyle{elsarticle-num} 
\bibliography{bibliography}

\end{document}